\documentclass{elsarticle}

\usepackage{lineno,hyperref}
\usepackage{amsmath,amssymb,amsfonts}
\usepackage[T1]{fontenc}
\usepackage[normalem]{ulem}
\usepackage{placeins}
\modulolinenumbers[5]

\usepackage{comment}

\usepackage{amsmath, upgreek}

\usepackage{caption}
\usepackage{subcaption}
\usepackage{multirow} 
\usepackage{bm}
\usepackage{xcolor}
\usepackage{multirow}
\usepackage{booktabs}

\makeatletter
\def\ps@pprintTitle{%
 \let\@oddhead\@empty
 \let\@evenhead\@empty
 \def\@oddfoot{\centerline{\thepage}}%
 \let\@evenfoot\@oddfoot}
\makeatother

\usepackage{numcompress}

\begin{document}

\begin{frontmatter}

\title{Self-organising maps in the analysis of strains of human abdominal wall to identify areas of similar mechanical behaviour
}



\author[wilis]{Mateusz Troka} 
\author[wilis]{Katarzyna Szepietowska}

\author[wilis]{Izabela Lubowiecka\corref{mycorrespondingauthor}}
\cortext[mycorrespondingauthor]{Corresponding author}
\ead{lubow@pg.edu.pl}

\address[wilis]{Faculty of Civil and Environmental Engineering, Gda\'nsk University of Technology, Gda\'nsk, Poland}

\begin{abstract}

The study refers to the application of a type of artificial neural network called the Self-Organising Map (SOM) for the identification of areas of the human abdominal wall that perform in a similar mechanical way. The research was based on data acquired during \textit{in vivo}  tests using the digital image correlation technique (DIC). The mechanical behaviour of the human abdominal wall was analysed during changing intra-abdominal pressure.
 SOM  allowed us to study simultaneously three variables in four time steps. The  variables referred to the principal strains and their directions. SOM classified  into clusters all the abdominal surface data points that behaved similarly in accordance with the 12 variables. The analysis of the clusters provided a better insight into  abdominal wall deformation  and its evolution under pressure than when observing a single mechanical variable.
The presented results may provide a better understanding of the mechanics of the living human abdominal wall. It might be particularly useful in the designing of surgical meshes for the treatment of abdominal hernias, which would 
 be mechanically compatible with identified regions of the human anterior abdominal wall, and possibly open the way for patient-specific implants.

\end{abstract}


\end{frontmatter}


\section{Introduction}

The principal motivation for this study was the problem of recurring abdominal hernias and the need to design medical implants that would effectively prevent this from happening. According to  \cite{Deeken2017}, the hernia recurrences affect around 30\% of patients. The same authors point out that most recurrences occur at the implant-tissue interface, indicating a gap in understanding how a mechanical mismatch between hernia repair materials and host tissue contribute to failure. In other words, there is a need to design surgical implants that are mechanically compatible with abdominal tissues \citep{bilsel2012search}. Despite the advancement of hernia implants over last years, searching for an ideal mesh remains a current problem \cite{ najm2023review}.

There are many surgical implants available on the medical market with quite a variety of material properties. The materials and techniques are extensively discussed in the literature (see e.g., \cite{Liu2021} where many more contributions are mentioned). Similarly, great diversity can be found in the mechanical performance of living tissues, and therefore also in the abdominal walls of different human subjects, both on the experimental  e.g., \cite{in_vivo_abdomen} and the computational level (see e.g., \cite{Karrech2023}).  
 Bearing in mind that different zones of the abdominal wall exhibit different deformation ranges 
 \cite{szymczak2012investigation}, and that there are also various commercial implants with various strain ranges,  \cite{Tomaszewska2013}, it may be assumed that the problem results from a lack of information regarding the correct selection of an implant for a given abdominal hernia location and patient. However,  there are no significant indications as to which implant should be applied to the given hernia location and also to the given patient. While stating the importance of location in treatment strategy \citep{bittner2014guidelines} and the current hernia classification, these significant indications are not included in surgical guidelines  \citep{muysoms2009classification}.

The ideal solution would be to design a specific implant for the specific hernia and patient, in accordance with what is today known as personalised/patient specific medicine. This is challenging because of the anisotropy and non-linearity in the mechanical behaviour of both surgical meshes and human abdominal wall tissues \cite{Deeken2017} as well as the lack of a validated model of the human living abdominal wall with an implanted surgical mesh.
 
Szymczak et al. \cite{ szymczak2017two} proposed a two-criteria optimisation procedure including forces calculated in the implant-tissue joints and the implant deflection in the objective function.  The optimisation was performed  for five different areas of the human abdominal wall to chose the best implant for each of them. Lubowiecka \cite{ Lubowiecka2015} and Lubowiecka et al. \cite{lubowiecka2016preliminary} show that various  optimal implant orientations  can be found for every zone. The influence of location on the suitable orientation was confirmed on a  rabbit model involving two different hernia locations \citep{he2023effect}. He et al. \cite{he2020numerical} have demonstrated by computer simulation that various implant elasticities may be appropriate for two different defect locations. The aforementioned studies investigating selected hernia locations did not consider the variability of living tissues in the human population, where \textit{in vivo} experiments could provide appropriate data. Therefore, the selection of a mechanically compatible surgical mesh and its orientation within the entire human living abdominal wall based  on experimental  data  collected \textit{in vivo}  is yet to be addressed.

\begin{figure}[ht]\centering
    \includegraphics[width=100mm]{ 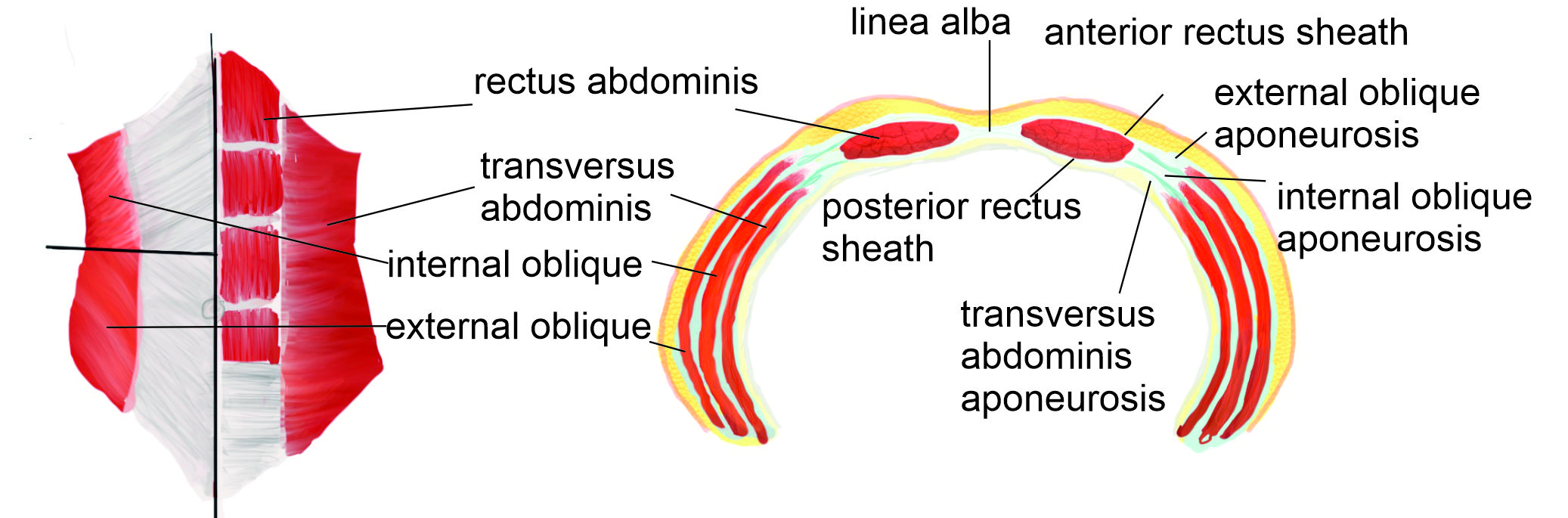}
    \caption{Abdominal wall scheme: anterior view with some layers of muscles partially removed (on the left), cross-section in region superior to arcuate line (on the right)}
    \label{Abdominal wall}
\end{figure}

 In addition, it is necessary to find the material parameters of the living human abdominal wall in order  to design the most mechanically compatible implant. To achieve that basing only on \textit{in vivo} non-invasive measurements is challenging, due to the heterogeneity, anisotropy and nonlinear mechanical behaviour of the abdominal wall (cf. assumed isotropic properties in \cite{simon2017towards}). Especially heterogeneity resulting from different muscle layers of different mechanical characteristics \cite{kriener2023mechanical}  in different parts of the abdominal wall (see Figure \ref{Abdominal wall}) complicates the process of designing appropriate implants. Indicating few similarly behaving regions can simplify the identification procedure by reducing the number of unknown parameters from one set in each data point to one set in a region. A priori knowledge of the zones with the same mechanical properties will reduce number of the material model parameters to be identified for the entire abdominal wall. Especially, since it is known that it is a structure with many layers of muscles and connective tissues that cannot be easily described in mechanical terms (see \cite{cmbbe2020}).
 
The main goal of this paper is to identify regions of the living human abdominal wall that may perform mechanically in a similar way under intra-abdominal pressure. The study is based on several human subjects, which also shows the diversity among analysed patients. 
Our approach is based on full-field data acquired with the digital image correlation technique (DIC) during \textit{in vivo} tests on patients undergoing peritoneal dialysis, described in detail in \cite{szepietowska2023fullfield}. However, an analysis based solely on experimental data is not easy because it concerns mechanical behaviour  under various conditions, e.g. during breathing, involving abdominal muscles \cite{MIKOLAJOWSKI2022106936}, and changes of intra-abdominal pressure as the abdominal cavity fills with fluid. Therefore, the question arises as to which quantities and which deformation states should be taken into account in order to identify similarly behaving regions. It is essential to analyse more than one state and more than one mechanical quantity (e.g., principal strain), difficult as it may be. The solution to analysing such a multidimensional data structure can be found in machine learning (ML). ML has indeed been used in predictive medicine \cite{Paramasivam2014}. ML has also already been applied to the topic of the human abdomen and hernias (see e.g. \cite{Kallinowski2021} in the subject of incisional hernia repair and hernia recurrences in \cite{Abbas2022}). However, to the authors' knowledge, no in-depth research has been conducted into identifying the  abdominal zones of similar properties with the use of ML. 

We propose to use an unsupervised machine learning technique called Self-Organising Maps (SOM) \cite{kohonen97} to cluster data and thus discover regions in the abdominal wall that present similar mechanical responses during the experiment. This method has been selected due to its ability to reduce the dimensionality of the analysed problem. SOM was applied before to assist the analysis of medical data, for example to identify malignant and non-malignant regions in liquid biopsies of thyroid nodules \cite{nobile2023unsupervised}. Another application was to analyse viscoelastic properties obtained by Atomic Force Microscopy on the breast cancer cell in order to compare cancer treatment methods \cite{weber2023application}. In this study we apply SOM to identify the clusters using full-field DIC deformation data acquired from non invasive \textit{in vivo} measurements in different time steps of the experiment. These clusters are groups of points on the abdominal walls of the human subjects based on data consisting of principal strains and principal strain directions. This novel approach using SOM enables us to analyse the strain field and consider deformation states in different time steps that correspond to respiration and the change of intra-abdominal pressure at different stages of peritoneal dialysis. Clustering is based here on the strain field because strain is a measurable quantity that can be easily obtained from a non-invasive full-field measurement. Moreover, the analysis of strain in different tissue regions may provide a valuable information in biomechanics, like it was shown e.g. in case of lungs strain analysis in \cite{nelson2023diseased}. In the literature, the strain field was also used to compare a healthy abdominal wall with one with an implant \cite{podwojewski2014mechanical} or to compare different methods of closing the abdominal wall \cite{le2020differences}.   

In this paper we propose an approach to acquire new information that would facilitate the identification of the mechanical properties of the human anterior abdominal wall with a view to improve surgical implant design and implementation. The paper presents new approach to full-field strain data analysis including various deformation states. This is an extension of our original idea presented in a conference paper \cite{lubowieckamechanical} in 2022.  {Recently Nguyen and Lejeune have proposed methodology of clustering strain, displacement  fields by unsupervised machine learning methods (k-means clustering, spectral clustering, iForest clustering and One-class Support Vector Machine)  \cite{nguyen2023segmenting}. Their study also raises the issue of analysing heterogeneous, highly deformable materials (such as soft tissues) and shows, however using artificially generated data, the effectiveness and limitations of such an approach.}

 {To sum up, SOM is used here to analyse and cluster data representing mechanical behaviour of the living human abdominal wall, which is relatively complex heterogeneous structure composed of tissues exhibiting anisotropic nonlinear behaviour with additional active response component due to muscles. }

\section{Materials and Methods}

\subsection{Experimental data}

\begin{figure}[ht]\centering
    \includegraphics[width=100mm]{ 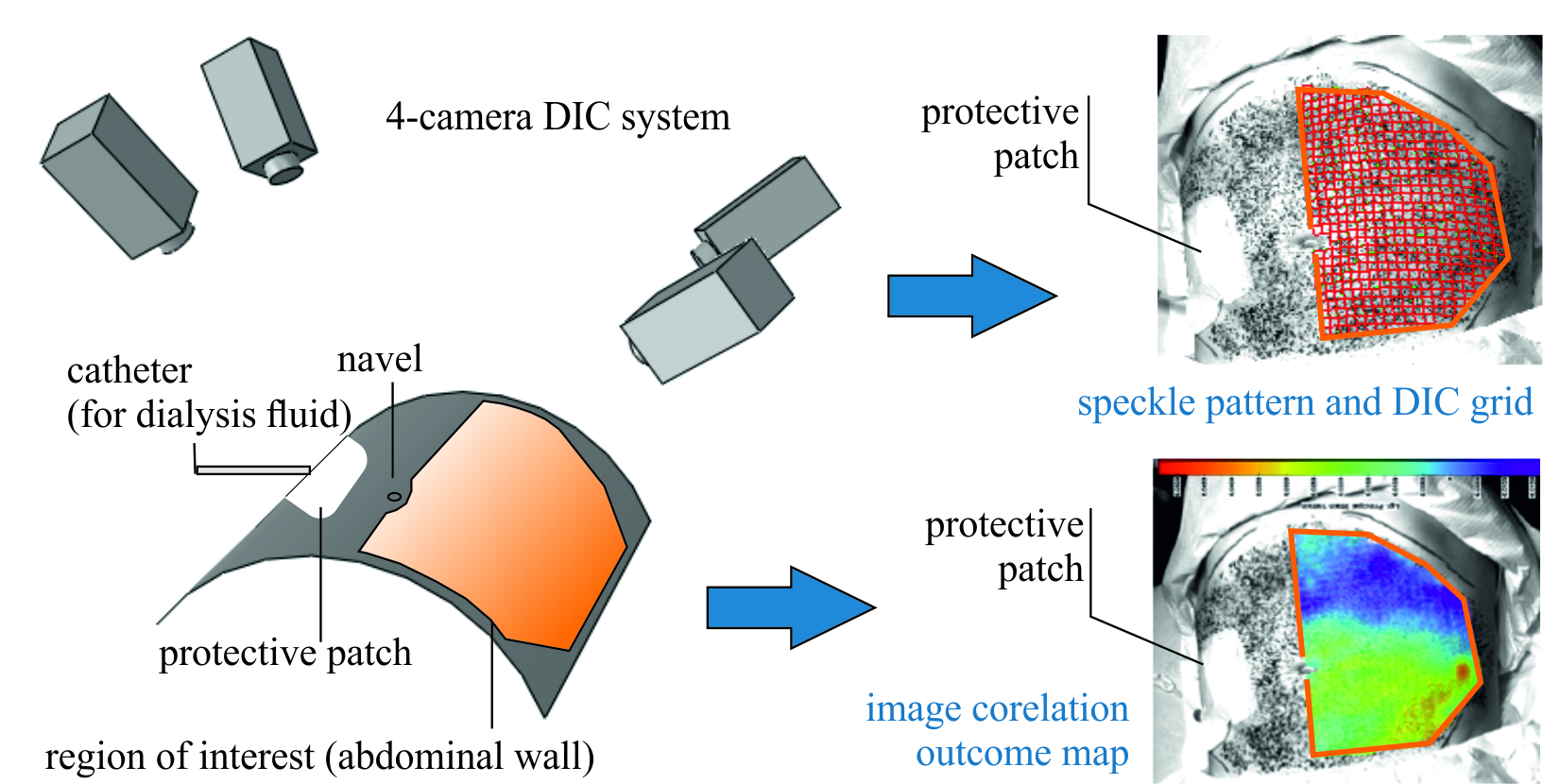}
    \caption{Input data acquisition: 4-camera Digital Image Correlation system (left); speckle pattern on abdominal wall (top right); outcome map resulting from image correlation (bottom right);  {DIC images of the abdominal wall are adapted from \cite{szepietowska2023fullfield}}}
    \label{DIC_eksp3}
\end{figure}

The study is based on experimental data acquired  {\textit{in vivo} on humans} and described in detail in \cite{szepietowska2023fullfield}. Twelve subjects, eight male and four female,  {suffering from end-stage kidney disease,} have been tested during peritoneal dialysis (PD) (see e.g., \cite{in_vivo_abdomen}),  {the procedure they undergo regularly. Only one of them has an operated hernia in history. Thus, the experiments have been conducted on subjects who did not have abdominal wall insufficiency, which is even more important when designing abdominal wall reconstructive implants}. 
  Four-Camera Digital Image Correlation system Dantec Q-400 has been applied to measure deformation of the subject abdominal wall during the introduction of the dialysis fluid,  {which lasts on average 7-8 minutes. The same amount of dialysis fluid was introduced in all patients, which resulted in different intraabdominal pressure (15.2$\pm$ 4 cmH$_2$O)}. The system was equipped with 4 digital cameras VCXU-23M with 2.3 Mpx matrix (resolution: 1920 x 1200 px) and lenses VS-1620HV (16 mm f/2.0–16). The  full-field measurements have been used to calculate 3D fields of displacements, strains and principal directions with the use of commercial correlation software Istra 4.7.6.580. These data, acquired as shown in Figure \ref{DIC_eksp3}, were later used as an input to the Self-Organising Maps.

The experiments were fully non-invasive and the measurements were contactless. All the subjects submitted a consent to participate in the study under a protocol approved by the Ethics Committee of Medical University of Gda\'nsk (NKBBN 314/2018).  {Details of the experiments and experimentally obtained data on which this study is based are extensively described in \cite{szepietowska2023fullfield}). Here, a  small sample of them is included to show the character of the strain fields.} 

\begin{figure}[ht!]\centering

   \includegraphics[width=0.9\textwidth]{ 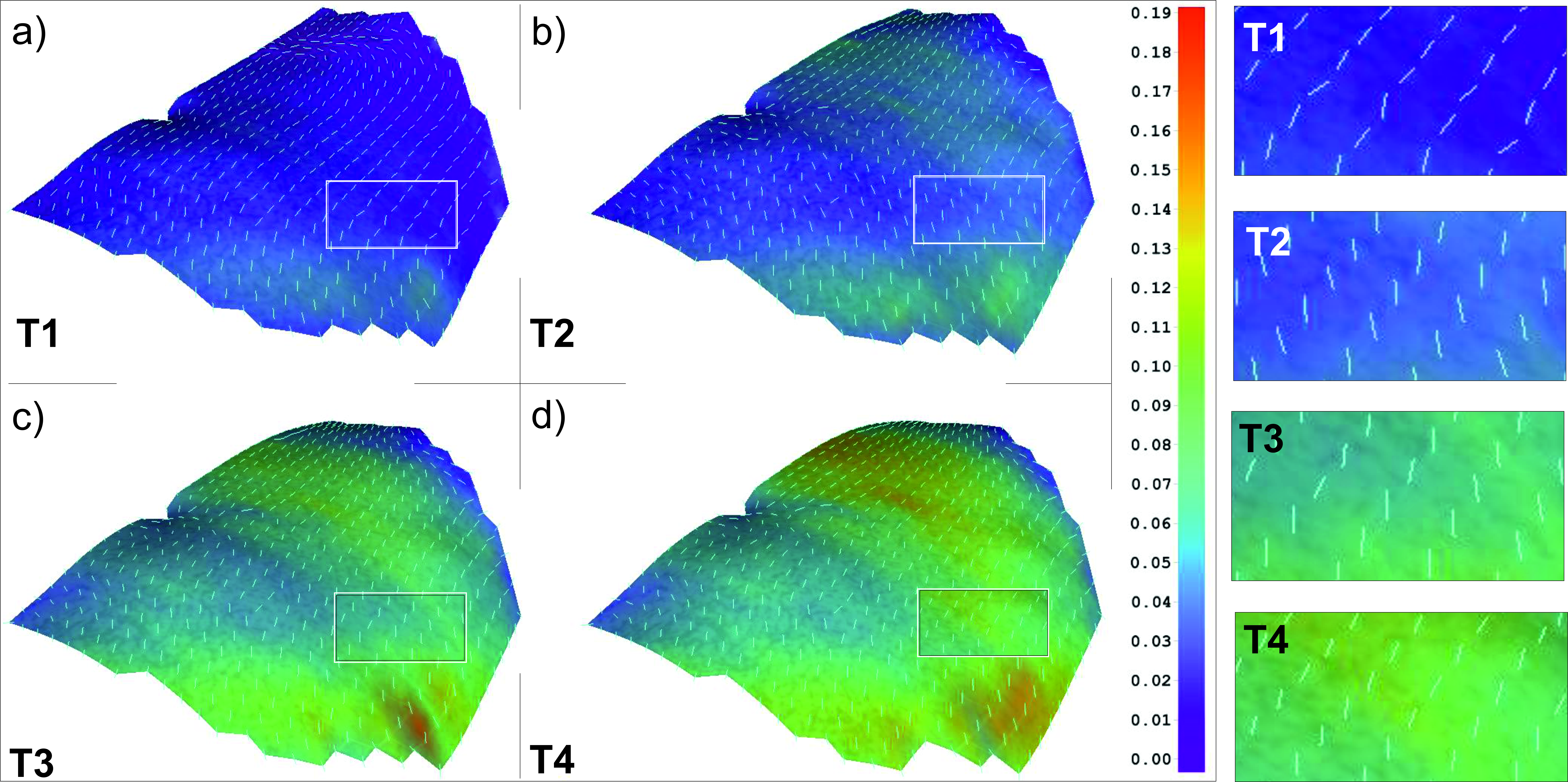}
    \caption{ Map of the principal Lagrangian strain $\varepsilon_1$  { (shown by colorscale in [-])} and its direction  {(shown as arrows)} in stages a) T1 b) T2 c) T3 d) T4 of subject D8; with zoomed marked parts of the abdominal surface with the directions of principal strains in T1--T4 on the right hand side  {(images of strain field adapted from \cite{szepietowska2023fullfield})}}
    \label{fig_dic8strain}
\end{figure}

As shown in Figure \ref{fig_dic8strain}, the values of strains in a tested abdominal wall change with the pressure level. This is normal in elastic material structures. 
  {However, it is interesting to notice that directions of the principal strains change during experiment, which may be related to the contribution of the active response of abdominal wall during breathing.} What is more, the principal strain directions on a given abdominal wall are not uniform. There are areas where the dominating principal direction changes together with the loading in  time steps T1, T2, T3 and T4 and there are areas that retain the same principal direction. Due to its complexity, different parts of the abdominal wall behave differently under pressure.

\subsection{Data analysis using Self-organising maps}

SOM is a method of data analysis in which, the nonlinear relations of multidimensional data and similarity in a dataset are visualised using a map of neurons. 
The map consists a single layer of neurons (nodes) set up in a two-dimensional grid \cite{kohonen97}. It can be interpreted as a projection of multidimensional data vectors on a 2D Euclidean space. The distances between the projections on the 2D plane are approximately the same as the distances in the original data input of the high dimensional space.
The similarity of the input data has the result of data points being placed close to each other. At the same time dissimilar data points are placed further apart \cite{kohonen97}.
Since SOM is an unsupervised learning algorithm, it does not require training data or a specifically expected output. Therefore, the analysis can be directly applied to the input data and the expected similar features can be extracted by grouping the data into clusters \cite{kohonen97}.

\subsubsection{SOM Algorithm}
\label{SOM_algorithm}
Let $\bm{x} \in \mathbb{R}^n$ be the $n$-dimensional  input vector which is selected from the input space, $\bm{x}= [x_1,x_2,\dots,x_n]$. Each neuron is a $n$-dimensional weight vector where $n$ is equal to the dimension of the input vectors. Thus the weight vector of neuron $j$ is described as $\bm{w}_j = [w_{j1},w_{j2},\dots,w_{jn}]$.

The SOM training algorithm moves the weight vectors so that they span across the data cloud, and the map of neurons is organised so that the neighbouring neurons on the grid have similar weight vectors \cite{matlabkohonen, toolbox_juha}. The training process is based on a criterion of minimising Euclidean distance, which enables finding the best matching unit (BMU), a wining neuron for an input vector in the output map.

Thus, BMU here denoted as $c$ is the node which minimises the Euclidean distance to input vector $\bm{x}$ according to the following equation (\ref{eq_1}):
\begin{equation}
   ||\bm{x}-\bm{w}_c||=\min_{j}(||\bm{x}-\bm{w}_j||)
   \label{eq_1}
\end{equation}

The weights of neighbouring nodes that are topographically adjacent are then updated according to the formula (\ref{eq_2}):
\begin{equation}
\bm{w}_j(t + 1) = \bm{w}_j(t)+h_{cj}(t)[\bm{x}(t)-\bm{w}_j(t)]
 \label{eq_2}
\end{equation}
distributing similar data locally around the BMU, where $t$ is the iteration number, $\bm{w}_j(t)$ is the weight vector, $\bm{w}_j(t+1)$ is updated weight vector and $h_{cj}(t)$ is a neighbourhood function. The formula (\ref{eq_2}) describes a sequential training algorithm. In this paper, however, we use a batch training algorithm, due to its better performance \cite{matlabkohonen}. In the case of the batch training algorithm, the whole dataset is presented to the map before any adjustments and then the new weight vectors are calculated as: 
\begin{equation}
\bm{w}_j(t + 1) =\frac{\sum_{i=1}^{n} h_{cj}(t) \bm{x}_i}{\sum_{i=1}^{n} h_{cj}(t)}
 \label{eq_3}
\end{equation}

Within the SOM analysis, an input data item $\bm{x}_i$ is presented to a set of nodes, of which the $BMU$, matches best with $\bm{x}_i$. All nodes that lie in the neighbourhood of a $BMU$ in the map will be updated during training iterations and finally match better with $\bm{x}_i$. 
 {The training is done in two phases: rough training with (initial) neighbourhood radius and large (initial) learning rate, and fine-tuning with small radius and learning rate. Either default Toolbox setting (based on a function of dataset size) or changing these parameters did not yield favourable results, thus arbitrary settings for all data sets were used, namely 300 iterations for rough training and 100 iterations for fine-tuning.}

For the SOM modelling, visualisation and clustering, MATLAB Toolbox was used \cite{toolbox_juha}.  {The number of neurons in the output map  is an important factor because it influences the final quality and topography of the SOM. Manual trial and error method suggested by \cite{matlabkohonen} did not bring satisfactory results. Therefore, a toolbox function was used to determine a sensible map grid size based on a heuristic formula for the total number of neurons, $N=5\sqrt{M}$ (see \cite{toolbox_juha}), where \textit{M} was the number of samples in the dataset. The ratios of the side lengths were based on the ratio between two biggest eigenvalues of the covariance matrix of the input data. Map sizes for all subjects are shown in Table \ref{Table_msize}}.


\begin{table}[ht]
\caption{ {SOM map sizes: numbers of rows $n_{row}$ and numbers of columns $n_{col}$ of each subject's map }}
\begin{tabular}
{ccccccccccccc}
\hline
 {No} &  {D1} &  {D2} &  {D3} &  { D4} &  {D5} &  { D6 } &  {D7} &  {D8 } &  { D9 } &  {D10 } &  {D11 } &  { D12 } \\
\hline
  {$n_{row}$} &  { 15 } &  { 14 } &  { 15  } &  { 16 } &  { 14 } &  {  16 } &  { 18 } &  { 14 } &  { 14 } &  { 13} &  { 13} &  { 12} \\
 {$n_{col}$} &  { 9 } &  { 8} &  { 8 } &  { 9 } &  { 8 } &  { 7 } &  { 8 } &  { 8 } &  { 10 } &  { 9 } &  { 8 } &  { 9} \\
\hline
\end{tabular}
 \label{Table_msize}
\end{table} 

Visualising SOM results is possible with the use of the U-matrix (unified distance) map, which shows the cluster structure of the map. Similar outputs are grouped in clusters marked by uniform areas of low values. The high values on the map indicate a cluster border. U-matrix can be presented along with component planes that show the dimensionless weighted average values for one variable of the input data (each value in each map unit). Then, a Toolbox function is used to find clusters based on the local minima of the U-matrix, and allocate each map unit to the clusters.
The best results of clustering was provided by a centroid (centre of the group of cases) cluster forming method that determines the average distance between cluster units. Other features of the SOM visualisation may be found in \cite{TMJ}.

\subsubsection{Input data}

\begin{figure}
 \centering
  \includegraphics[width=0.9\textwidth]{ 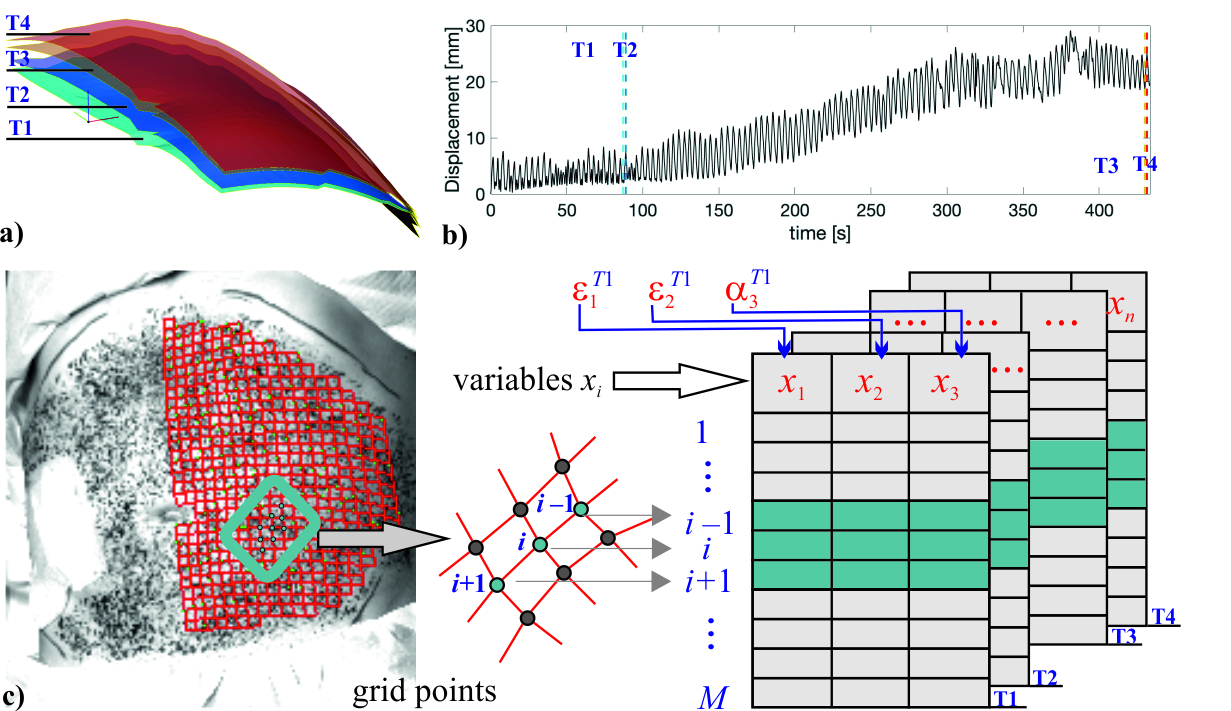}
  \caption{Input to Self-Organising Map: a) deformation states of the abdominal wall in chosen time steps (T1--T4); b) displacement of point on abdominal wall in time, with  time steps T1--T4 marked ( {adapted from \cite{szepietowska2023fullfield}}); c) dataset and $n$-dimensional input vector to SOM ( {abdomen image adapted from \cite{szepietowska2023fullfield}})} 
  \label{brzuchy_som_grid_T1T4}
\end{figure}

Green-Lagrange strain tensor, described by formula (\ref{GL_strain_tensor})
\begin{equation}
    \bm{E}=\frac{1}{2}\left( \bm{F}^\top \bm{F} - \bm{I}\right),
    \label{GL_strain_tensor}
\end{equation}
where $\bm{F}$ stands for deformation gradient, used as representation of the strain field.  
The principal strains, $\varepsilon_1$ and $\varepsilon_2$, and directions represented by  angle $\alpha$ in  grid points computed at four time steps, were used as entries to the SOM analysis of each subject. Here $\alpha$ is the angle of the first principal direction to the transverse axis of the abdominal wall, so $\alpha=0$ means that the principal direction of $\varepsilon_1$ is aligned to the transverse direction of the abdominal wall.
The four time steps representing four different deformation states of the abdominal wall were selected for exhalation and inhalation in the early \textit{ca.} for the initial 20\% of the duration  {of the procedure of peritoneal dialysis,} denoted as T1 and T2, respectively) and final stages (T3 for exhalation and T4 for inhalation) of the experiment (Figure \ref{brzuchy_som_grid_T1T4}a,b).  {The steps T1 and T2 were selected at the beginning of the dialysis when a little amount of the fluid was introduced that would allow one to observe some non-zero deformations to compare with the state of the filled abdominal cavity (T3 and T4) and to observe the effect of breathing.}  {The reference timestep T0 is the time of first subject’s exhalation during experiment. Therefore, the reference configuration is the one corresponding to the drained abdominal wall at exhalation. The abdominal wall deformation in timesteps T1, T2, T3 and T4 is calculated in reference to T0.} Only half of the abdominal wall was considered, as in \cite{szepietowska2023fullfield}.

Then the dataset of the entries for the four time points, structured as in Figure  \ref{brzuchy_som_grid_T1T4}c, is applied as 12-dimensional input vector  $\bm{x}$= [$\varepsilon^{\textrm{T}1}_{1}$, $\varepsilon^{\textrm{T}1}_{2}$, $\alpha^{\textrm{T}1}$,$\varepsilon^{\textrm{T}2}_{1}$, $\varepsilon^{\textrm{T}2}_{2}$,$\alpha^{\textrm{T}2}$,  $\varepsilon^{\textrm{T}3}_{1}$, $\varepsilon^{\textrm{T}3}_{2}$, $\alpha^{\textrm{T}3}$, $\varepsilon^{\textrm{T}4}_{1}$, $\varepsilon^{\textrm{T}4}_{2}$, $\alpha^{\textrm{T}4}$], where the superscript denotes the time step. 

The vector length $M$ of each subject refers to the number of grid points generated on the tested surface. Therefore, the value of $M$ can vary.  In this way, the entire analysed area of the abdominal wall is represented in the dataset. Thus, the constructed input vector allows for the simultaneous analysis of  abdominal wall mechanics during inhalation and exhalation and with  {abdominal cavity fully} filled and  {at the initial stage of the dialysis meaning with little liquid inside}.

In our study, SOM reduces the 12-dimensional data space to a two-dimensional result space by clustering the data points on a 2D map expressed in U-matrix. Clusters, as SOM output, represent grouped points whose mechanical behaviour is similar. 
After grouping, these points can be mapped onto the abdominal wall, identifying  areas characterised by similar behaviour due to Lagrangian strains.

\subsubsection{SOM results evaluation}
 \label{SOM_evaluation}
 
 {To evaluate SOM quality of a given resultant map we can use two dimensionless measures. Average quantization error gives map resolution and shows average distance between each data vector and its BMU. Topographic error measures topology preservation and is the ratio of data vectors for which first and second BMUs are not adjacent (neighbouring units) on resultant map to the total number of vectors \cite{matlabkohonen}. But,  both measures give the best results when the map has over-fitted the data. This may happen when the number of map units is as large or larger dataset. Therefore, direct cluster's quality assessment might give more suited information about quality.} 

While many data mining techniques search and detect similarities in data, one should also evaluate results provided by clustering algorithms to avoid an incorrect assignation of data objects to clusters. The silhouette value is a common index for visual evaluation of clustering, introduced by  \cite{Kaufman1990}. The silhouette value $s_i$ for the $i^{th}$ data point can be defined by the formula
\begin{equation}
   s_i=\frac{{b_i}-{a_i}}{\max({a_i},{b_i})},
   \label{eq_4}
\end{equation}
where $a_i$ is the average distance from the $i^{th}$ data point to the other data points in the same cluster; $b_i$ is the minimum average distance from the $i^{th}$ data point to points in other clusters.  {Thus, the silhouette value can vary from -1 to 1. When $s_i$ has a value close to 1, the object is said to be 'well classified'. Conversely, when $s_i$ is close to -1 the opposite happens, and the object is believed to be 'incorrectly classified'. When the index is near zero it is an object that lies in between clusters, and it is not clear where it should be classified. In the silhouette plot the $s_i$ are graphed as horizontal bars \cite{Everitt2011}. Following this, the average of the $s_i$ over the entire set can be examined. According to \cite{Kaufman1990}, the averaged silhouette widths for the entire dataset should be greater than 0.5. The average widths below 0.2 indicate a lack of substantial cluster structure and thus an additional methods should be used on this dataset}. 

\section{ {Results}}
\subsection{Identification of areas on abdominal wall with similar deformation state under pressure}

\begin{figure}[ht!]\centering
    \includegraphics[width=\textwidth]{ 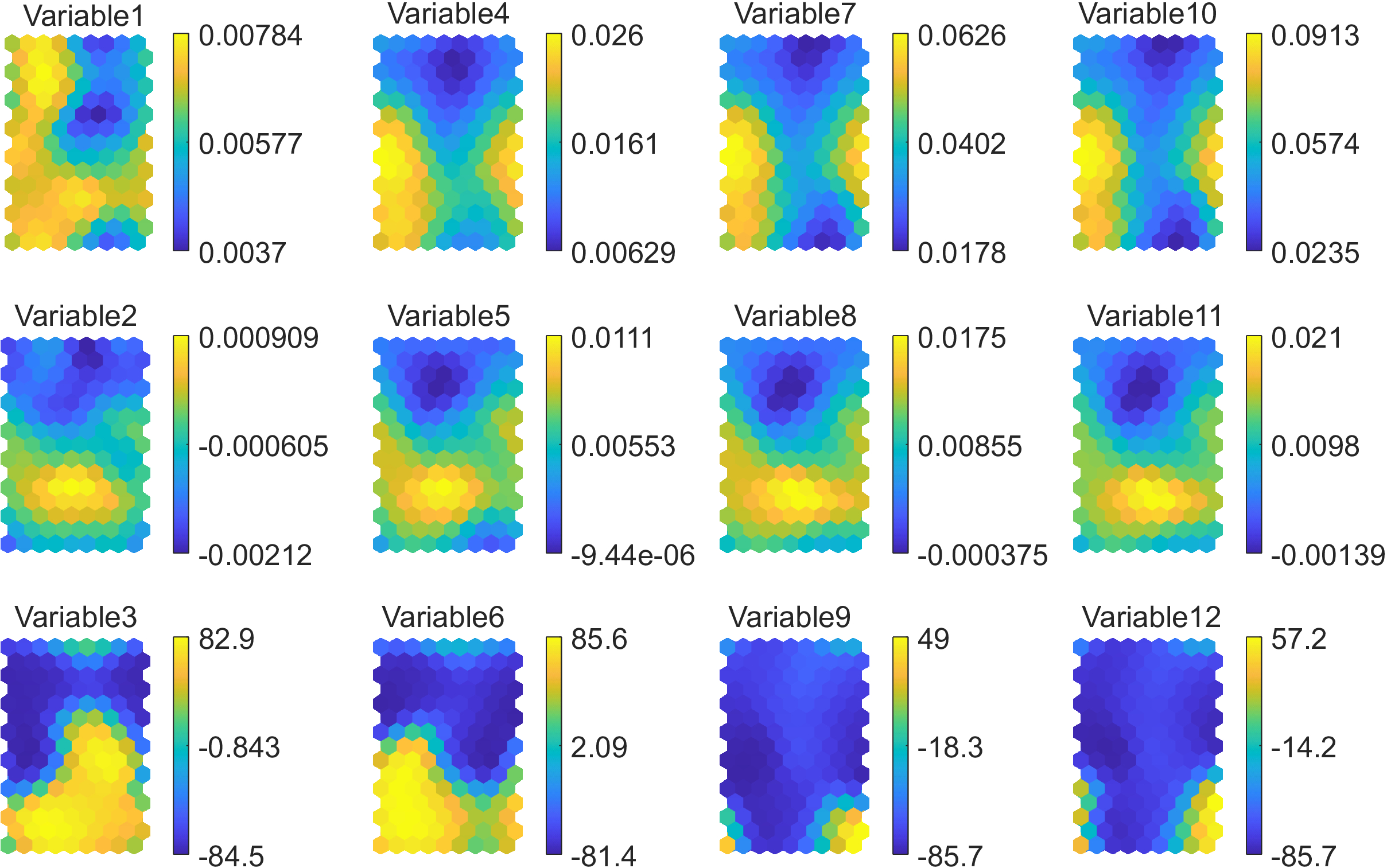}
    \caption{Component planes for subject D1. The following variables are the elements of  input vector $\bm{x}$. The component planes of all the variables are shown, those in the first column correspond to time step T1, in the second T2, in the third T3 and in the fourth T4. The first row corresponds to $\varepsilon_1$, the second row to $\varepsilon_2$ and the third one to $\alpha$.}
    \label{comp_planes_DIC1}
\end{figure}

 {The results of SOM analysis can be visualised by individual component planes, which are the arrays of scalar values representing the $i^{th}$ components of all the weight vectors $\bm{w}_j$ (projections of a single variable on the neuron map). By plotting the component planes for all dimensions, all information about the neuron vectors can be displayed. They reveal how impact-full data samples of which variable were on the map of neurons. Exemplary component planes, for subject D1, are shown in Figure \ref{comp_planes_DIC1}. Colorbar scale is denormalised and is showing the range of values for neurons of the respective maps.}  

\begin{figure}[ht]\centering
\begin{subfigure}[t]{0.45\textwidth}  
\setcounter{subfigure}{0} 
    \includegraphics[width=\textwidth]{ 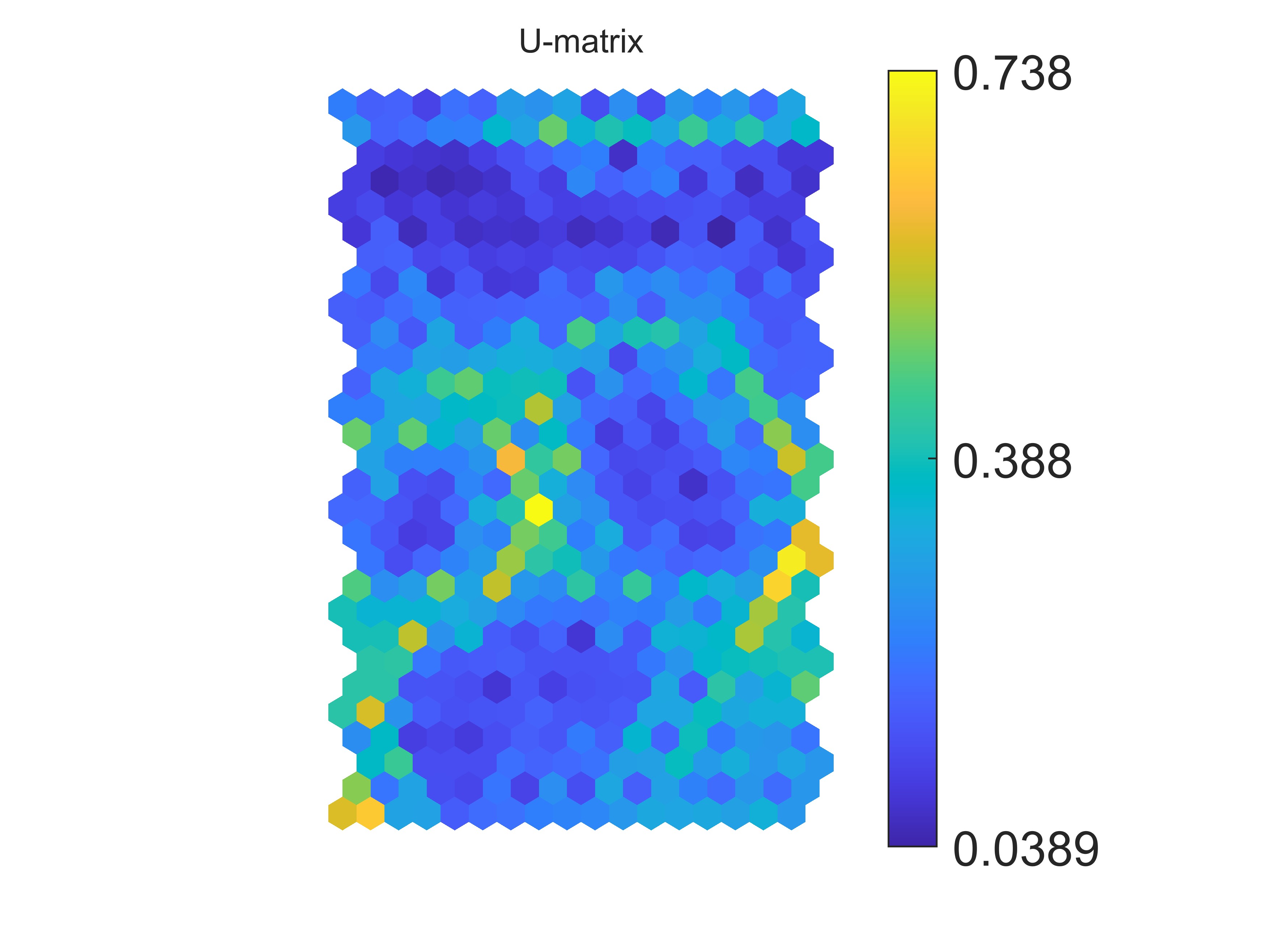}
    \caption{U-matrix}
    \label{boxplot_max}
    \end{subfigure}
\begin{subfigure}[t]{0.45\textwidth}  
\setcounter{subfigure}{3} 
    \includegraphics[width=\textwidth]{ 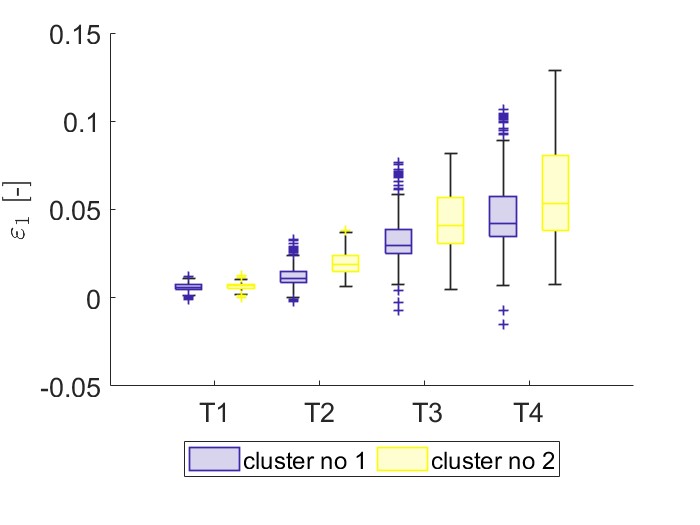}
   \caption{ Boxplot of the maximum principal strain  $\varepsilon_1$ for each cluster}
    \end{subfigure}

\begin{subfigure}[t]{0.45\textwidth}  
\setcounter{subfigure}{1} 
    \includegraphics[width=\textwidth]{ 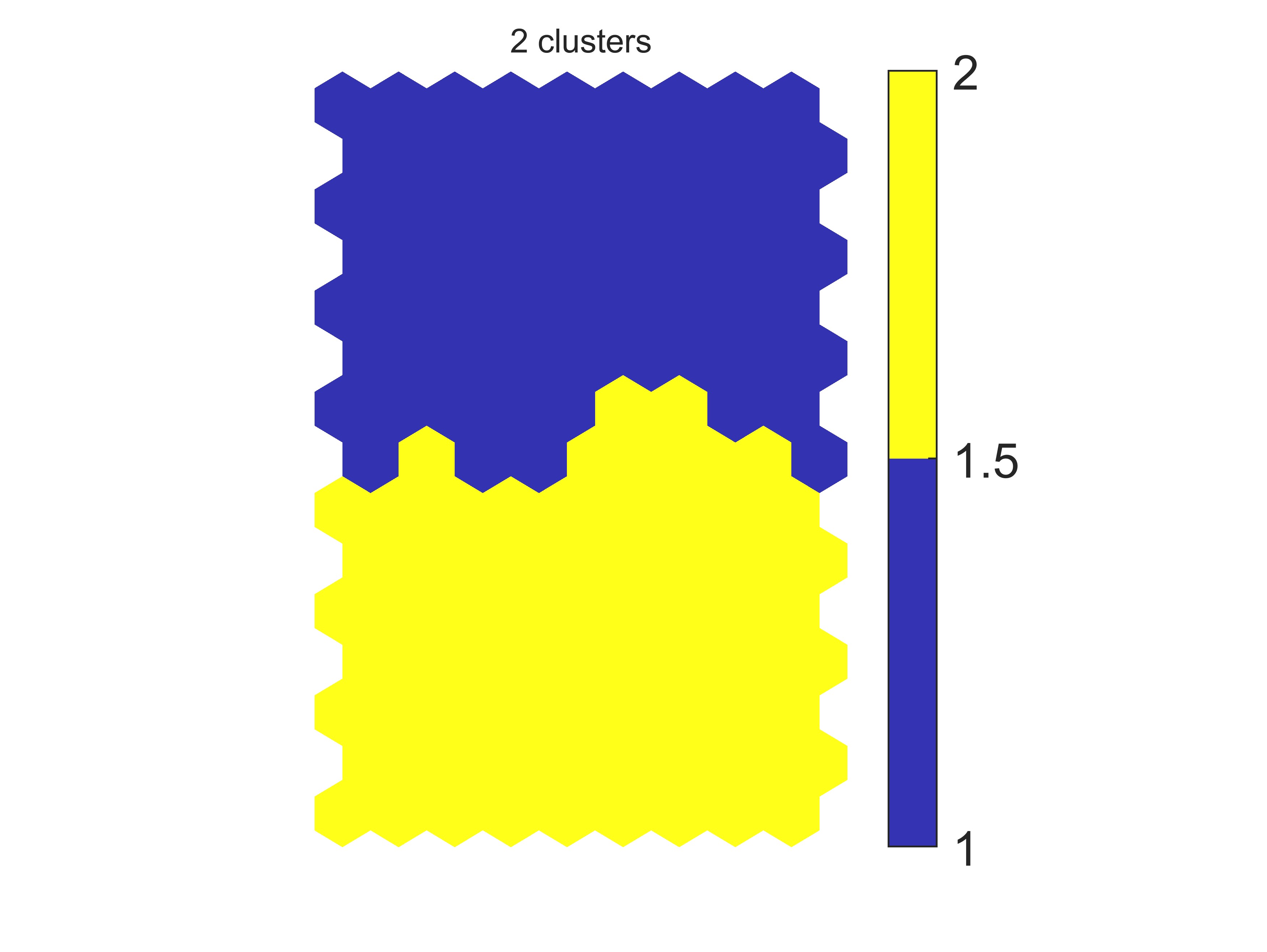}
    \caption{SOM clustering}
    \label{boxplot_max}
\end{subfigure}
\begin{subfigure}[t]{0.45\textwidth}  
\setcounter{subfigure}{4} 
    \includegraphics[width=\textwidth]{ 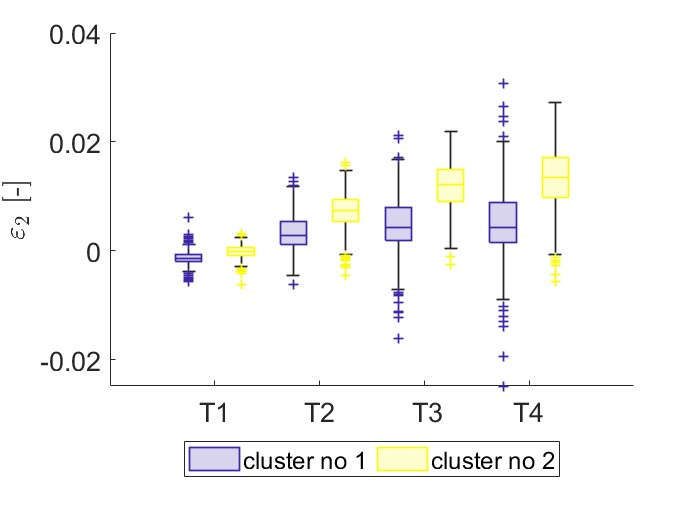}
    \caption{ Boxplot of the minimum principal strain  $\varepsilon_2$ for each cluster}
    \label{boxplot_min}
    \end{subfigure}
\begin{subfigure}[t]{0.45\textwidth}  
\setcounter{subfigure}{2}
    \includegraphics[width=\textwidth]{ 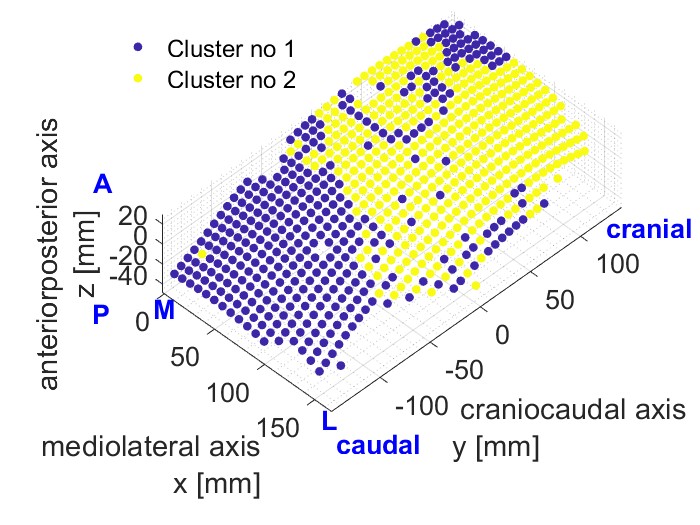}
    \caption{ {Clusters found by SOM marked on the abdominal wall surface}}
    \label{boxplot_max}
\end{subfigure}
\begin{subfigure}[t]{0.45\textwidth}  
\setcounter{subfigure}{5}
    \includegraphics[width=\textwidth]{ 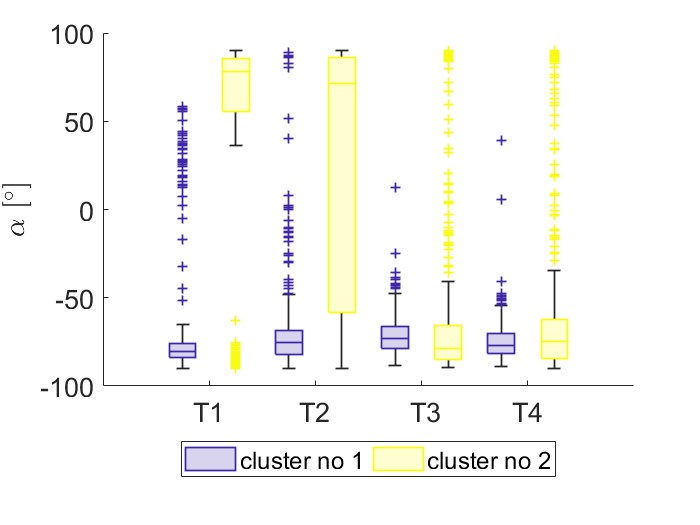}
    \caption{Boxplot of the principal direction $\alpha$ for each cluster  }
    \end{subfigure}
  
    \caption{Cluster results obtained by SOM in case of subject D1 (male, 78 years old, BMI 30.1 kg/m$^2$, intra-abdominal pressure 11 cmH$_2$O) }\label{fig_som_result_dic1}
\end{figure}

\begin{figure}[ht]\centering
\begin{subfigure}[t]{0.45\textwidth}  
\setcounter{subfigure}{0} 
    \includegraphics[width=\textwidth]{ 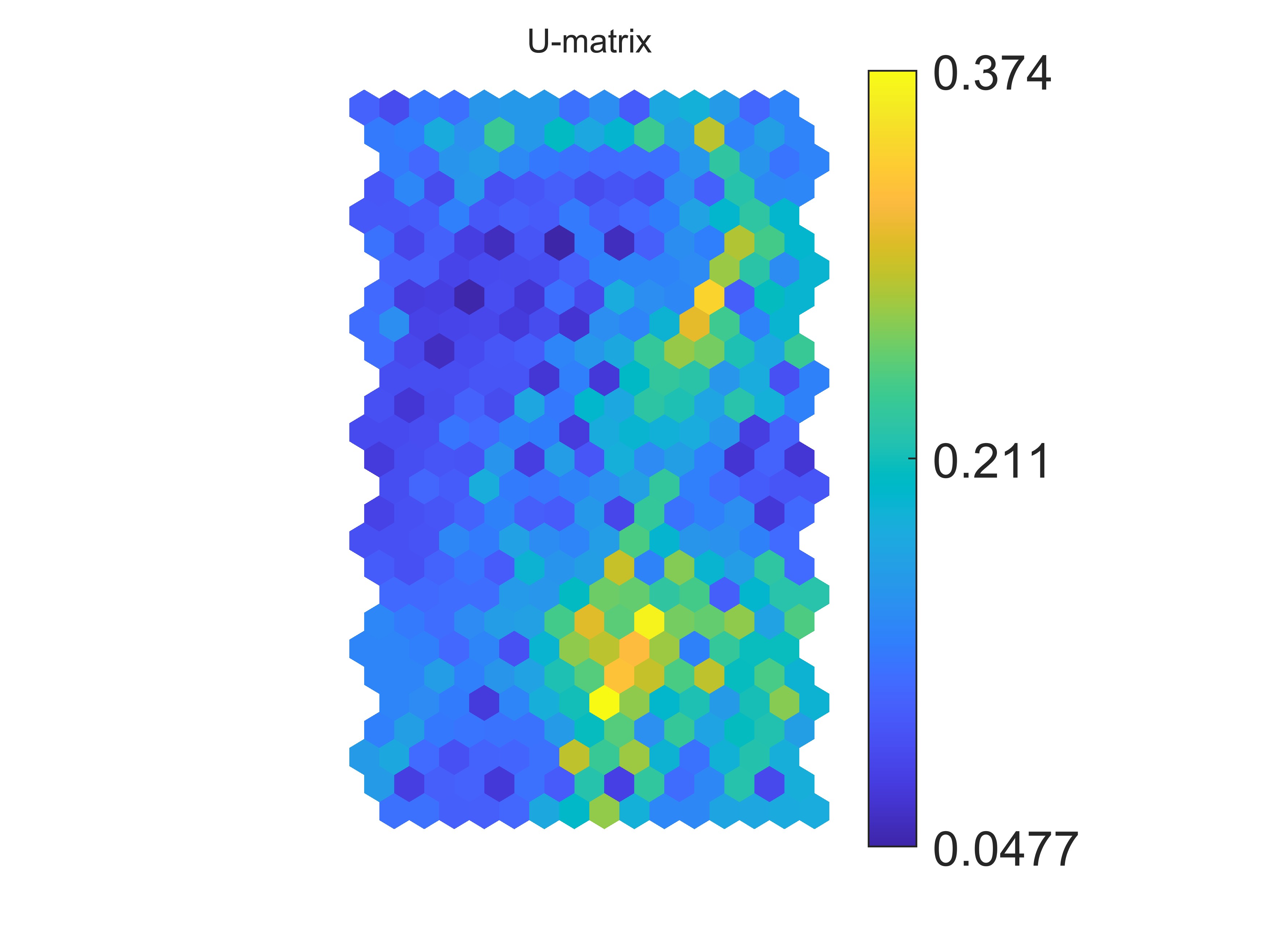}
    \caption{U-matrix}
    \label{boxplot_max}
    \end{subfigure}
\begin{subfigure}[t]{0.45\textwidth}  
\setcounter{subfigure}{3} 
    \includegraphics[width=\textwidth]{ 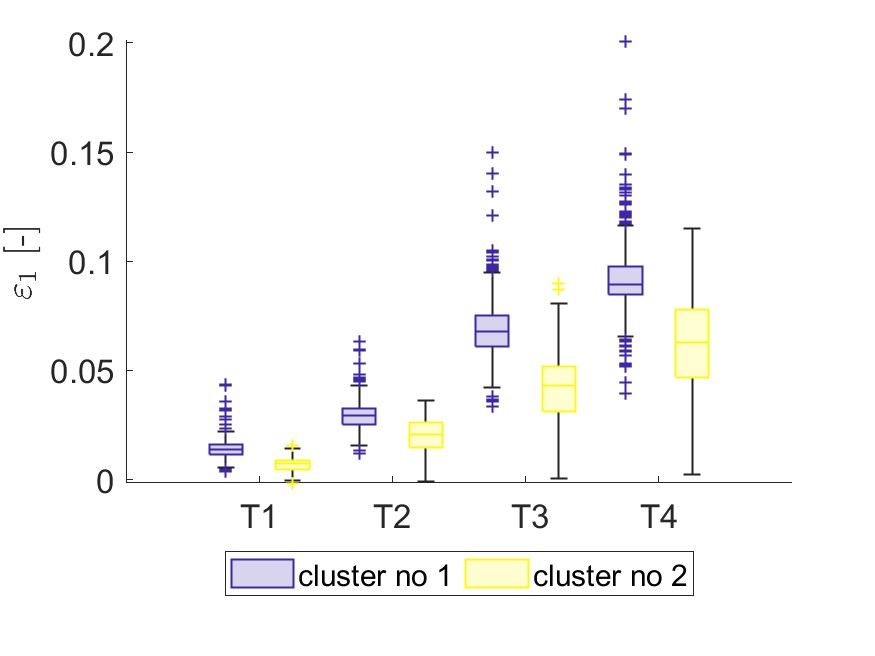}
   \caption{ Boxplot of the maximum principal strain  $\varepsilon_1$ for each cluster}
    \end{subfigure}

\begin{subfigure}[t]{0.45\textwidth}  
\setcounter{subfigure}{1} 
    \includegraphics[width=\textwidth]{ 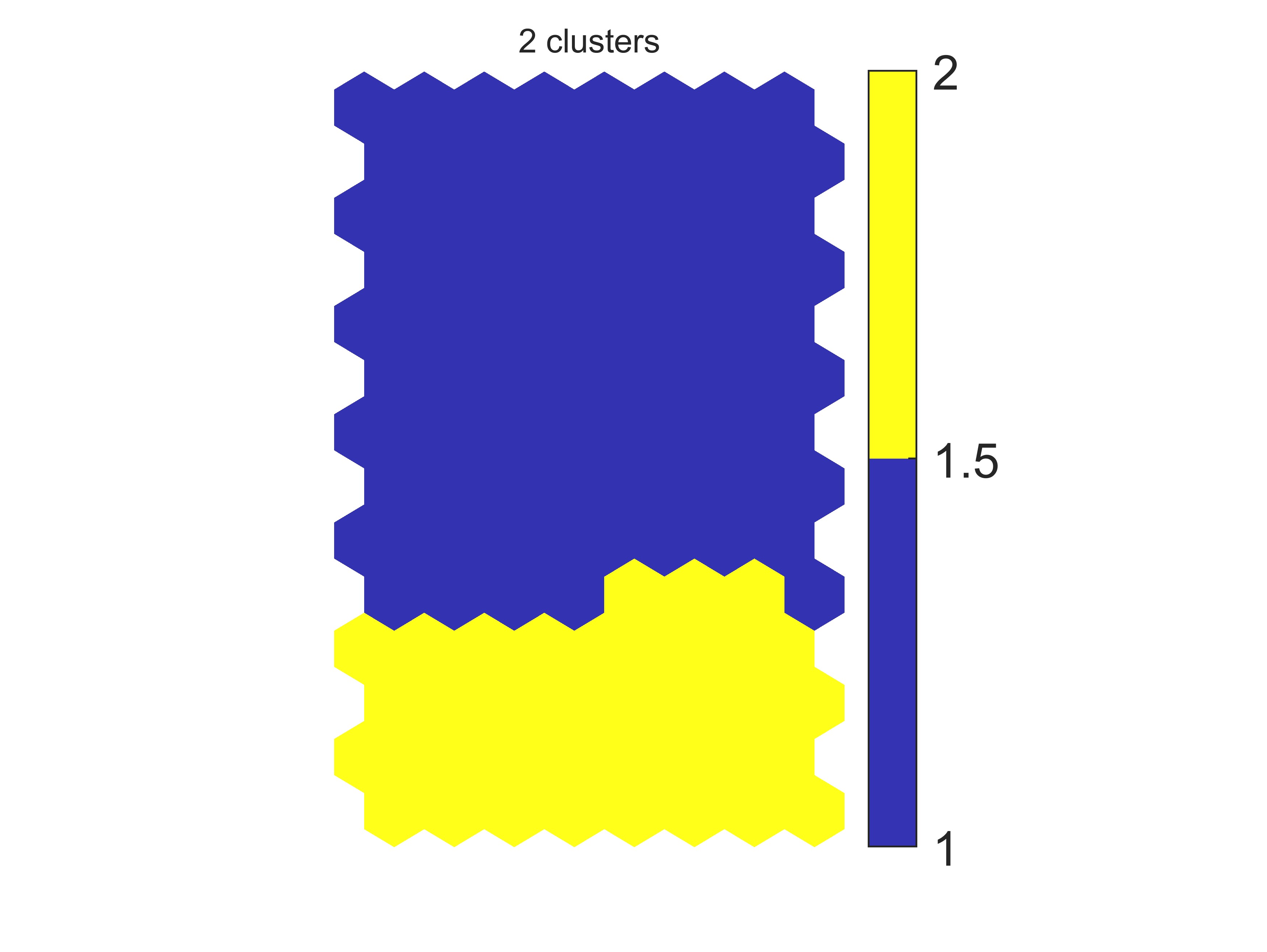}
    \caption{SOM clustering}
    \label{boxplot_max}
\end{subfigure}
\begin{subfigure}[t]{0.45\textwidth}  
\setcounter{subfigure}{4} 
    \includegraphics[width=\textwidth]{ 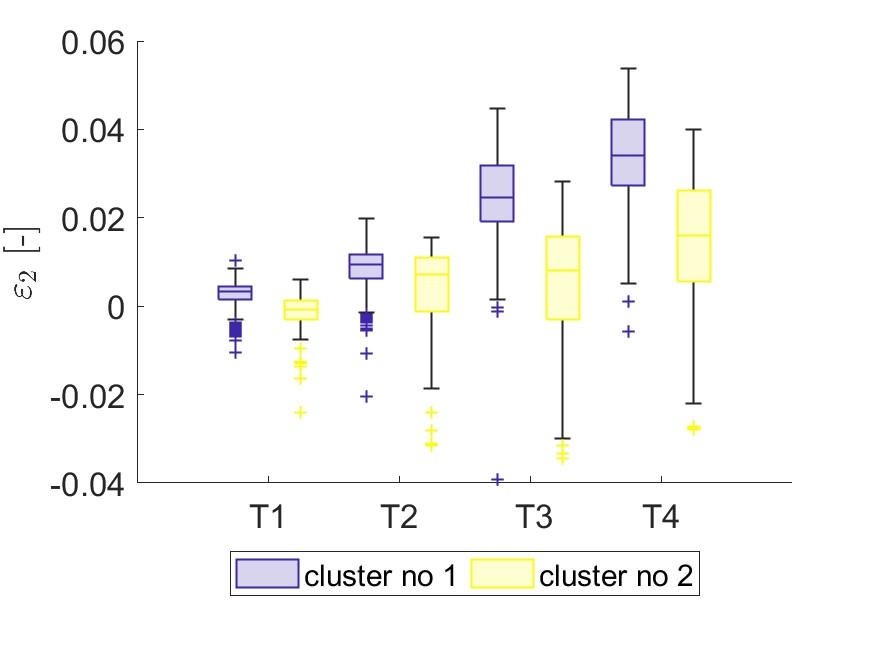}
    \caption{ Boxplot of the minimum principal strain  $\varepsilon_2$ for each cluster}
    \label{boxplot_min}
    \end{subfigure}
\begin{subfigure}[t]{0.45\textwidth}  
\setcounter{subfigure}{2}
    \includegraphics[width=\textwidth]{ 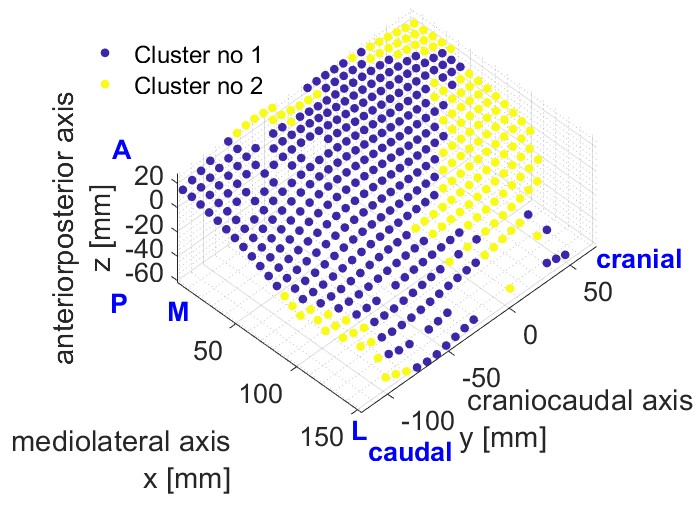}
    \caption{ {Clusters found by SOM marked on the abdominal wall surface}}
    \label{boxplot_max}
\end{subfigure}
\begin{subfigure}[t]{0.45\textwidth}  
\setcounter{subfigure}{5}
    \includegraphics[width=\textwidth]{ 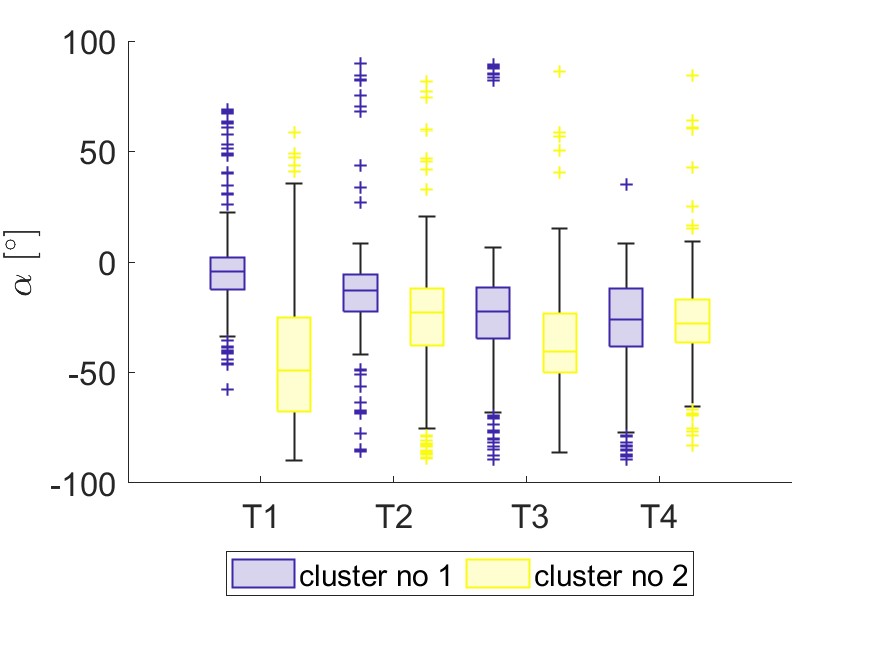}
    \caption{Boxplot of the principal direction $\alpha$ for each cluster  }
    \end{subfigure}
  
    \caption{Cluster results obtained by SOM in case of subject D2 (female, 48 years old, BMI 21.6 kg/m$^2$, intra-abdominal pressure 15 cmH$_2$O)} \label{fig_som_result_dic2}
\end{figure}

\begin{figure}[ht]\centering
\begin{subfigure}[t]{0.45\textwidth}  
\setcounter{subfigure}{0} 
    \includegraphics[width=\textwidth]{ 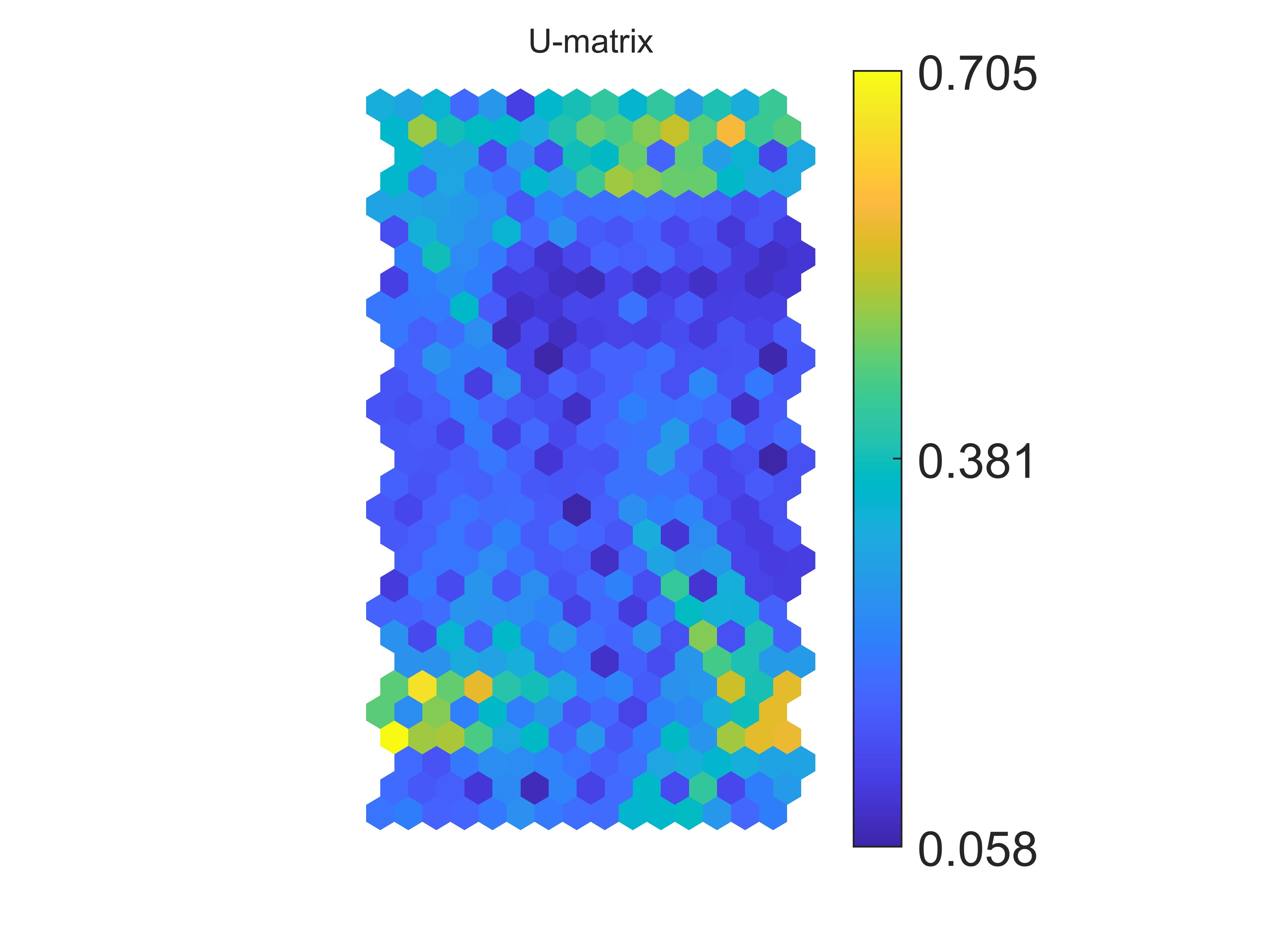}
    \caption{U-matrix}
    \label{boxplot_max}
    \end{subfigure}
\begin{subfigure}[t]{0.45\textwidth}  
\setcounter{subfigure}{3} 
    \includegraphics[width=\textwidth]{ 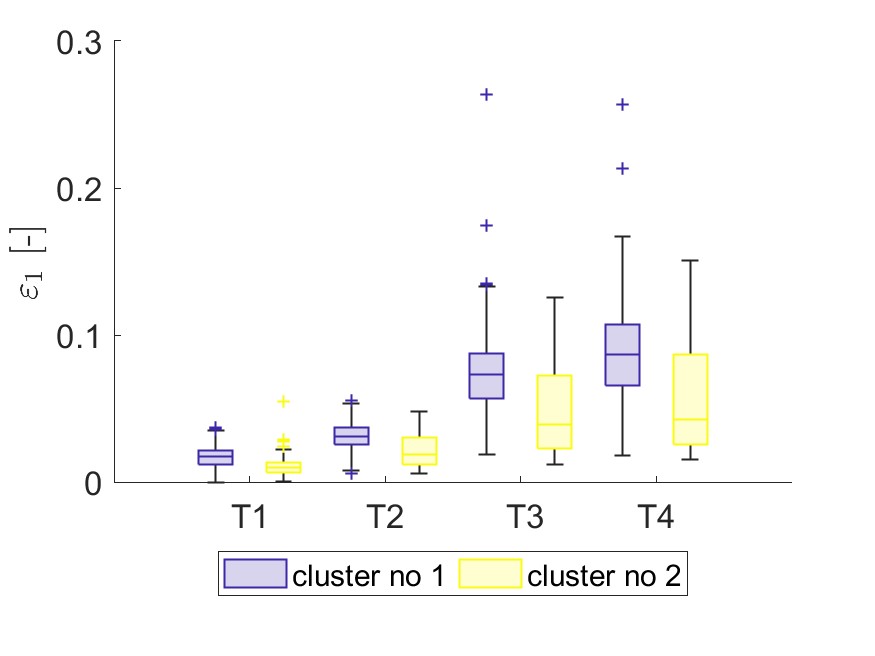}
   \caption{ Boxplot of the maximum principal strain  $\varepsilon_1$ for each cluster}
    \end{subfigure}

\begin{subfigure}[t]{0.45\textwidth}  
\setcounter{subfigure}{1} 
    \includegraphics[width=\textwidth]{ 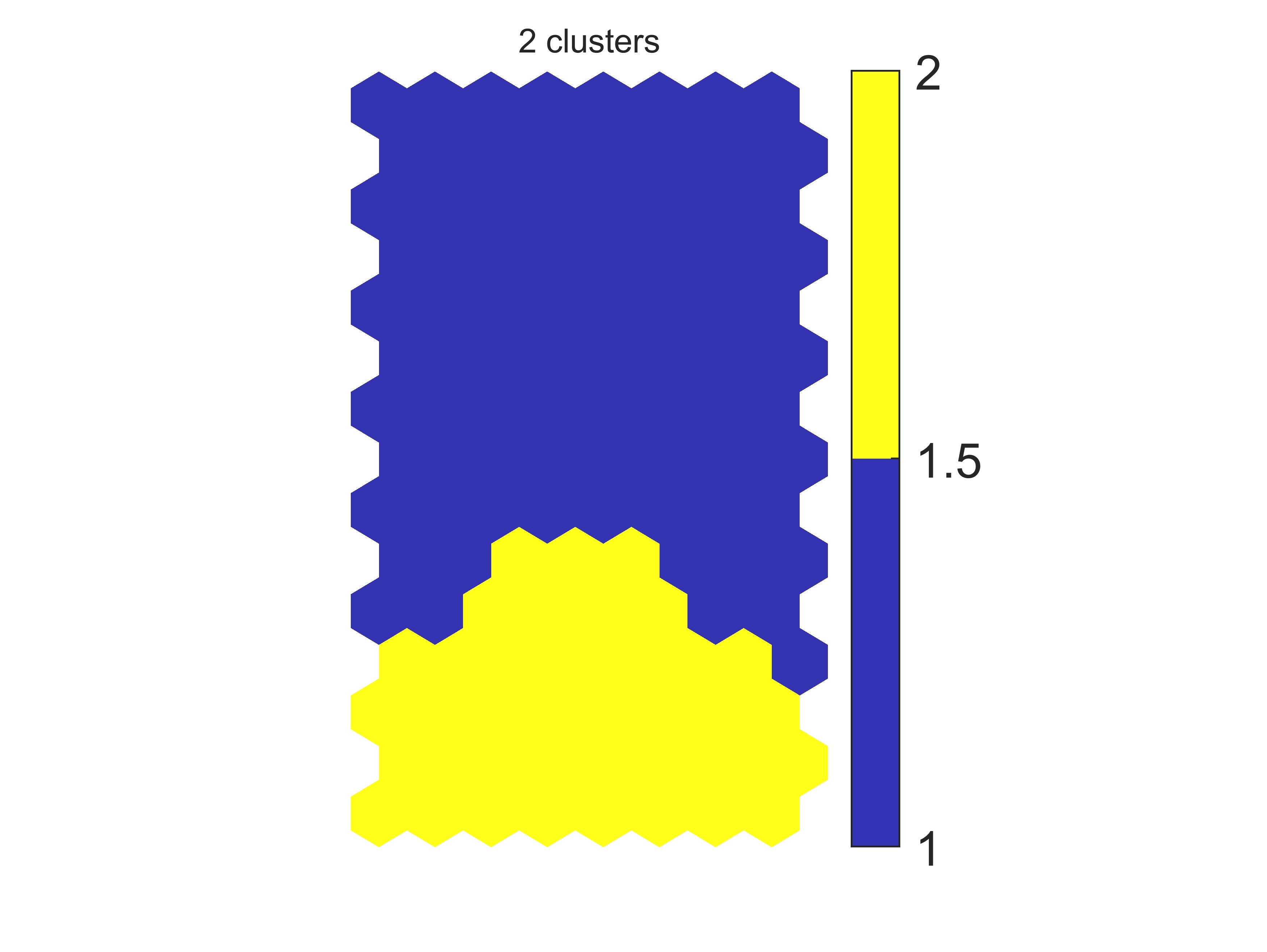}
    \caption{SOM clustering}
    \label{boxplot_max}
\end{subfigure}
\begin{subfigure}[t]{0.45\textwidth}  
\setcounter{subfigure}{4} 
    \includegraphics[width=\textwidth]{ 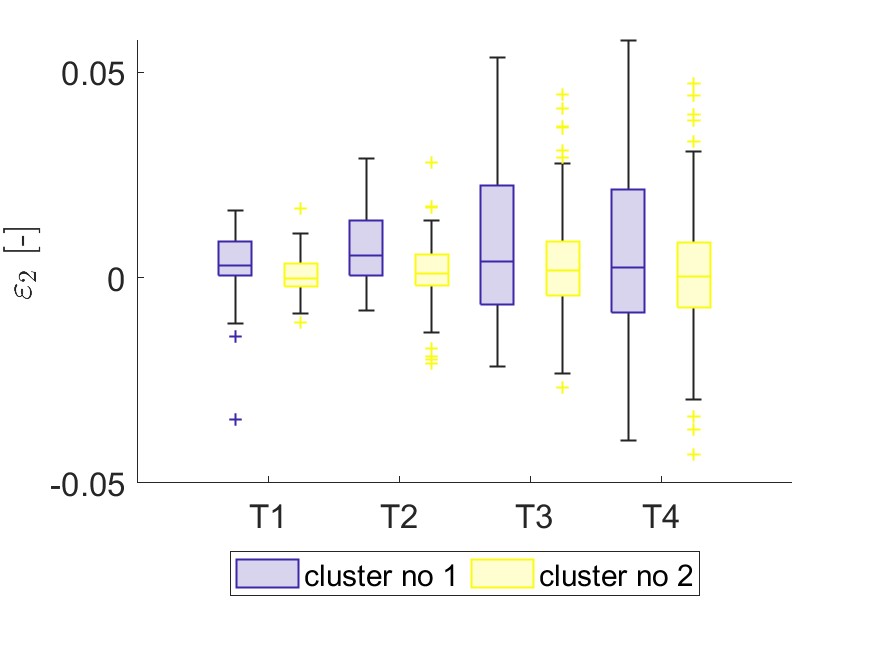}
    \caption{ Boxplot of the minimum principal strain  $\varepsilon_2$ for each cluster}
    \label{boxplot_min}
    \end{subfigure}
\begin{subfigure}[t]{0.45\textwidth}  
\setcounter{subfigure}{2}
    \includegraphics[width=\textwidth]{ 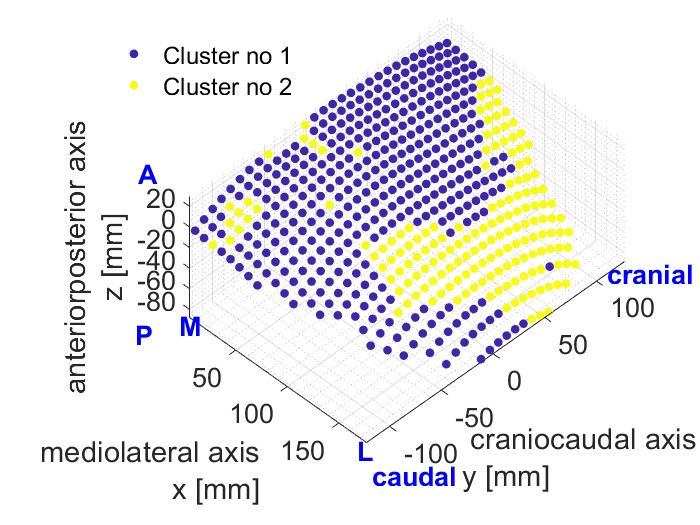}
    \caption{ {Clusters found by SOM marked on the abdominal wall surface}}
    \label{boxplot_max}
\end{subfigure}
\begin{subfigure}[t]{0.45\textwidth}  
\setcounter{subfigure}{5}
    \includegraphics[width=\textwidth]{ 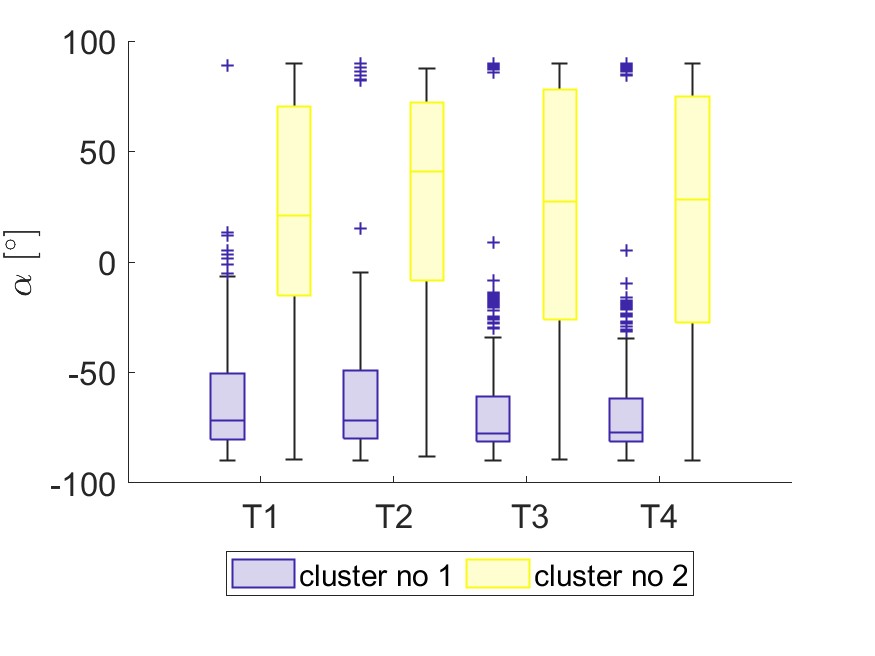}
    \caption{Boxplot of the principal direction $\alpha$ for each cluster  }
    \end{subfigure}
  
    \caption{Cluster results obtained by SOM in case of subject D3 (female, 73 years old, BMI 26.6 kg/m$^2$, intra-abdominal pressure 11 cmH$_2$O)} \label{fig_som_result_dic3}
\end{figure}

\begin{figure}[ht]\centering
\begin{subfigure}[t]{0.45\textwidth}  
\setcounter{subfigure}{0} 
    \includegraphics[width=\textwidth]{ 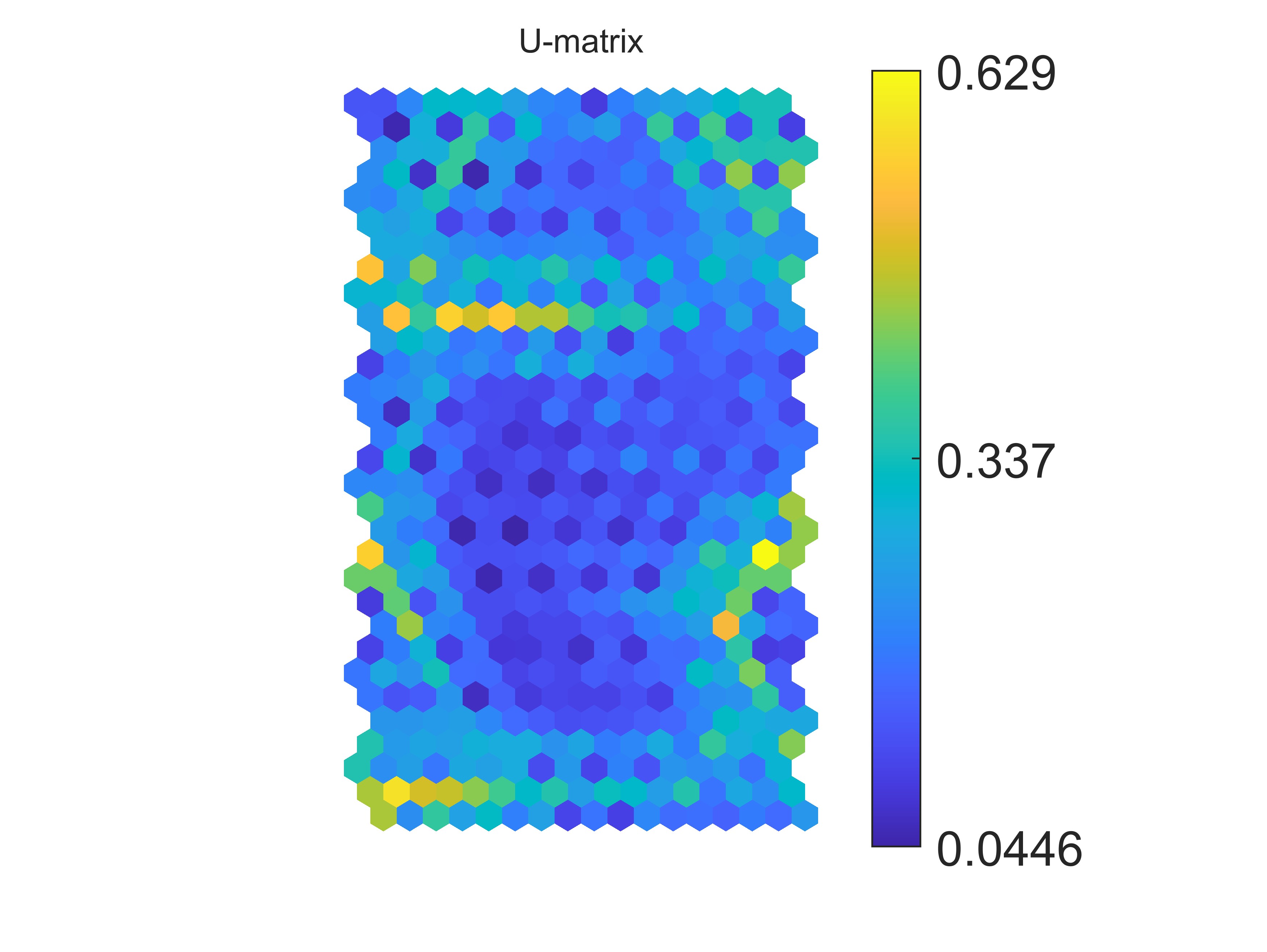}
    \caption{U-matrix}
    \label{boxplot_max}
    \end{subfigure}
\begin{subfigure}[t]{0.45\textwidth}  
\setcounter{subfigure}{3} 
    \includegraphics[width=\textwidth]{ 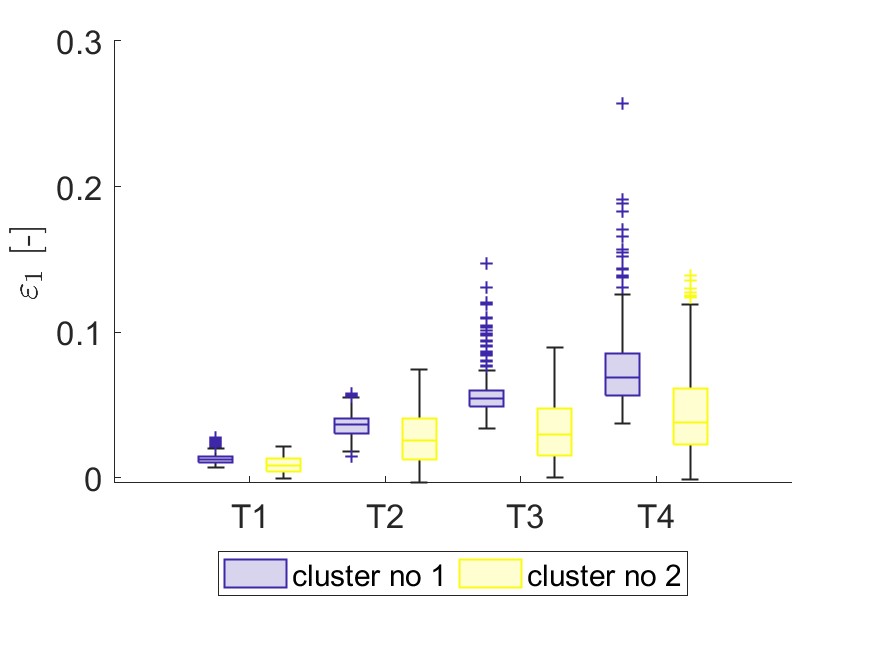}
   \caption{ Boxplot of the maximum principal strain  $\varepsilon_1$ for each cluster}
    \end{subfigure}

\begin{subfigure}[t]{0.45\textwidth}  
\setcounter{subfigure}{1} 
    \includegraphics[width=\textwidth]{ 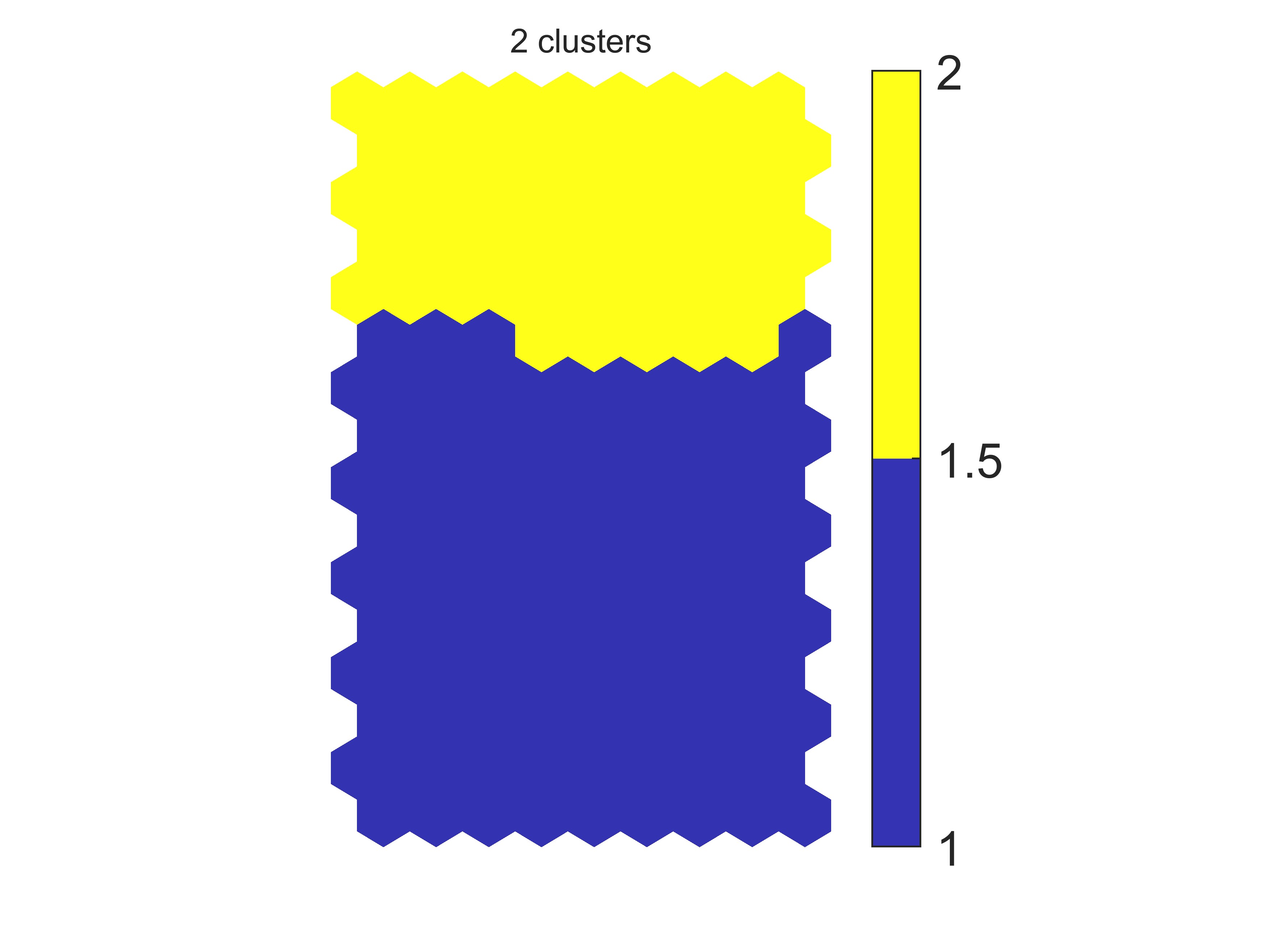}
    \caption{SOM clustering}
    \label{boxplot_max}
\end{subfigure}
\begin{subfigure}[t]{0.45\textwidth}  
\setcounter{subfigure}{4} 
    \includegraphics[width=\textwidth]{ 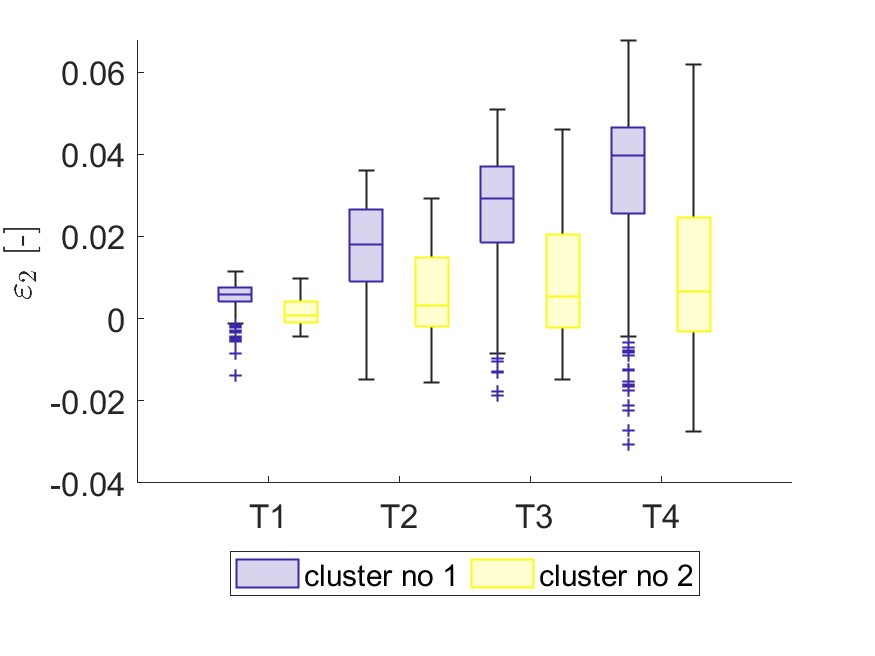}
    \caption{ Boxplot of the minimum principal strain  $\varepsilon_2$ for each cluster}
    \label{boxplot_min}
    \end{subfigure}
\begin{subfigure}[t]{0.45\textwidth}  
\setcounter{subfigure}{2}
    \includegraphics[width=\textwidth]{ 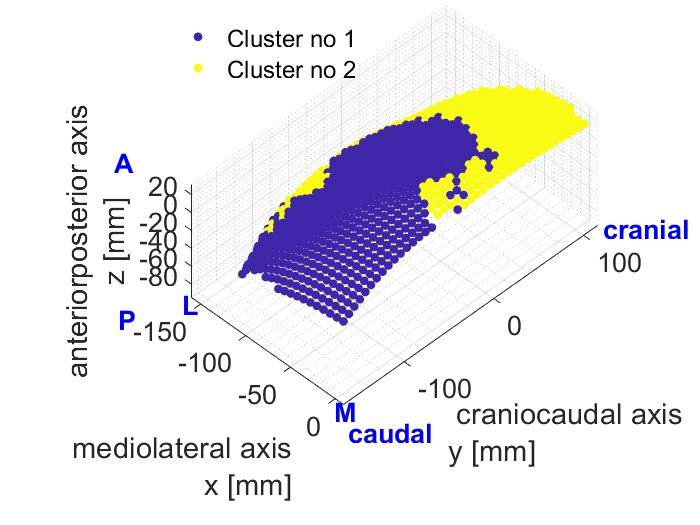}
    \caption{ {Clusters found by SOM marked on the abdominal wall surface}}
    \label{boxplot_max}
\end{subfigure}
\begin{subfigure}[t]{0.45\textwidth}  
\setcounter{subfigure}{5}
    \includegraphics[width=\textwidth]{ 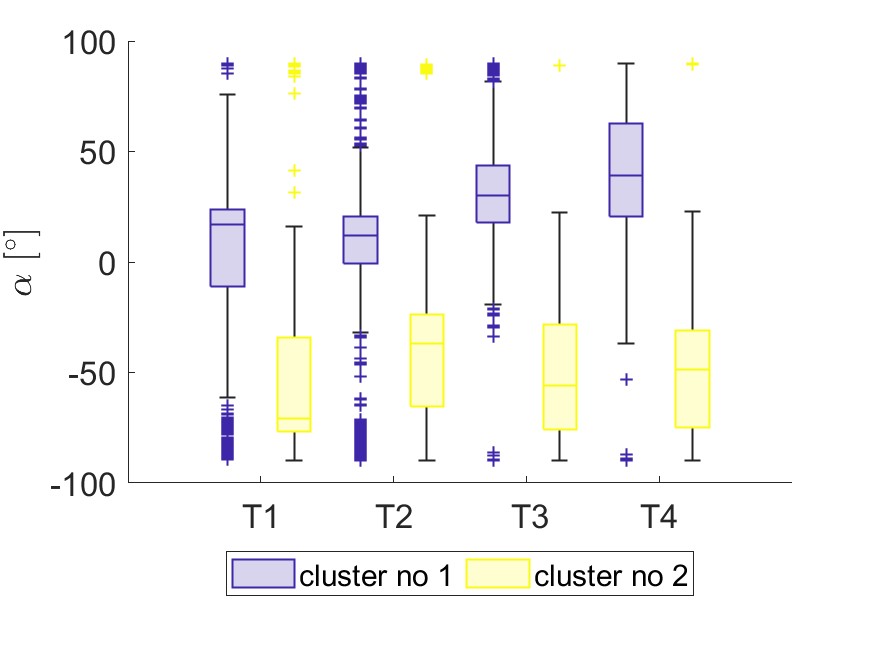}
    \caption{Boxplot of the principal direction $\alpha$ for each cluster  }
    \end{subfigure}
  
    \caption{Cluster results obtained by SOM in case of subject D4 (male, 70 years old, BMI 27.9 kg/m$^2$, intra-abdominal pressure 21 cmH$_2$O)} \label{fig_som_result_dic4}
\end{figure}

\begin{figure}[ht]\centering
\begin{subfigure}[t]{0.45\textwidth}  
\setcounter{subfigure}{0} 
    \includegraphics[width=\textwidth]{ 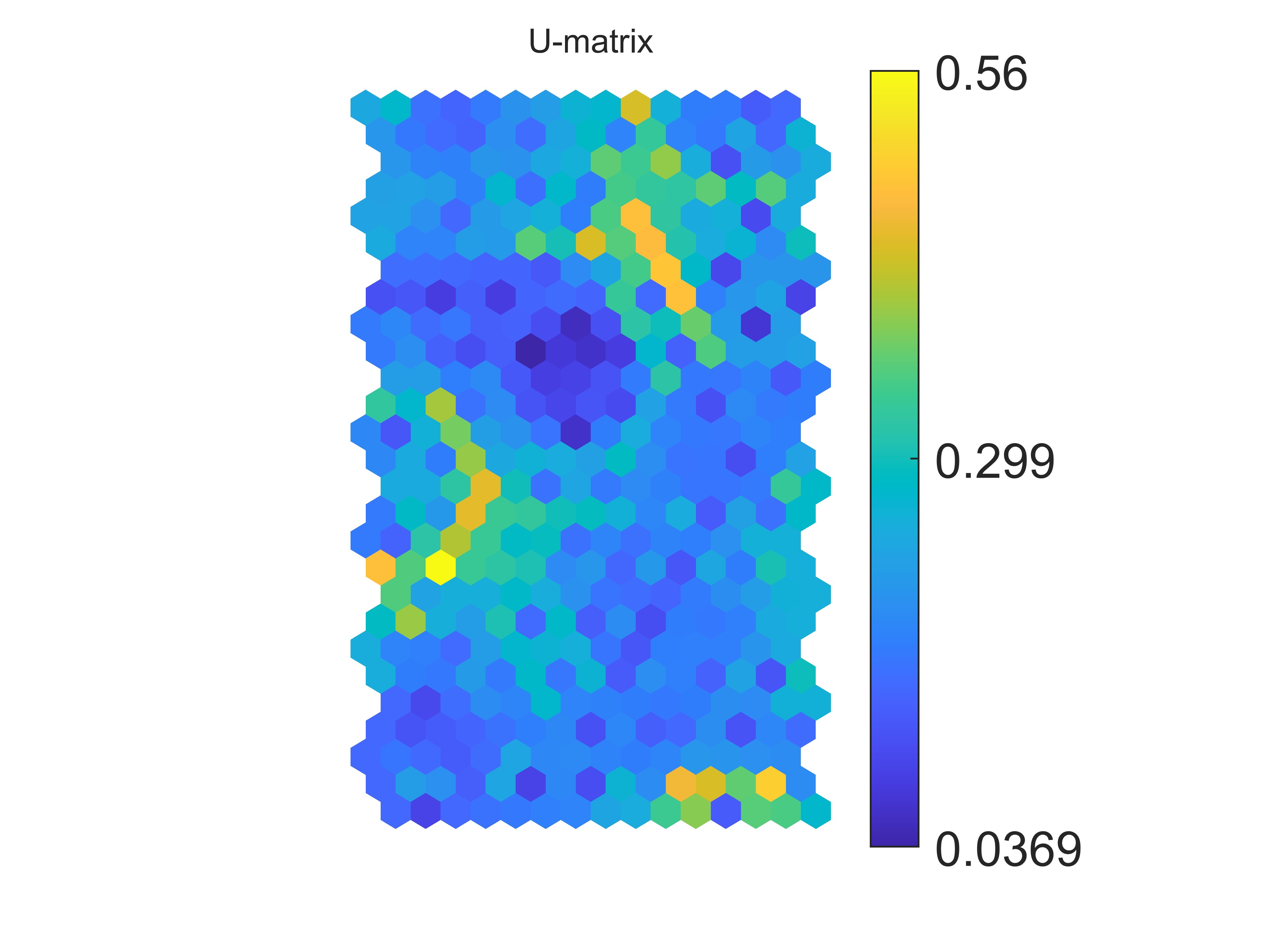}
    \caption{U-matrix}
    \label{boxplot_max}
    \end{subfigure}
\begin{subfigure}[t]{0.45\textwidth}  
\setcounter{subfigure}{3} 
    \includegraphics[width=\textwidth]{ 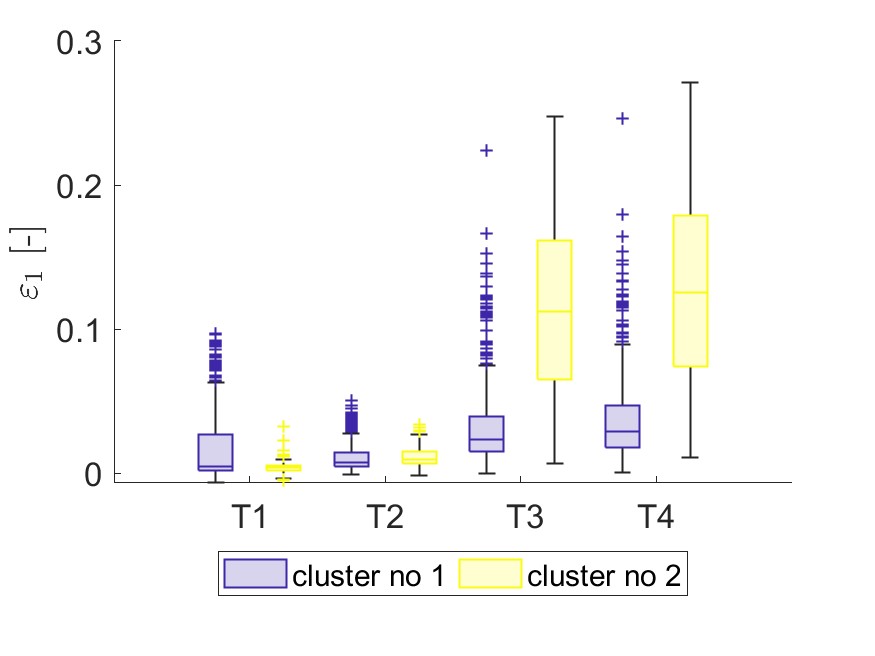}
   \caption{ Boxplot of the maximum principal strain  $\varepsilon_1$ for each cluster}
    \end{subfigure}

\begin{subfigure}[t]{0.45\textwidth}  
\setcounter{subfigure}{1} 
    \includegraphics[width=\textwidth]{ 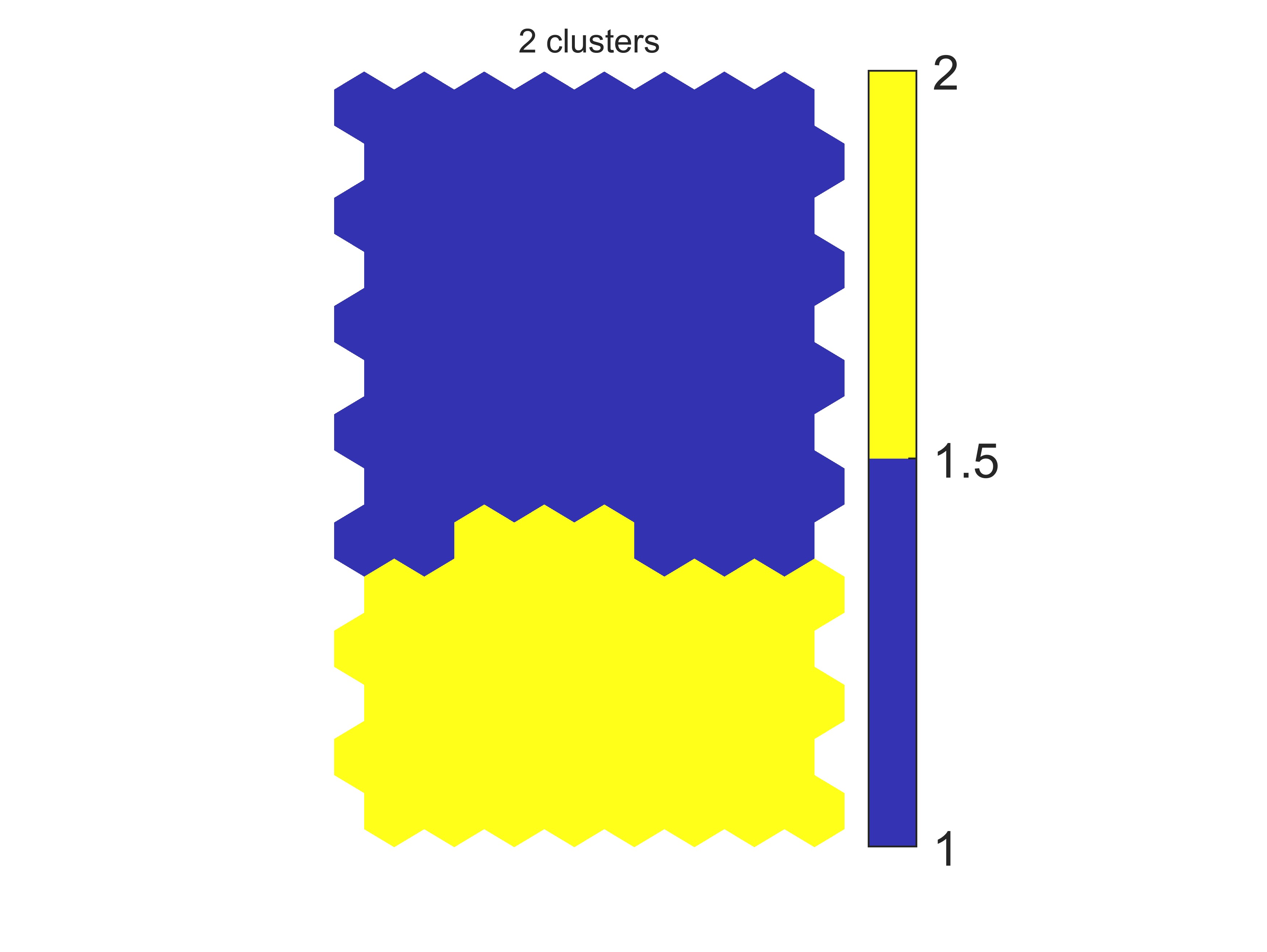}
    \caption{SOM clustering}
    \label{boxplot_max}
\end{subfigure}
\begin{subfigure}[t]{0.45\textwidth}  
\setcounter{subfigure}{4} 
    \includegraphics[width=\textwidth]{ 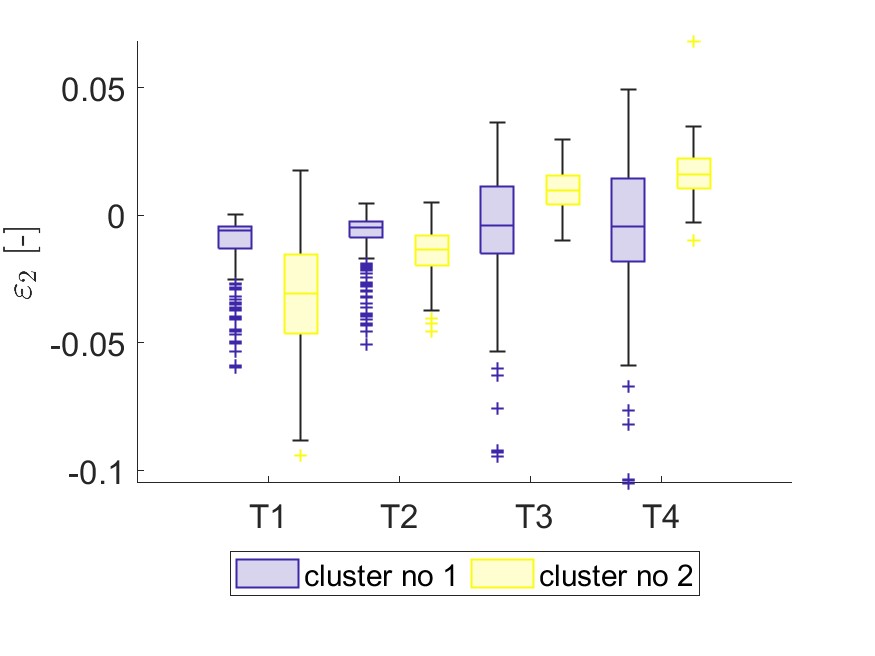}
    \caption{ Boxplot of the minimum principal strain  $\varepsilon_2$ for each cluster}
    \label{boxplot_min}
    \end{subfigure}
\begin{subfigure}[t]{0.45\textwidth}  
\setcounter{subfigure}{2}
    \includegraphics[width=\textwidth]{ 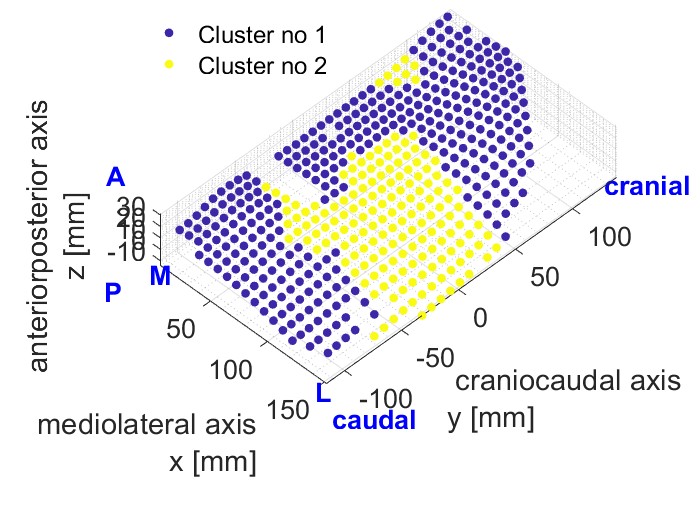}
    \caption{ {Clusters found by SOM marked on the abdominal wall surface}}
    \label{boxplot_max}
\end{subfigure}
\begin{subfigure}[t]{0.45\textwidth}  
\setcounter{subfigure}{5}
    \includegraphics[width=\textwidth]{ 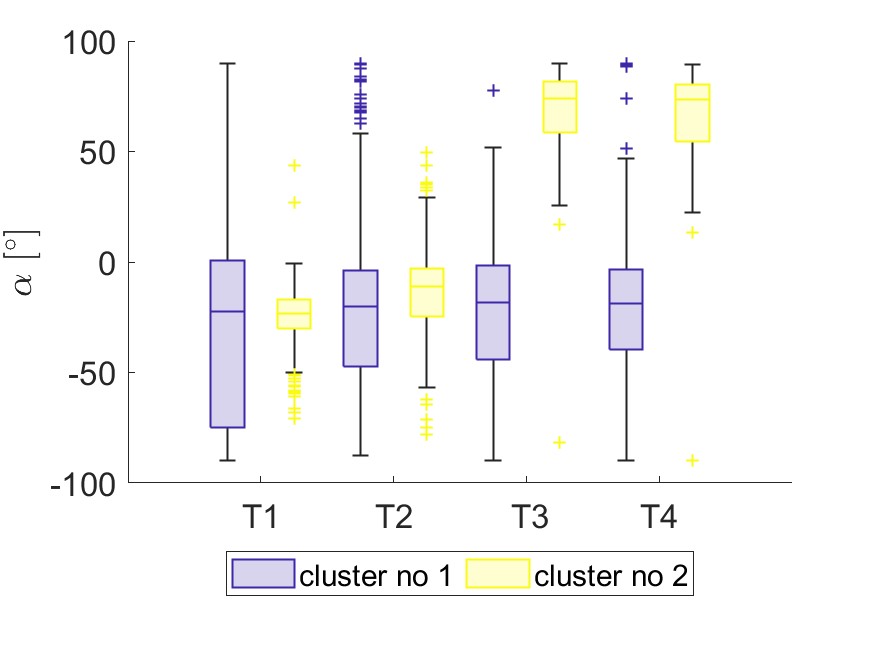}
    \caption{Boxplot of the principal direction $\alpha$ for each cluster  }
    \end{subfigure}
  
    \caption{Cluster results obtained by SOM in case of subject D5 (male, 74 years old, BMI 30.1 kg/m$^2$, intra-abdominal pressure 12 cmH$_2$O)} \label{fig_som_result_dic5}
\end{figure}

\begin{figure}[ht]\centering
\begin{subfigure}[t]{0.45\textwidth}  
\setcounter{subfigure}{0} 
    \includegraphics[width=\textwidth]{ 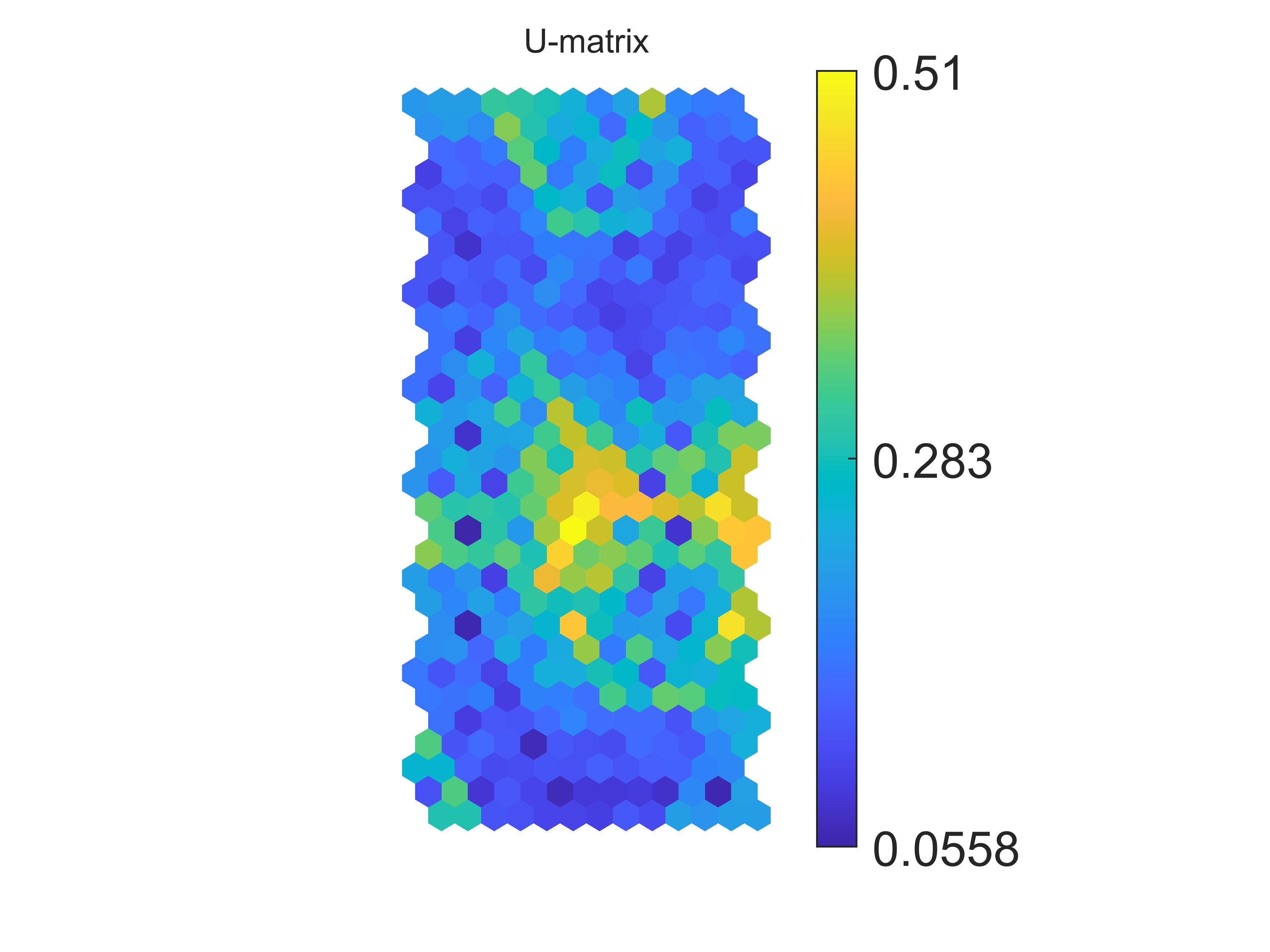}
    \caption{U-matrix}
    \label{boxplot_max}
    \end{subfigure}
\begin{subfigure}[t]{0.45\textwidth}  
\setcounter{subfigure}{3} 
    \includegraphics[width=\textwidth]{ 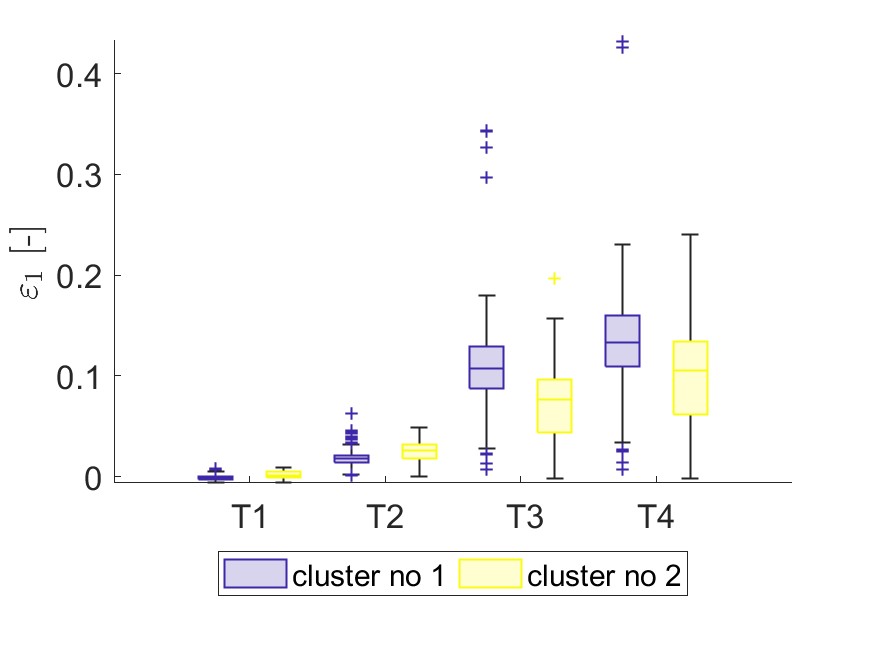}
   \caption{ Boxplot of the maximum principal strain  $\varepsilon_1$ for each cluster}
    \end{subfigure}

\begin{subfigure}[t]{0.45\textwidth}  
\setcounter{subfigure}{1} 
    \includegraphics[width=\textwidth]{ 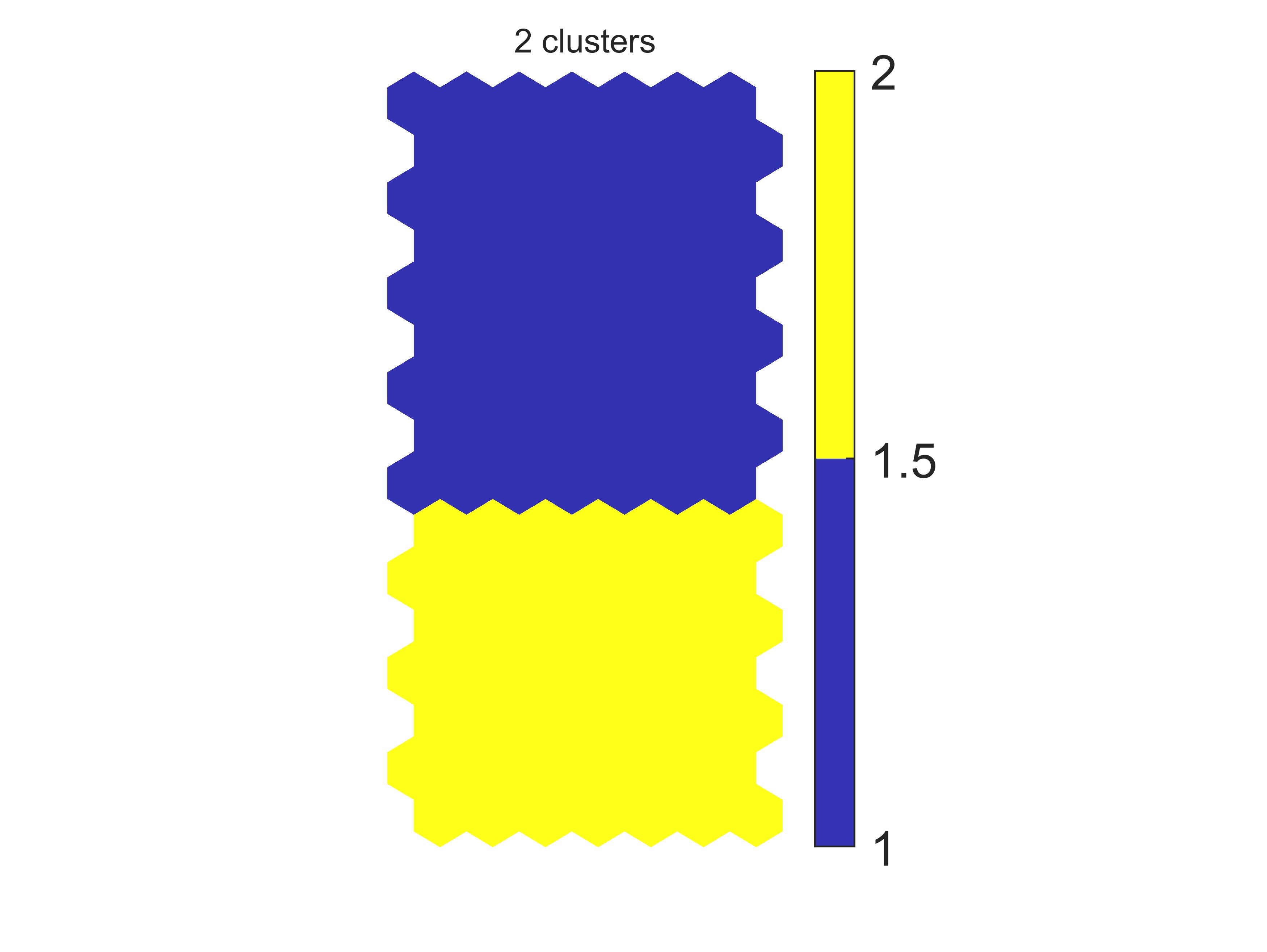}
    \caption{SOM clustering}
    \label{boxplot_max}
\end{subfigure}
\begin{subfigure}[t]{0.45\textwidth}  
\setcounter{subfigure}{4} 
    \includegraphics[width=\textwidth]{ 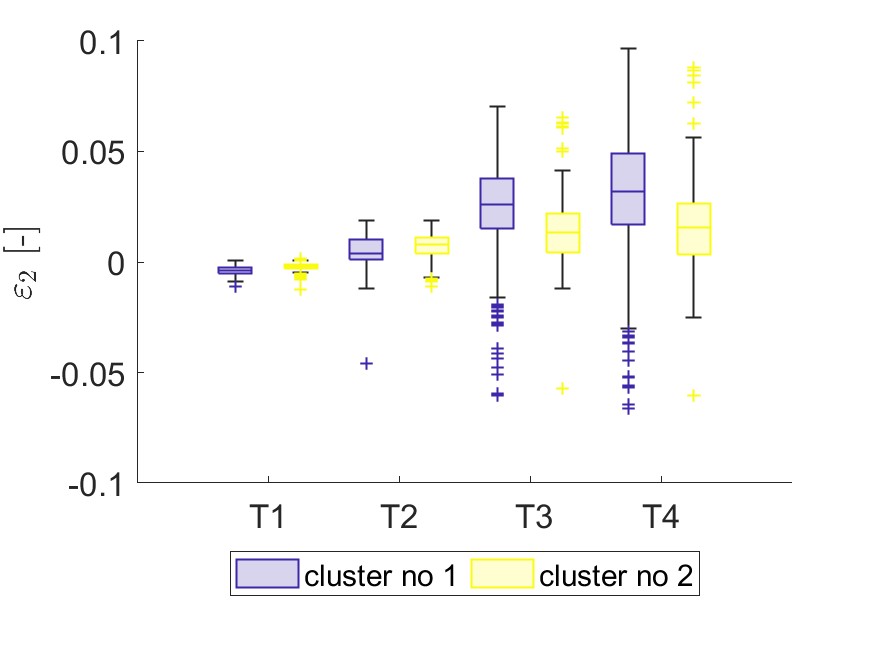}
    \caption{ Boxplot of the minimum principal strain  $\varepsilon_2$ for each cluster}
    \label{boxplot_min}
    \end{subfigure}
\begin{subfigure}[t]{0.45\textwidth}  
\setcounter{subfigure}{2}
    \includegraphics[width=\textwidth]{ 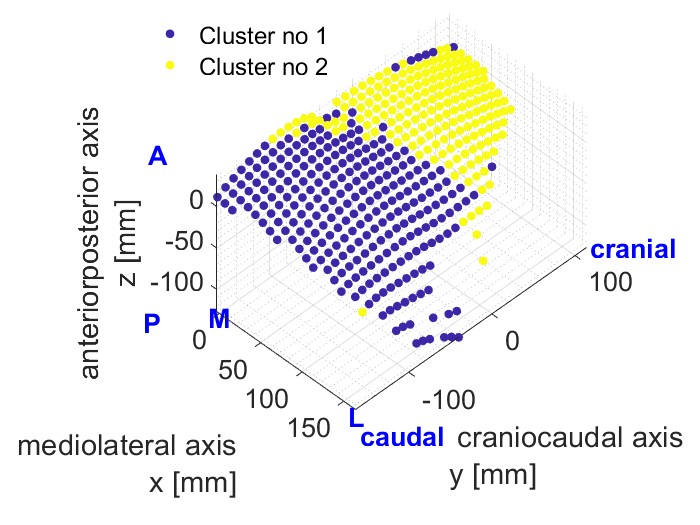}
    \caption{ {Clusters found by SOM marked on the abdominal wall surface}}
    \label{boxplot_max}
\end{subfigure}
\begin{subfigure}[t]{0.45\textwidth}  
\setcounter{subfigure}{5}
    \includegraphics[width=\textwidth]{ 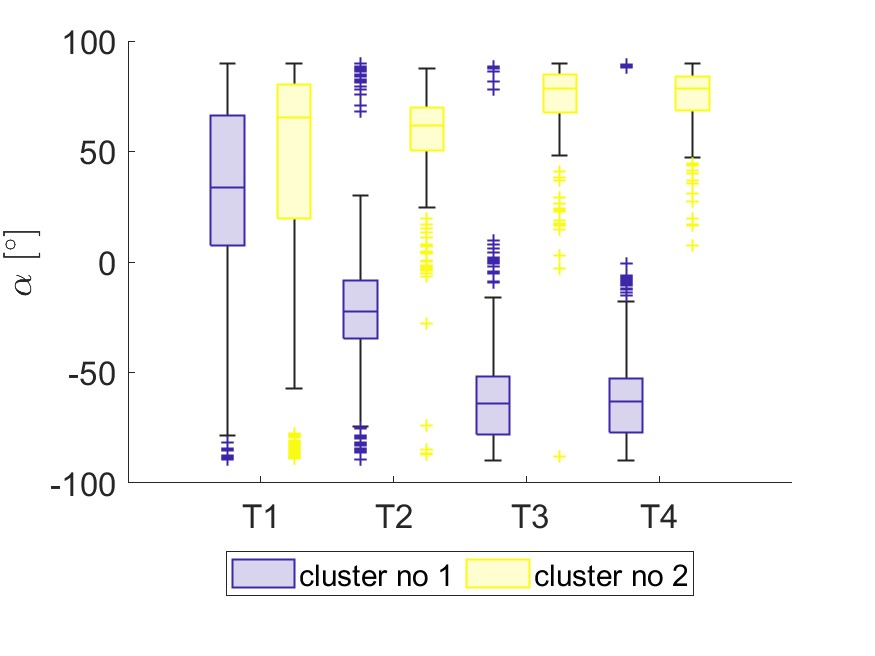}
    \caption{Boxplot of the principal direction $\alpha$ for each cluster  }
    \end{subfigure}
  
    \caption{Cluster results obtained by SOM in case of subject D6 (female, 65 years old, BMI 26.2 kg/m$^2$, intra-abdominal pressure 15 cmH$_2$O)} \label{fig_som_result_dic6}
\end{figure}

\begin{figure}[ht]\centering
\begin{subfigure}[t]{0.45\textwidth}  
\setcounter{subfigure}{0} 
    \includegraphics[width=\textwidth]{ 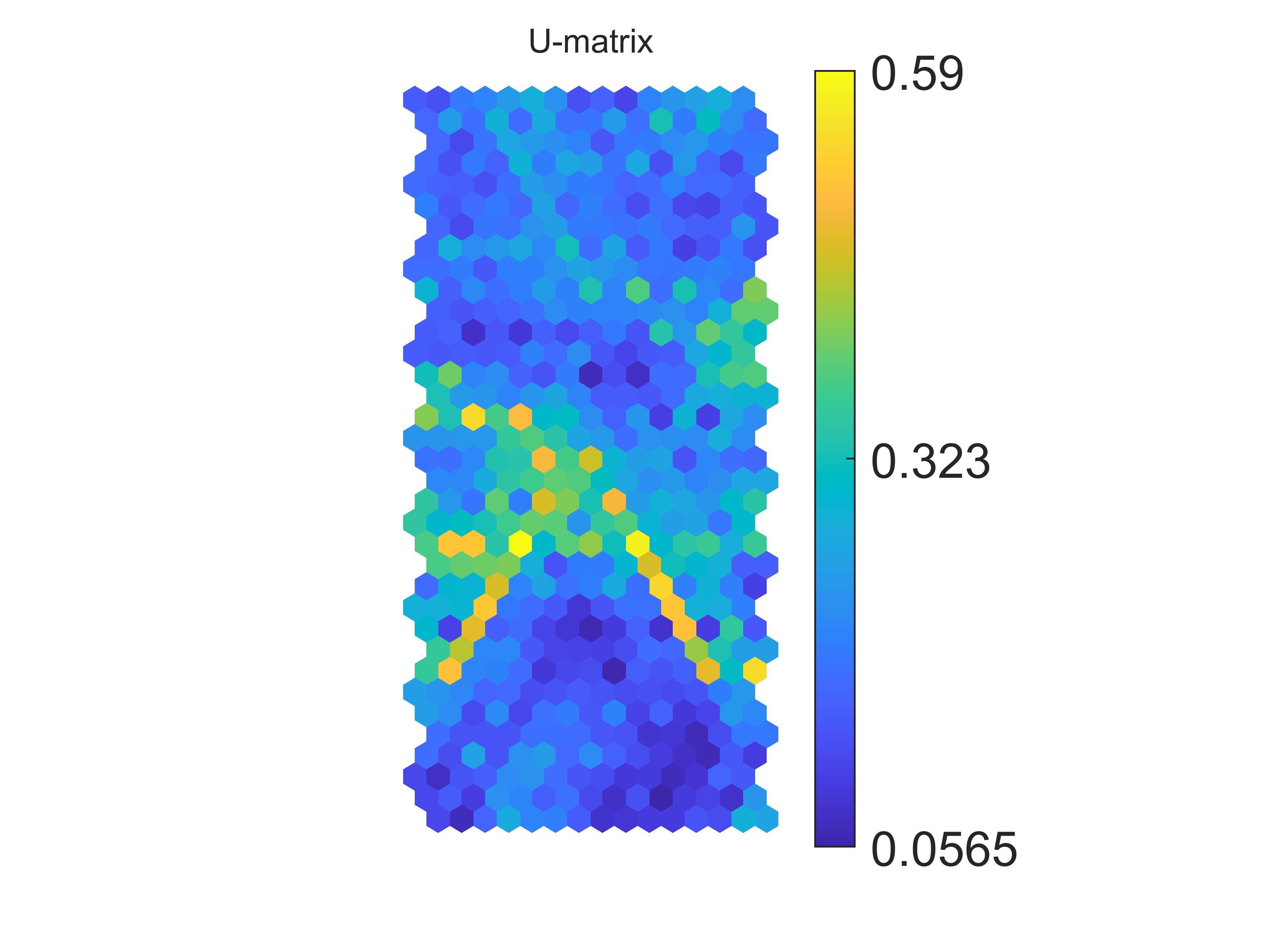}
    \caption{U-matrix}
    \label{boxplot_max}
    \end{subfigure}
\begin{subfigure}[t]{0.45\textwidth}  
\setcounter{subfigure}{3} 
    \includegraphics[width=\textwidth]{ 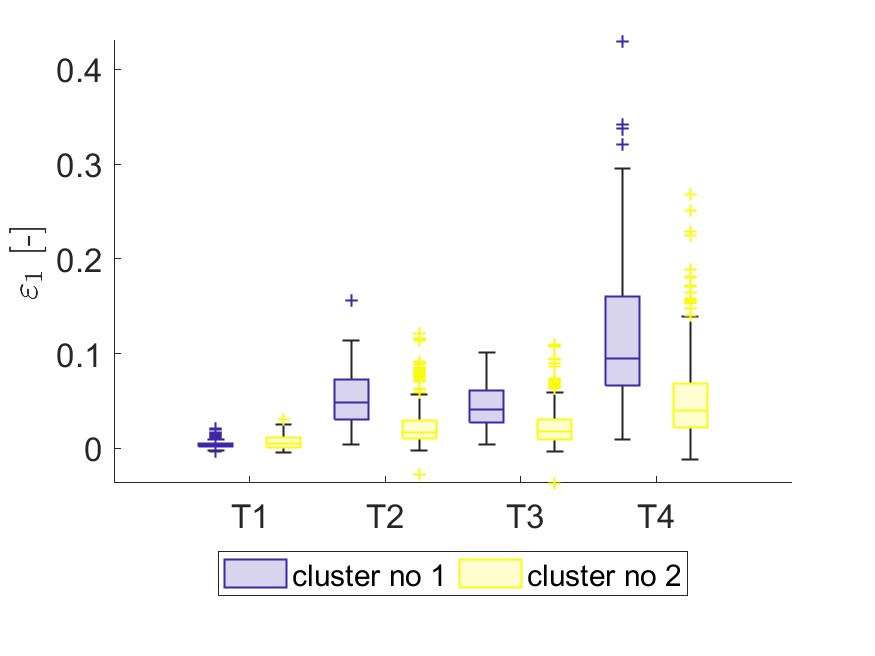}
   \caption{ Boxplot of the maximum principal strain  $\varepsilon_1$ for each cluster}
    \end{subfigure}

\begin{subfigure}[t]{0.45\textwidth}  
\setcounter{subfigure}{1} 
    \includegraphics[width=\textwidth]{ 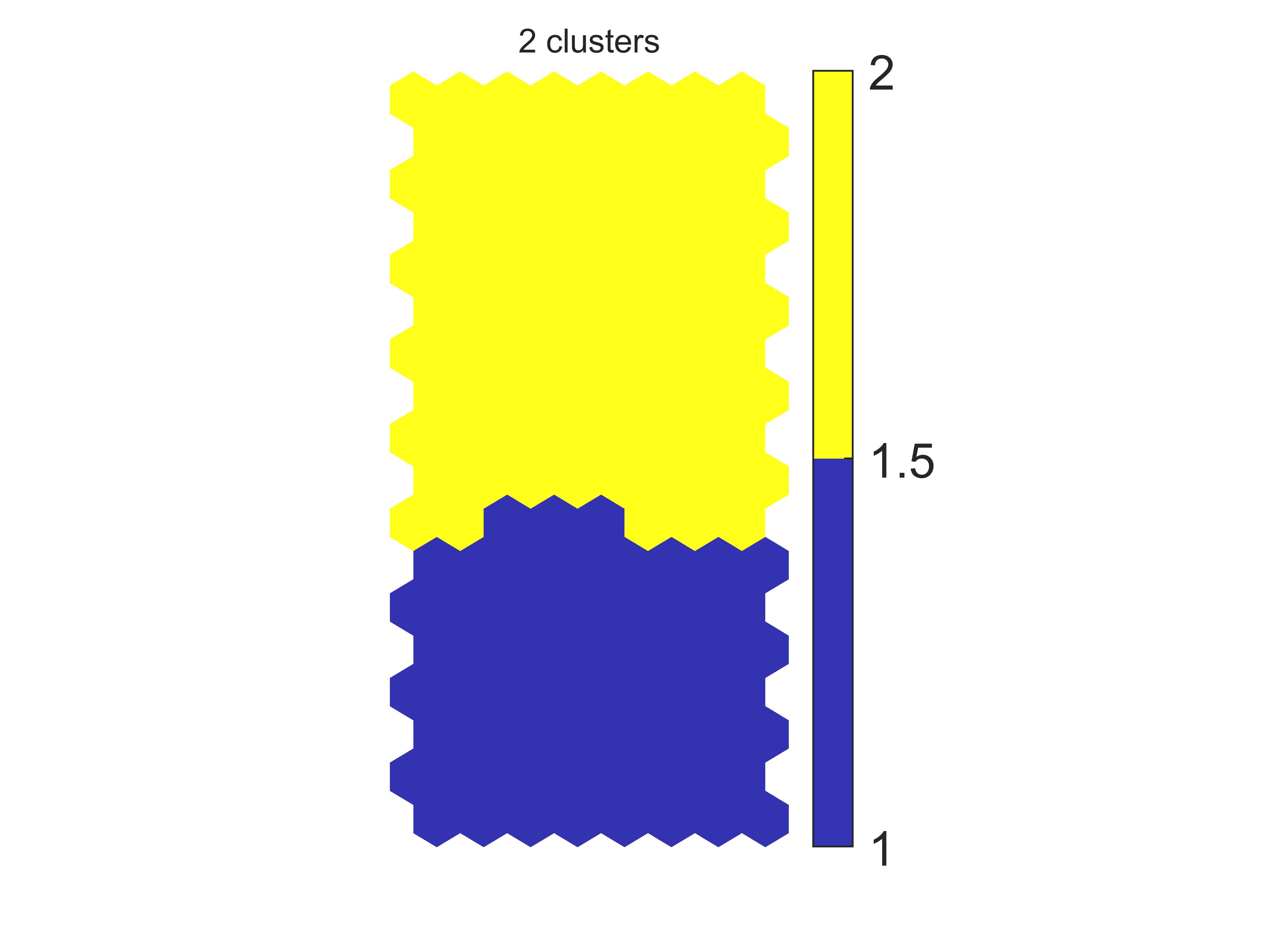}
    \caption{SOM clustering}
    \label{boxplot_max}
\end{subfigure}
\begin{subfigure}[t]{0.45\textwidth}  
\setcounter{subfigure}{4} 
    \includegraphics[width=\textwidth]{ 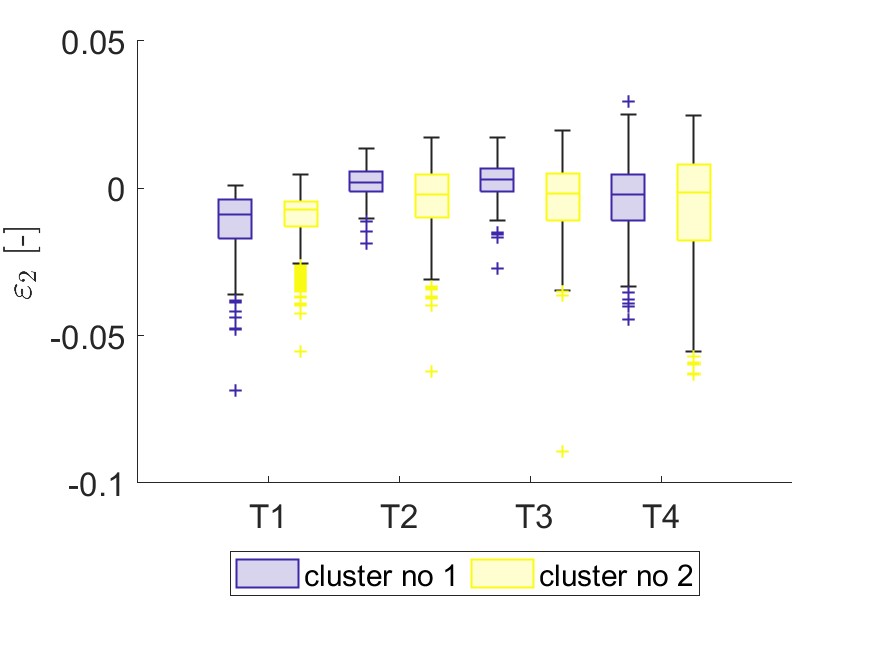}
    \caption{ Boxplot of the minimum principal strain  $\varepsilon_2$ for each cluster}
    \label{boxplot_min}
    \end{subfigure}
\begin{subfigure}[t]{0.45\textwidth}  
\setcounter{subfigure}{2}
    \includegraphics[width=\textwidth]{ 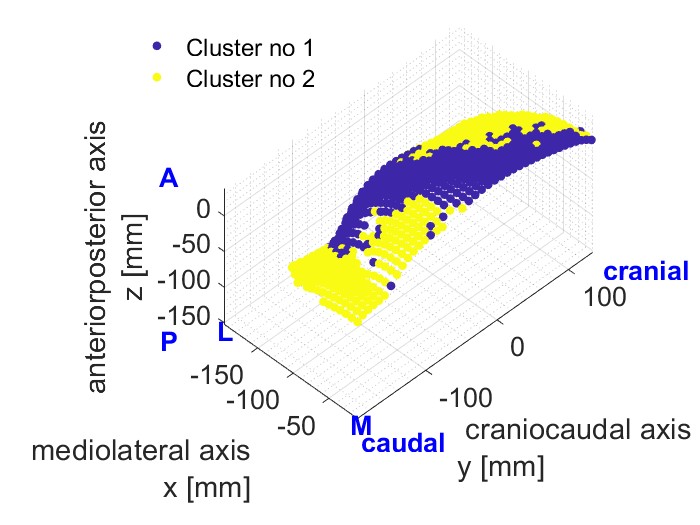}
    \caption{ {Clusters found by SOM marked on the abdominal wall surface}}
    \label{boxplot_max}
\end{subfigure}
\begin{subfigure}[t]{0.45\textwidth}  
\setcounter{subfigure}{5}
    \includegraphics[width=\textwidth]{ 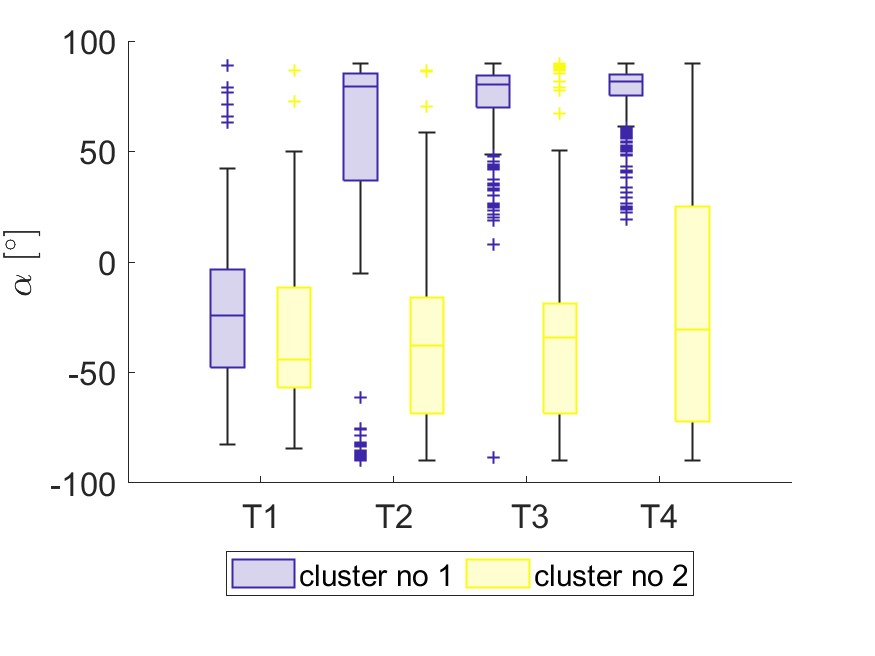}
    \caption{Boxplot of the principal direction $\alpha$ for each cluster  }
    \end{subfigure}
  
    \caption{Cluster results obtained by SOM in case of subject D7 (male, 88 years old, BMI 30.1 kg/m$^2$, intra-abdominal pressure 21 cmH$_2$O)} \label{fig_som_result_dic7}
\end{figure}

\begin{figure}[ht]\centering
\begin{subfigure}[t]{0.45\textwidth}  
\setcounter{subfigure}{0} 
    \includegraphics[width=\textwidth]{ 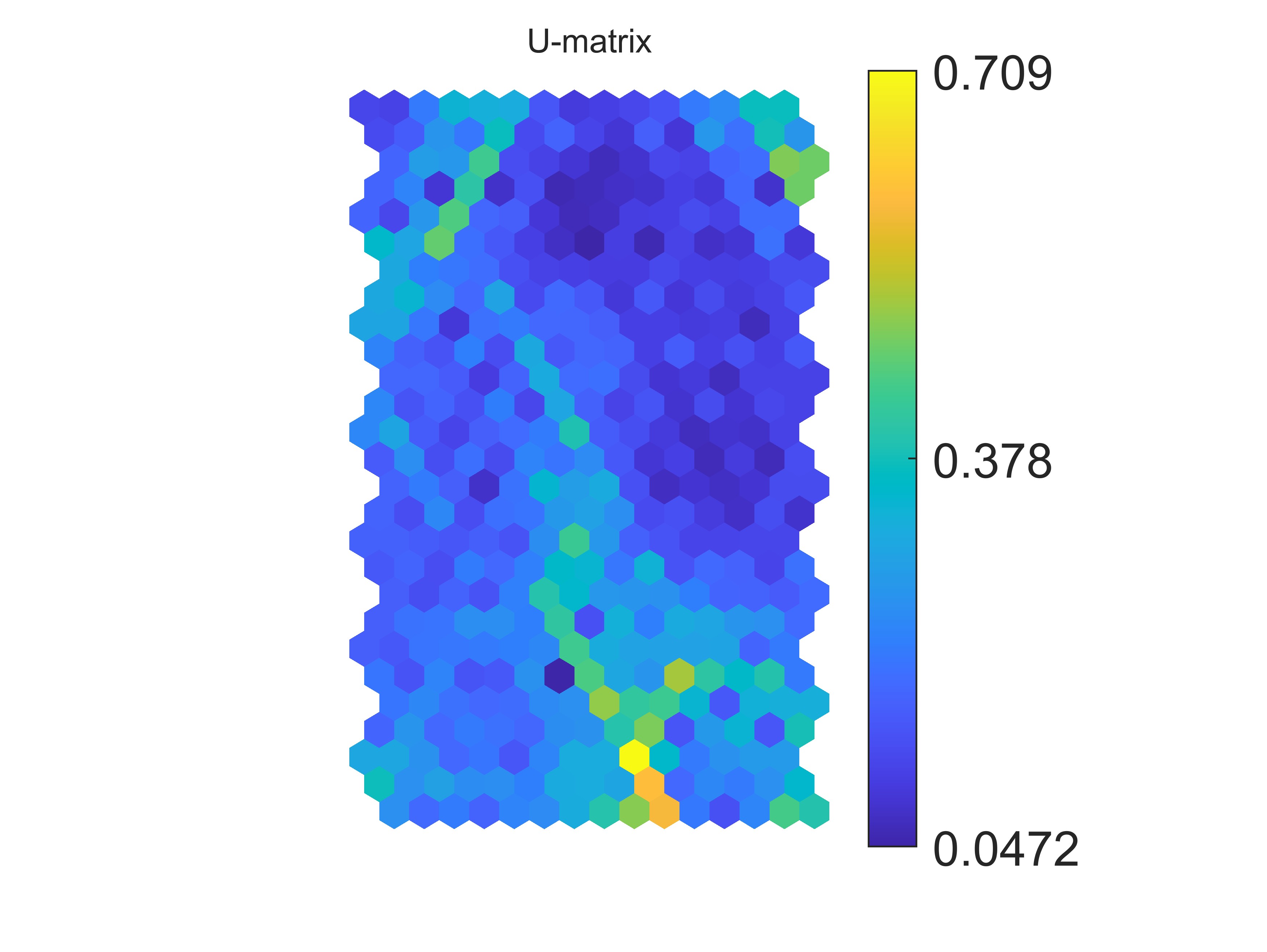}
    \caption{U-matrix}
    \label{boxplot_max}
    \end{subfigure}
\begin{subfigure}[t]{0.45\textwidth}  
\setcounter{subfigure}{3} 
    \includegraphics[width=\textwidth]{ 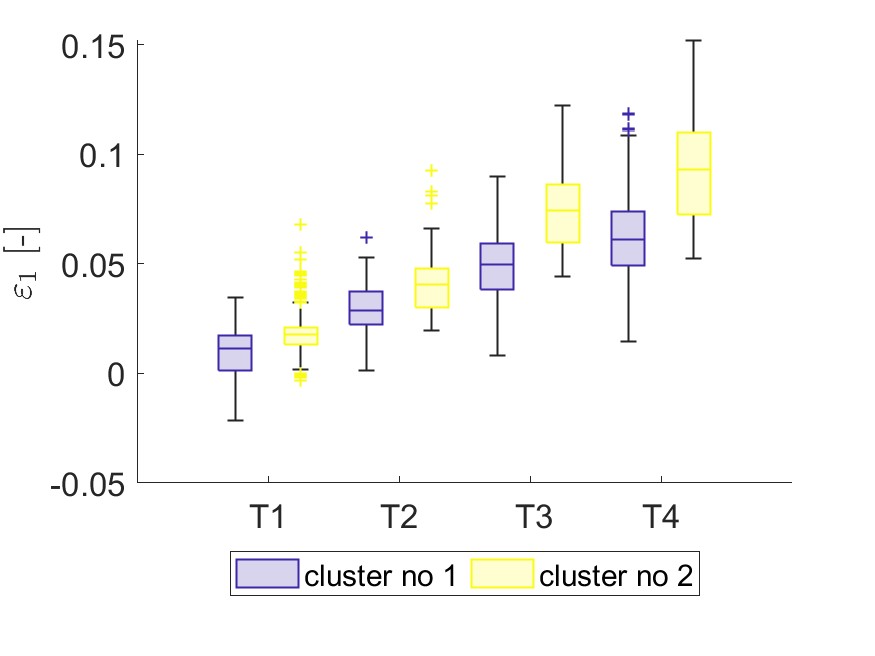}
   \caption{ Boxplot of the maximum principal strain  $\varepsilon_1$ for each cluster}
    \end{subfigure}

\begin{subfigure}[t]{0.45\textwidth}  
\setcounter{subfigure}{1} 
    \includegraphics[width=\textwidth]{ 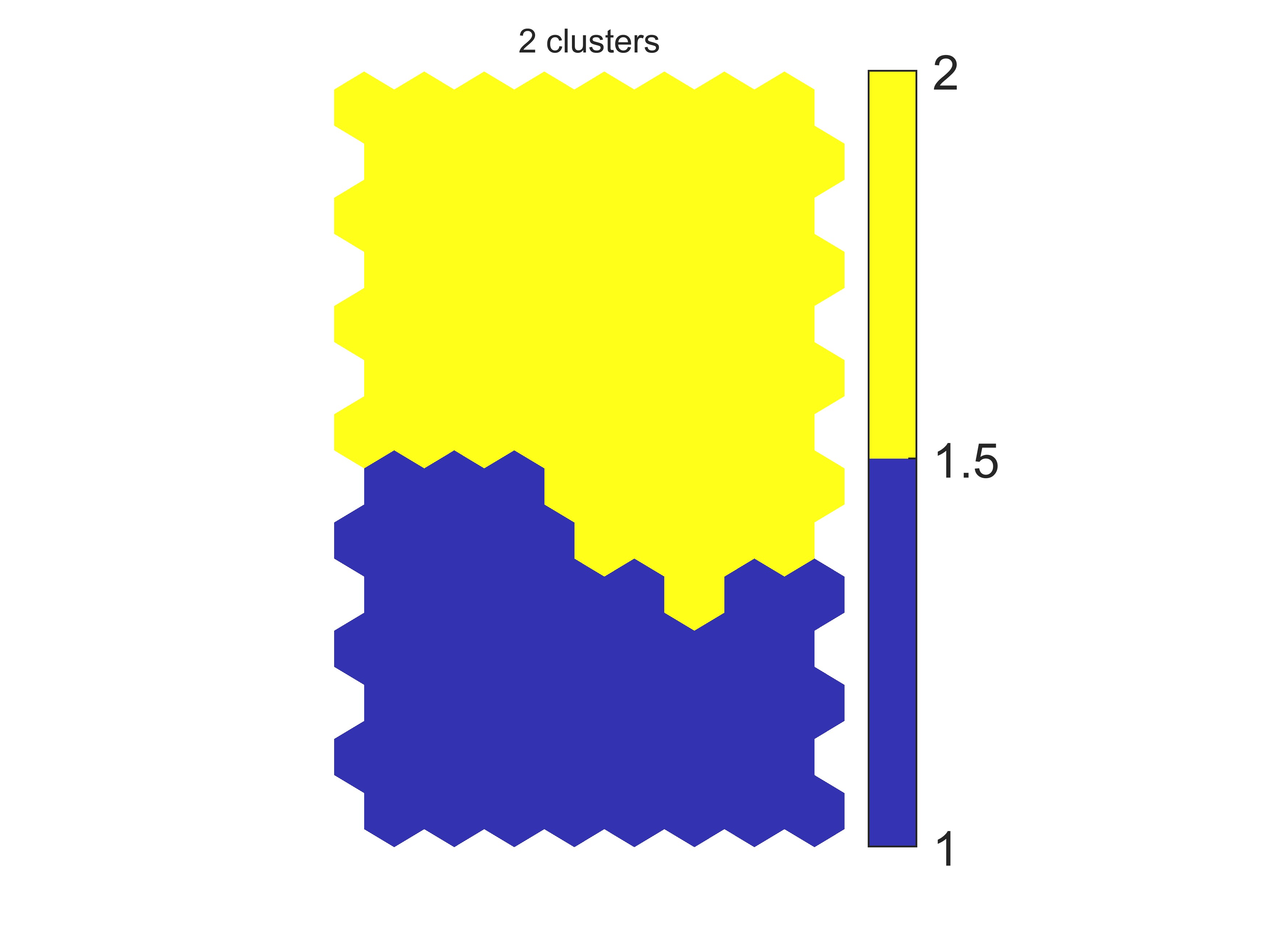}
    \caption{SOM clustering}
    \label{boxplot_max}
\end{subfigure}
\begin{subfigure}[t]{0.45\textwidth}  
\setcounter{subfigure}{4} 
    \includegraphics[width=\textwidth]{ 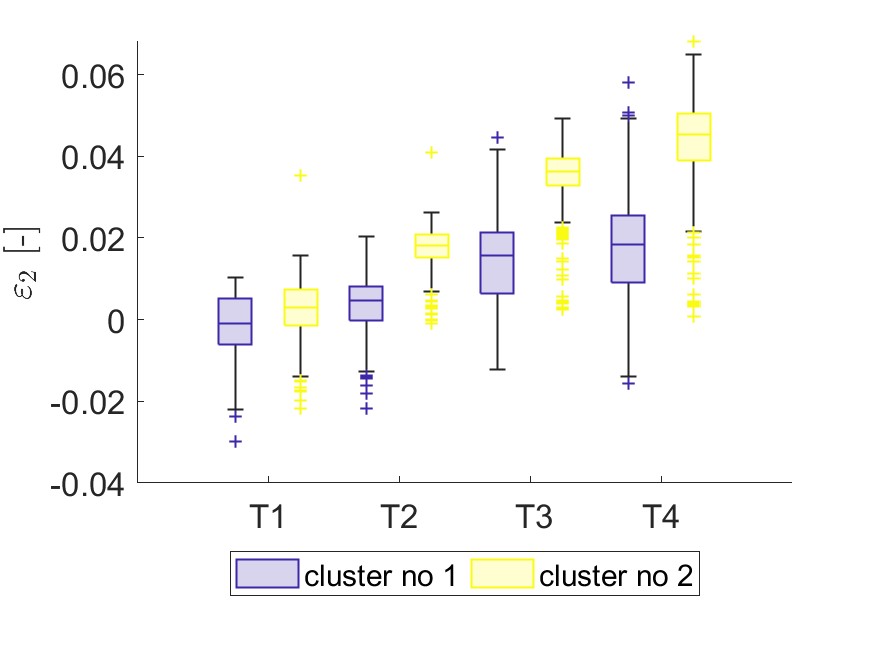}
    \caption{ Boxplot of the minimum principal strain  $\varepsilon_2$ for each cluster}
    \label{boxplot_min}
    \end{subfigure}
\begin{subfigure}[t]{0.45\textwidth}  
\setcounter{subfigure}{2}
    \includegraphics[width=\textwidth]{ 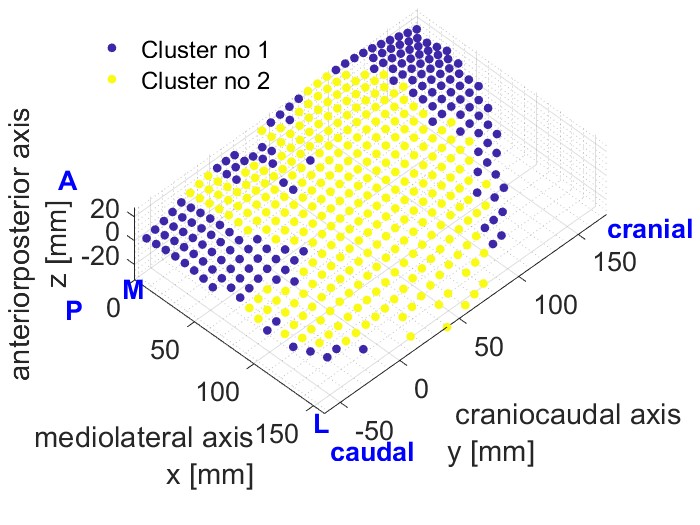}
    \caption{ {Clusters found by SOM marked on the abdominal wall surface}}
    \label{boxplot_max}
\end{subfigure}
\begin{subfigure}[t]{0.45\textwidth}  
\setcounter{subfigure}{5}
    \includegraphics[width=\textwidth]{ 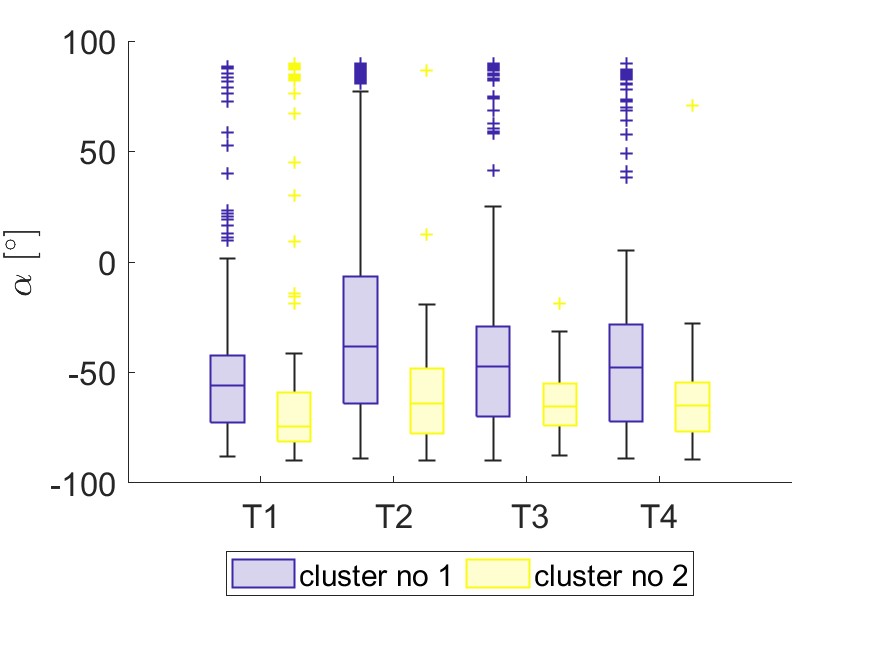}
    \caption{Boxplot of the principal direction $\alpha$ for each cluster  }
    \end{subfigure}
  
    \caption{Cluster results obtained by SOM in case of subject D8 (male, 61 years old, BMI 20.5 kg/m$^2$, intra-abdominal pressure 16 cmH$_2$O)} \label{fig_som_result_dic8}
\end{figure}

\begin{figure}[ht]\centering
\begin{subfigure}[t]{0.45\textwidth}  
\setcounter{subfigure}{0} 
    \includegraphics[width=\textwidth]{ 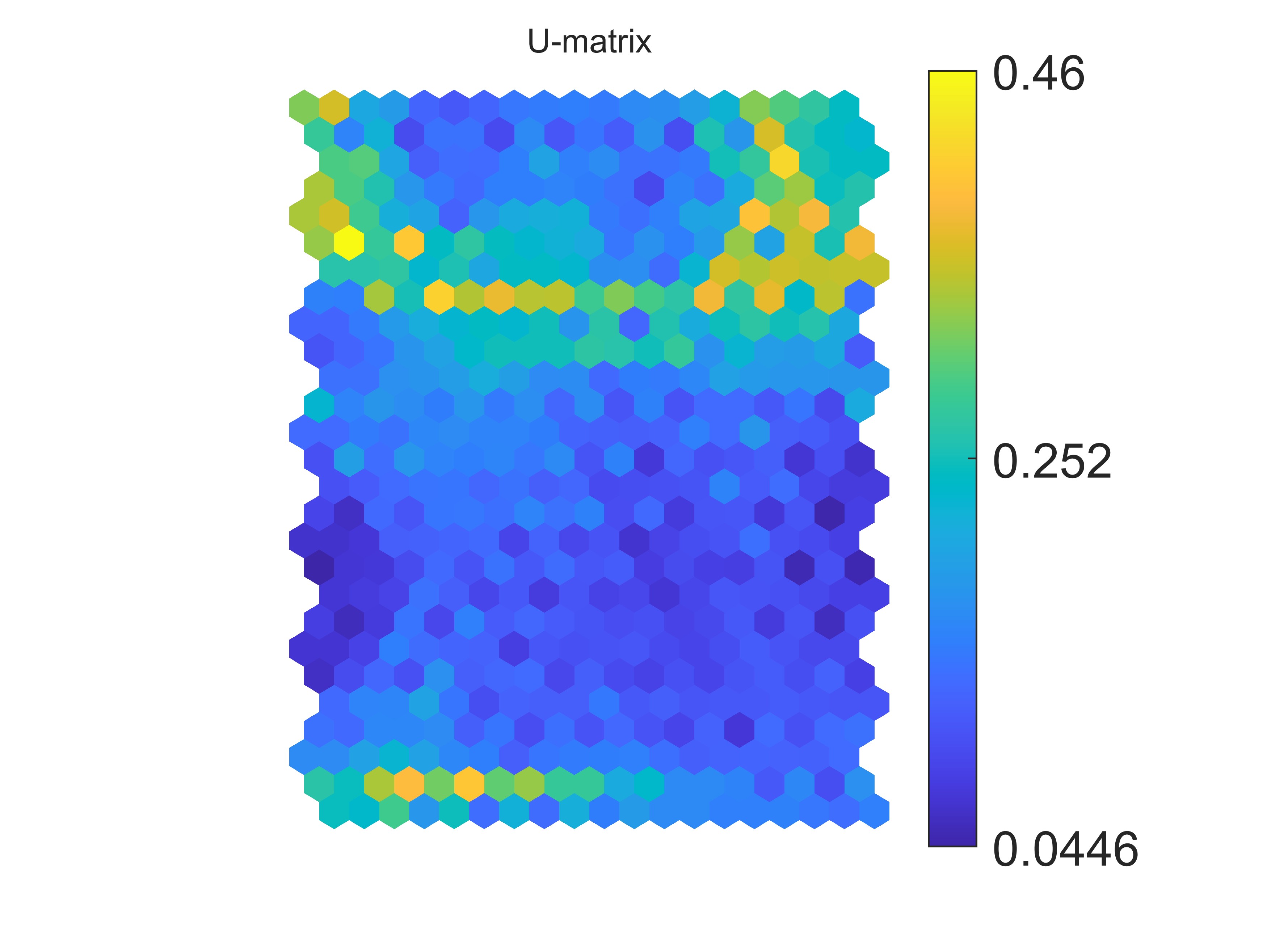}
    \caption{U-matrix}
    \label{boxplot_max}
    \end{subfigure}
\begin{subfigure}[t]{0.45\textwidth}  
\setcounter{subfigure}{3} 
    \includegraphics[width=\textwidth]{ 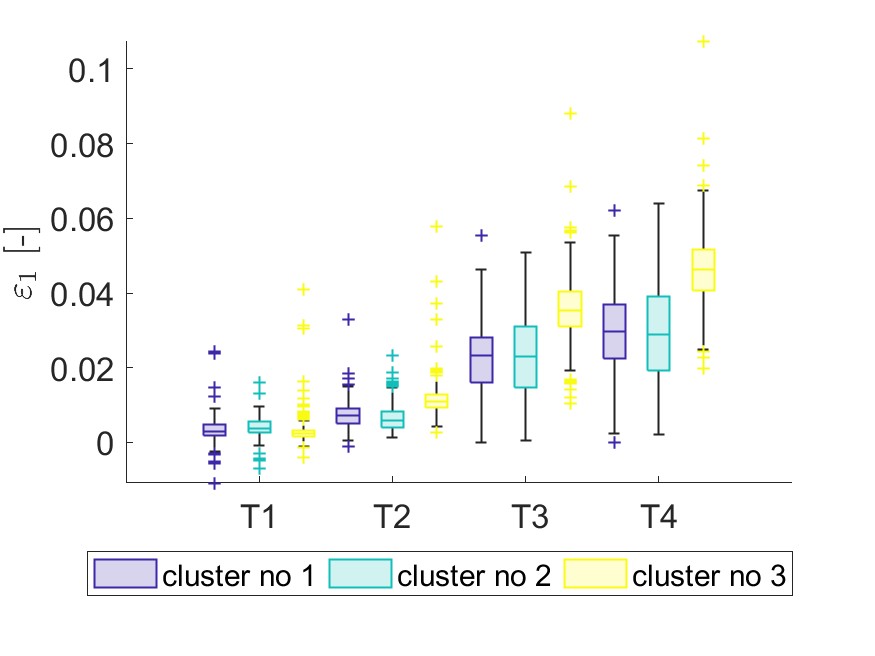}
   \caption{ Boxplot of the maximum principal strain  $\varepsilon_1$ for each cluster}
    \end{subfigure}

\begin{subfigure}[t]{0.45\textwidth}  
\setcounter{subfigure}{1} 
    \includegraphics[width=\textwidth]{ 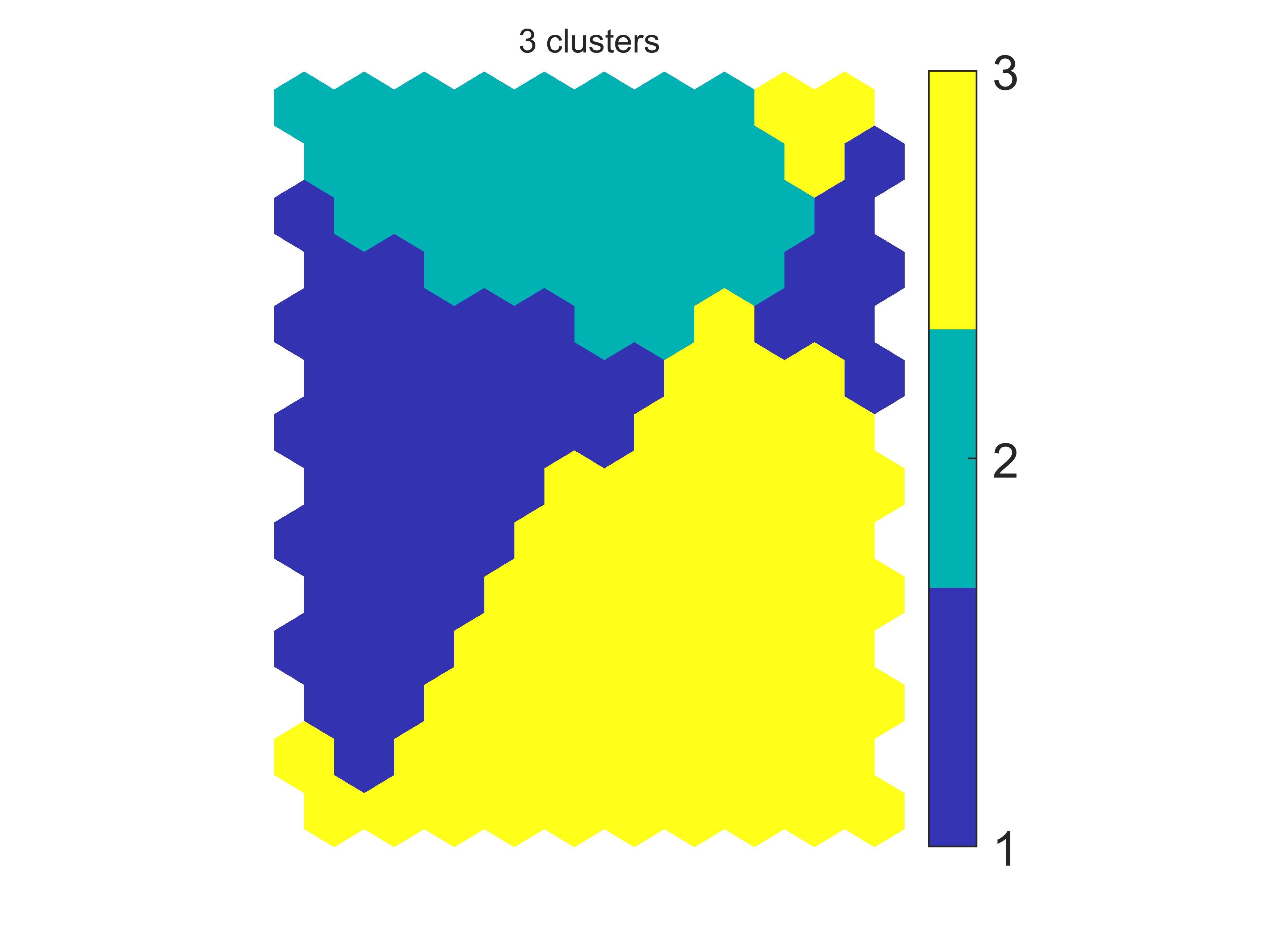}
    \caption{SOM clustering}
    \label{boxplot_max}
\end{subfigure}
\begin{subfigure}[t]{0.45\textwidth}  
\setcounter{subfigure}{4} 
    \includegraphics[width=\textwidth]{ 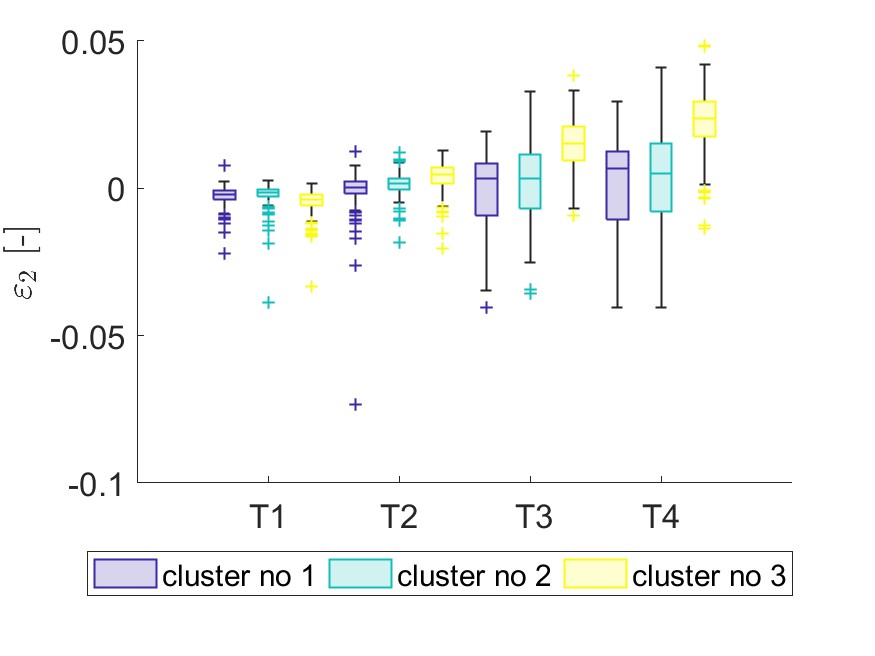}
    \caption{ Boxplot of the minimum principal strain  $\varepsilon_2$ for each cluster}
    \label{boxplot_min}
    \end{subfigure}
\begin{subfigure}[t]{0.45\textwidth}  
\setcounter{subfigure}{2}
    \includegraphics[width=\textwidth]{ 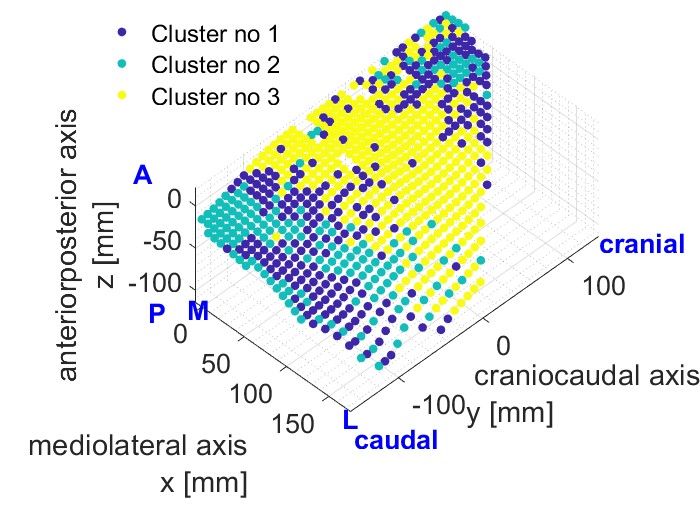}
    \caption{ {Clusters found by SOM marked on the abdominal wall surface}}
    \label{boxplot_max}
\end{subfigure}
\begin{subfigure}[t]{0.45\textwidth}  
\setcounter{subfigure}{5}
    \includegraphics[width=\textwidth]{ 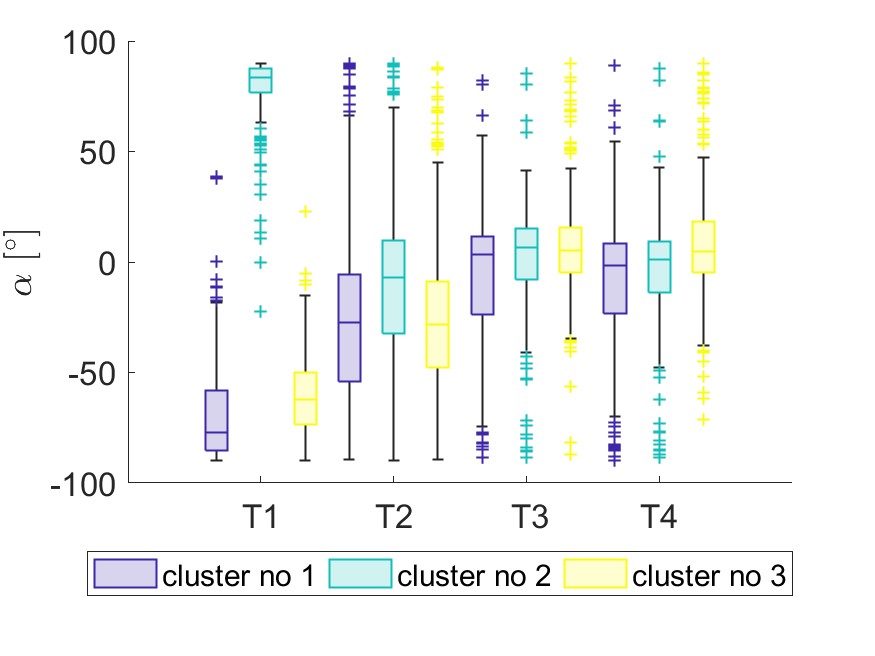}
    \caption{Boxplot of the principal direction $\alpha$ for each cluster  }
    \end{subfigure}
  
    \caption{Cluster results obtained by SOM in case of subject D9 (male, 46 years old, BMI 27.4 kg/m$^2$, intra-abdominal pressure 20 cmH$_2$O)} \label{fig_som_result_dic9}
\end{figure}

\begin{figure}[ht]\centering
\begin{subfigure}[t]{0.45\textwidth}  
\setcounter{subfigure}{0} 
    \includegraphics[width=\textwidth]{ 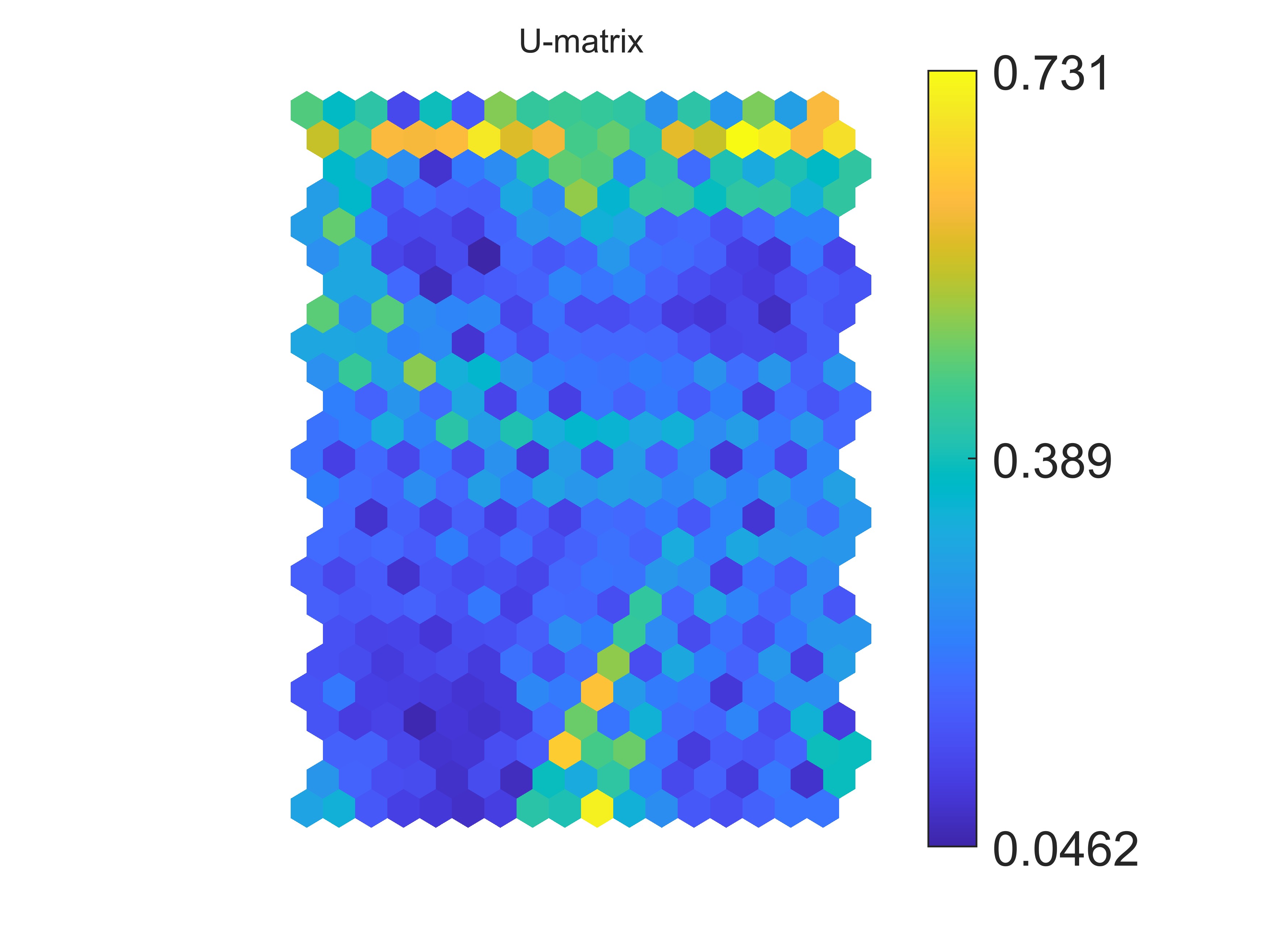}
    \caption{U-matrix}
    \label{boxplot_max}
    \end{subfigure}
\begin{subfigure}[t]{0.45\textwidth}  
\setcounter{subfigure}{3} 
    \includegraphics[width=\textwidth]{ 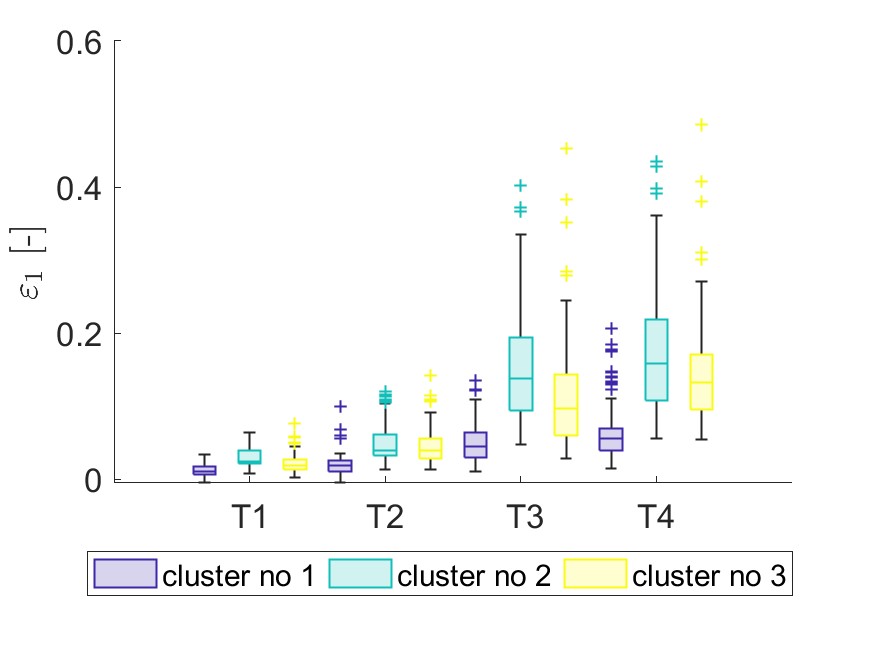}
   \caption{ Boxplot of the maximum principal strain  $\varepsilon_1$ for each cluster}
    \end{subfigure}

\begin{subfigure}[t]{0.45\textwidth}  
\setcounter{subfigure}{1} 
    \includegraphics[width=\textwidth]{ 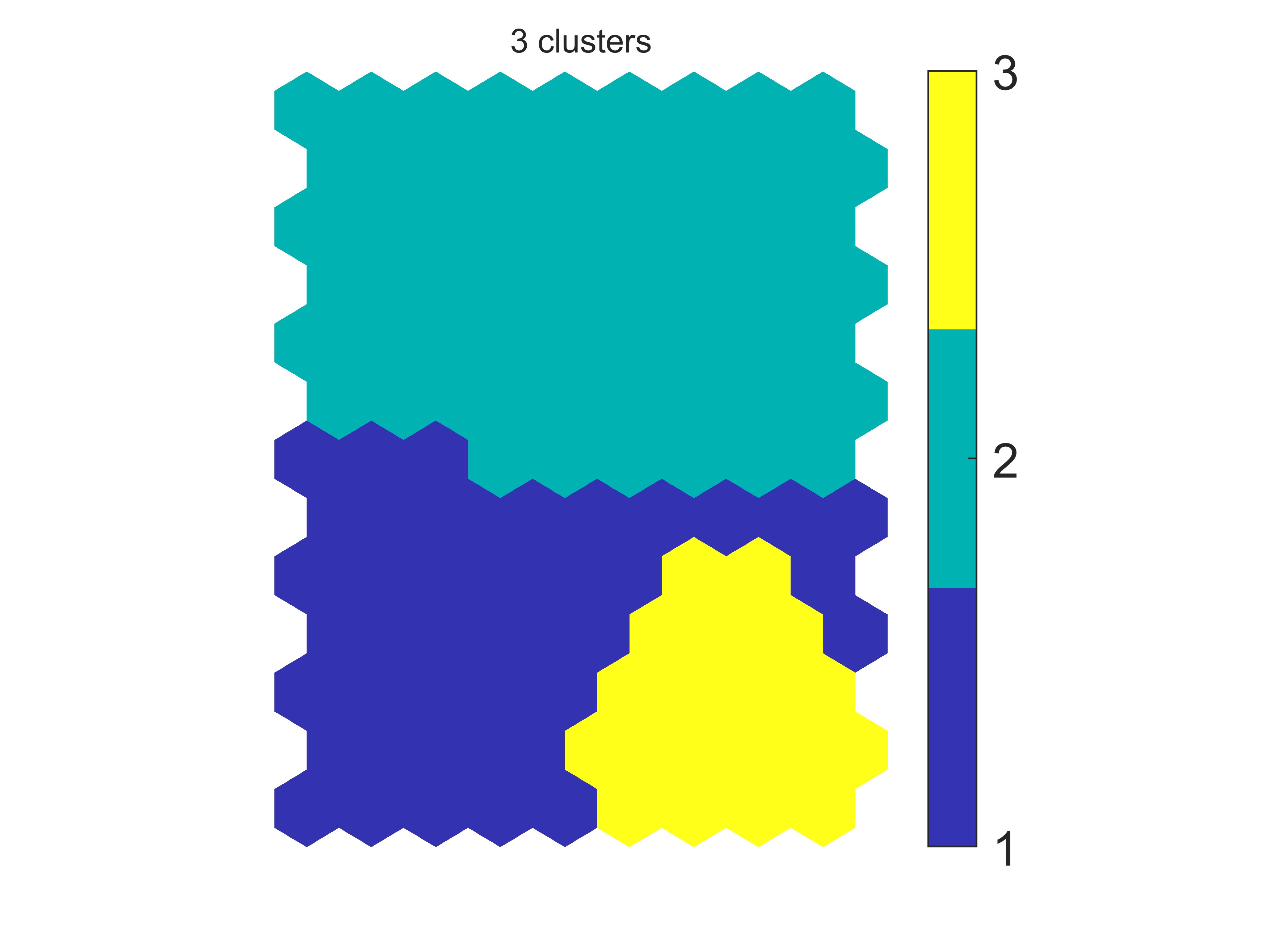}
    \caption{SOM clustering}
    \label{boxplot_max}
\end{subfigure}
\begin{subfigure}[t]{0.45\textwidth}  
\setcounter{subfigure}{4} 
    \includegraphics[width=\textwidth]{ 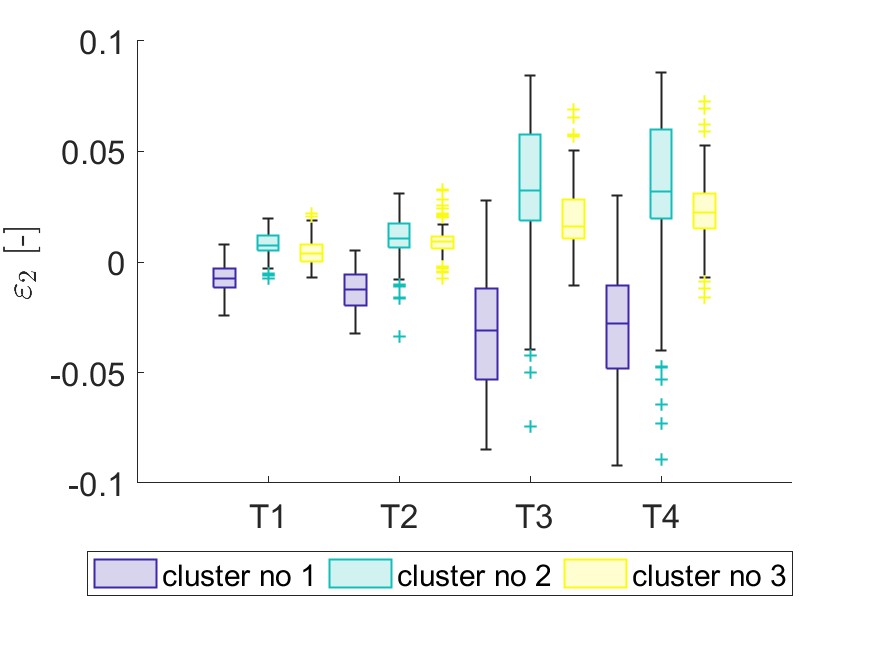}
    \caption{ Boxplot of the minimum principal strain  $\varepsilon_2$ for each cluster}
    \label{boxplot_min}
    \end{subfigure}
\begin{subfigure}[t]{0.45\textwidth}  
\setcounter{subfigure}{2}
    \includegraphics[width=\textwidth]{ 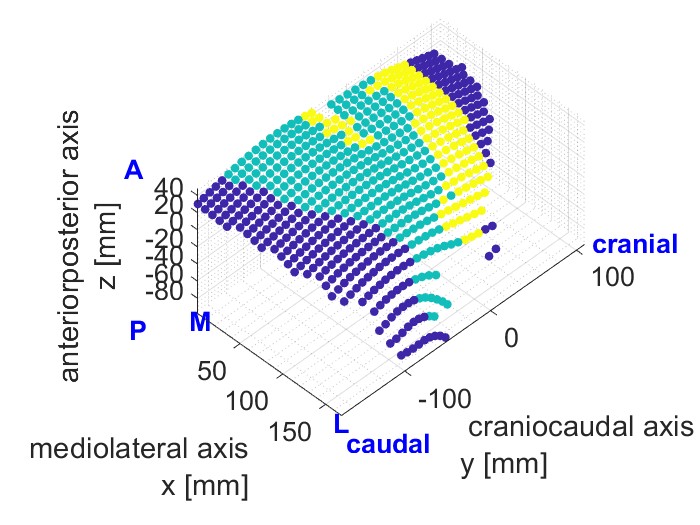}
    \caption{ {Clusters found by SOM marked on the abdominal wall surface}}
    \label{boxplot_max}
\end{subfigure}
\begin{subfigure}[t]{0.45\textwidth}  
\setcounter{subfigure}{5}
    \includegraphics[width=\textwidth]{ 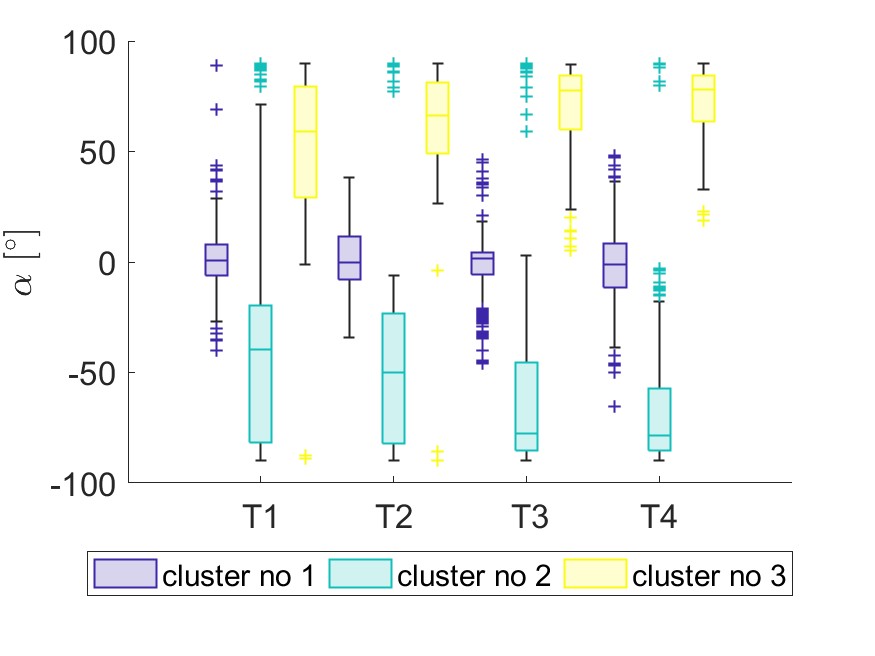}
    \caption{Boxplot of the principal direction $\alpha$ for each cluster  }
    \end{subfigure}
  
    \caption{Cluster results obtained by SOM in case of subject D10 (female, 72 years old, BMI 25.7 kg/m$^2$, intra-abdominal pressure 12 cmH$_2$O)} \label{fig_som_result_dic10}
\end{figure}

\begin{figure}[ht]\centering
\begin{subfigure}[t]{0.45\textwidth}  
\setcounter{subfigure}{0} 
    \includegraphics[width=\textwidth]{ 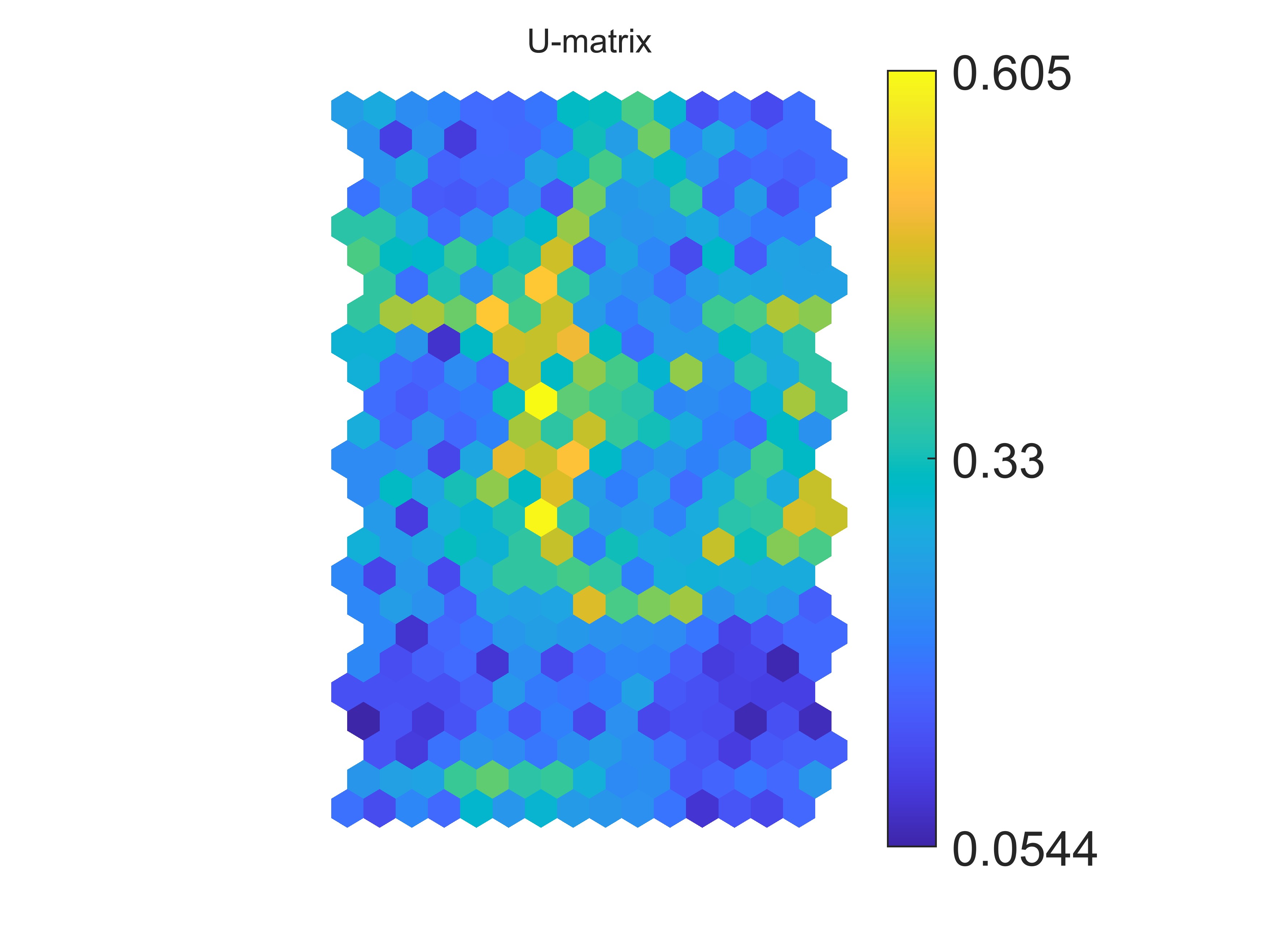}
    \caption{U-matrix}
    \label{boxplot_max}
    \end{subfigure}
\begin{subfigure}[t]{0.45\textwidth}  
\setcounter{subfigure}{3} 
    \includegraphics[width=\textwidth]{ 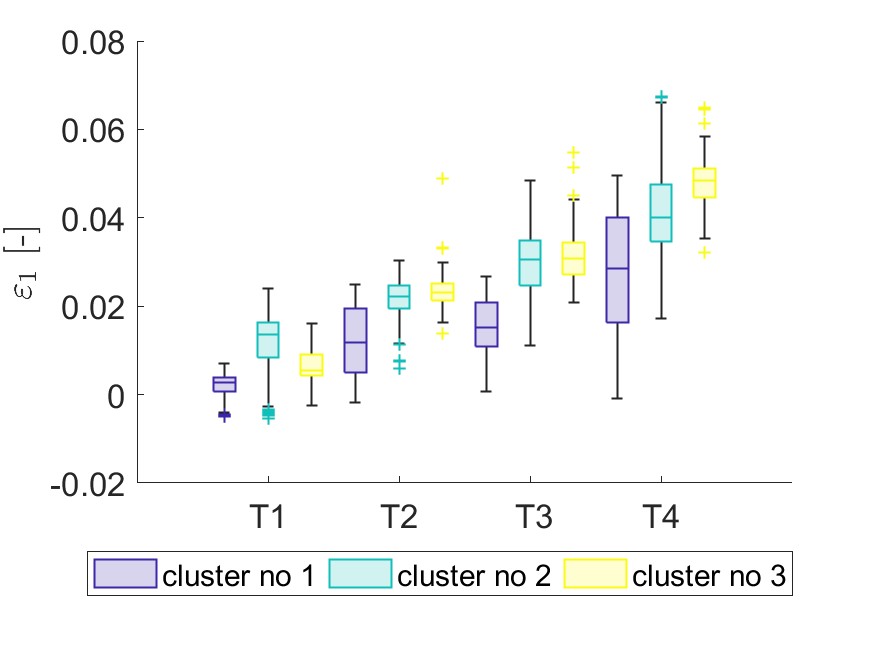}
   \caption{ Boxplot of the maximum principal strain  $\varepsilon_1$ for each cluster}
    \end{subfigure}

\begin{subfigure}[t]{0.45\textwidth}  
\setcounter{subfigure}{1} 
    \includegraphics[width=\textwidth]{ 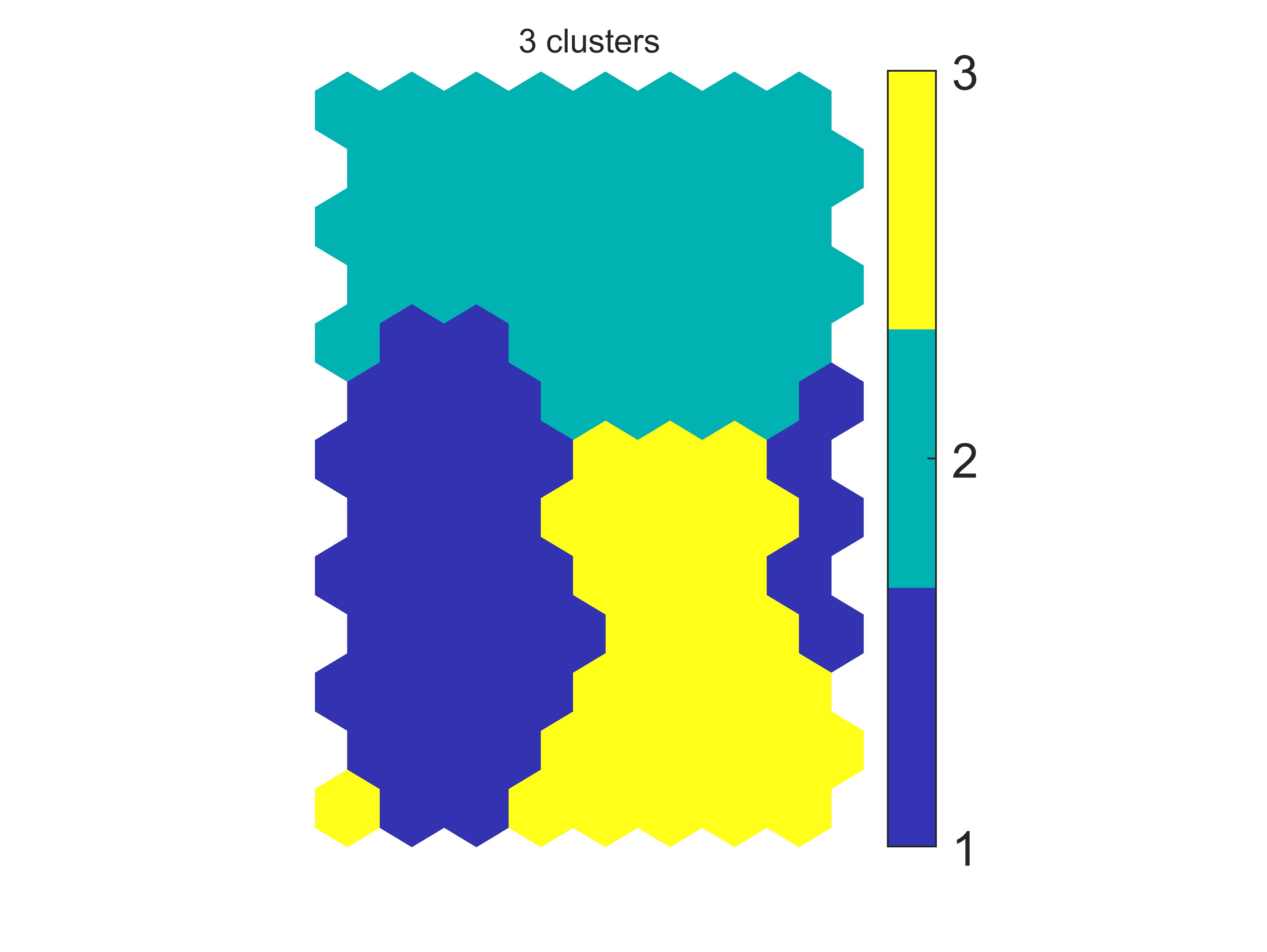}
    \caption{SOM clustering}
    \label{boxplot_max}
\end{subfigure}
\begin{subfigure}[t]{0.45\textwidth}  
\setcounter{subfigure}{4} 
    \includegraphics[width=\textwidth]{ 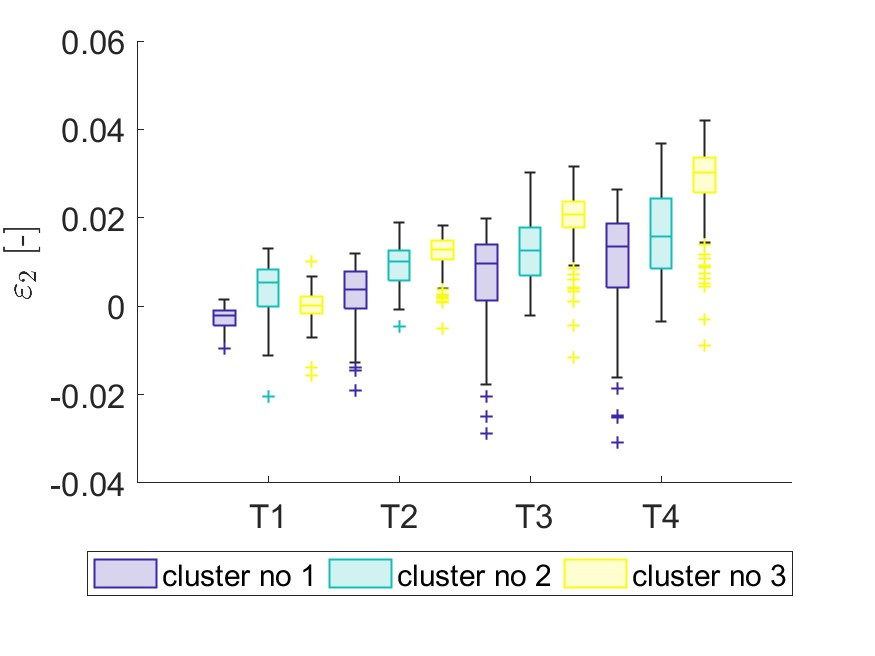}
    \caption{ Boxplot of the minimum principal strain  $\varepsilon_2$ for each cluster}
    \label{boxplot_min}
    \end{subfigure}
\begin{subfigure}[t]{0.45\textwidth}  
\setcounter{subfigure}{2}
    \includegraphics[width=\textwidth]{ 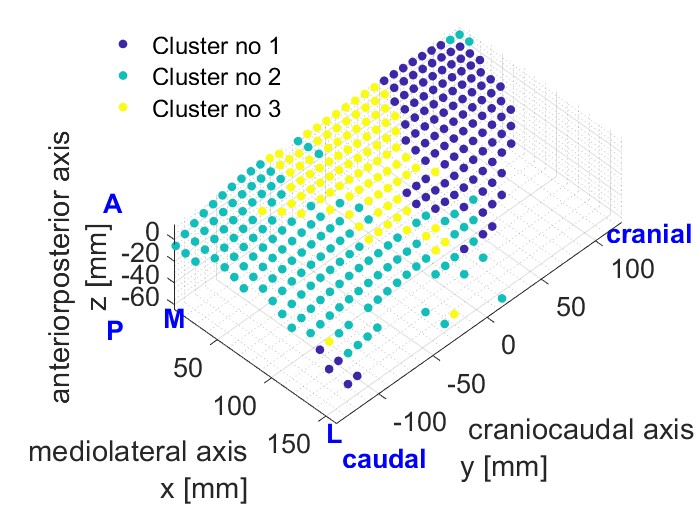}
    \caption{ {Clusters found by SOM marked on the abdominal wall surface}}
    \label{boxplot_max}
\end{subfigure}
\begin{subfigure}[t]{0.45\textwidth}  
\setcounter{subfigure}{5}
    \includegraphics[width=\textwidth]{ 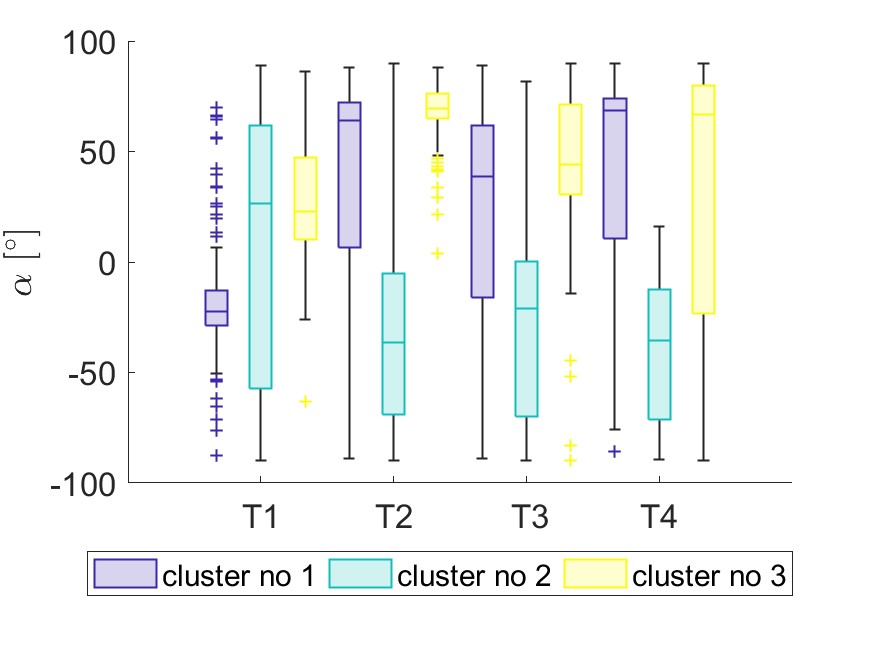}
    \caption{Boxplot of the principal direction $\alpha$ for each cluster  }
    \end{subfigure}
  
    \caption{Cluster results obtained by SOM in case of subject D11 (male, 36 years old, BMI 27.8 kg/m$^2$, intra-abdominal pressure 18 cmH$_2$O)} \label{fig_som_result_dic11}
\end{figure}

\begin{figure}[ht]\centering
\begin{subfigure}[t]{0.45\textwidth}  
\setcounter{subfigure}{0} 
    \includegraphics[width=\textwidth]{ 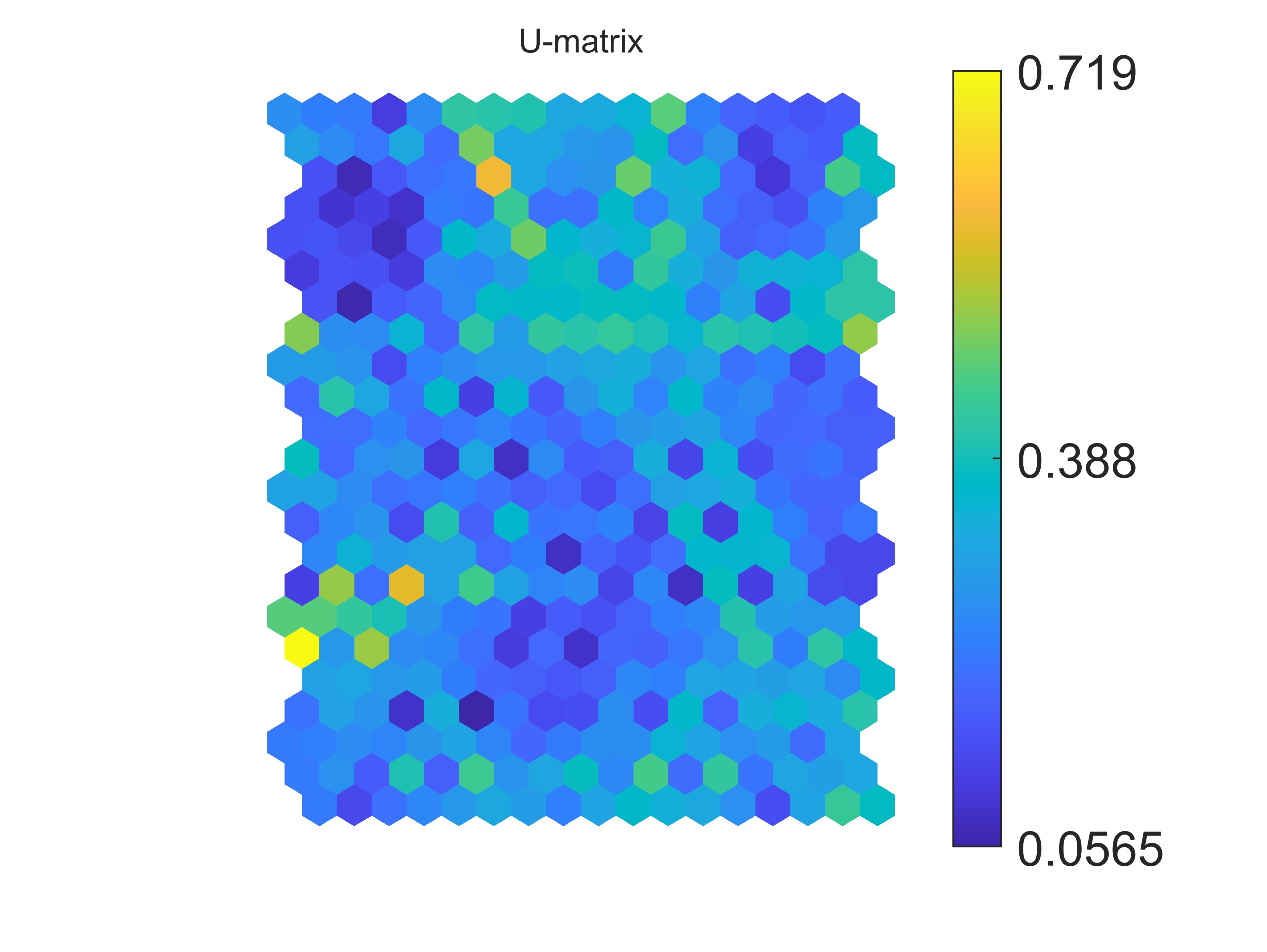}
    \caption{U-matrix}
    \label{boxplot_max}
    \end{subfigure}
\begin{subfigure}[t]{0.45\textwidth}  
\setcounter{subfigure}{3} 
    \includegraphics[width=\textwidth]{ 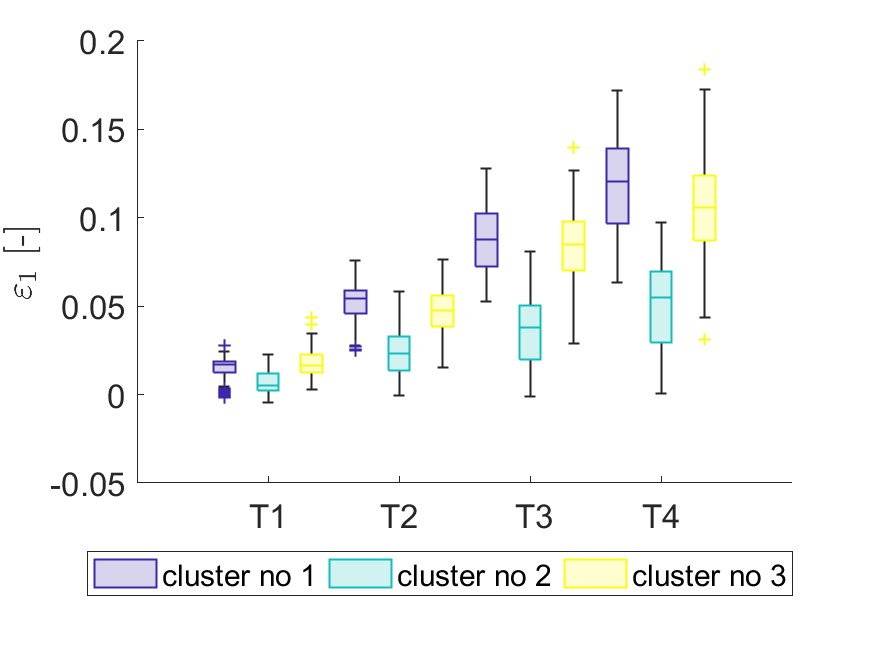}
   \caption{ Boxplot of the maximum principal strain  $\varepsilon_1$ for each cluster}
    \end{subfigure}

\begin{subfigure}[t]{0.45\textwidth}  
\setcounter{subfigure}{1} 
    \includegraphics[width=\textwidth]{ 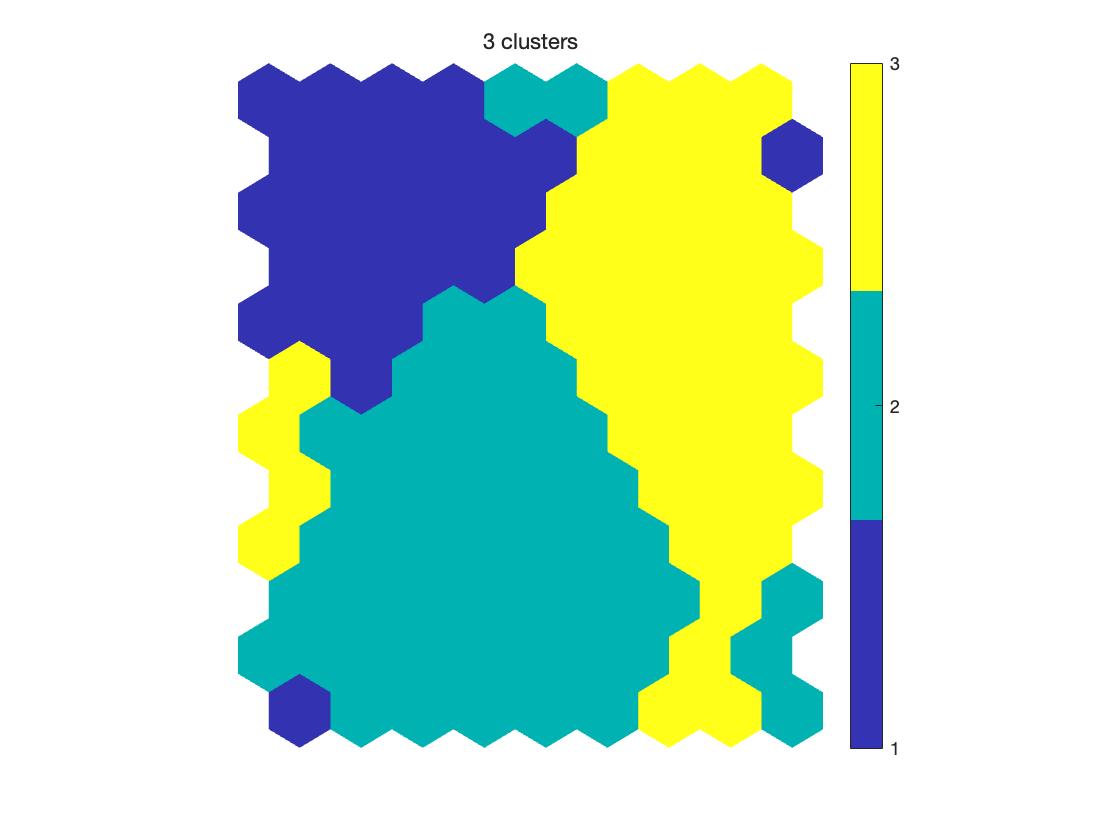}
    \caption{SOM clustering}
    \label{boxplot_max}
\end{subfigure}
\begin{subfigure}[t]{0.45\textwidth}  
\setcounter{subfigure}{4} 
    \includegraphics[width=\textwidth]{ 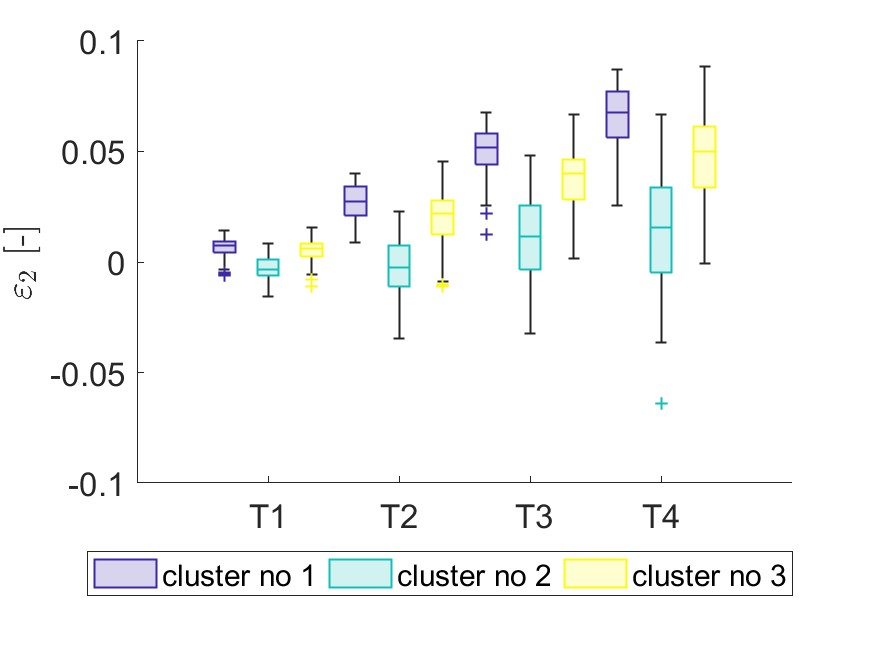}
    \caption{ Boxplot of the minimum principal strain  $\varepsilon_2$ for each cluster}
    \label{boxplot_min}
    \end{subfigure}
\begin{subfigure}[t]{0.45\textwidth}  
\setcounter{subfigure}{2}
    \includegraphics[width=\textwidth]{ 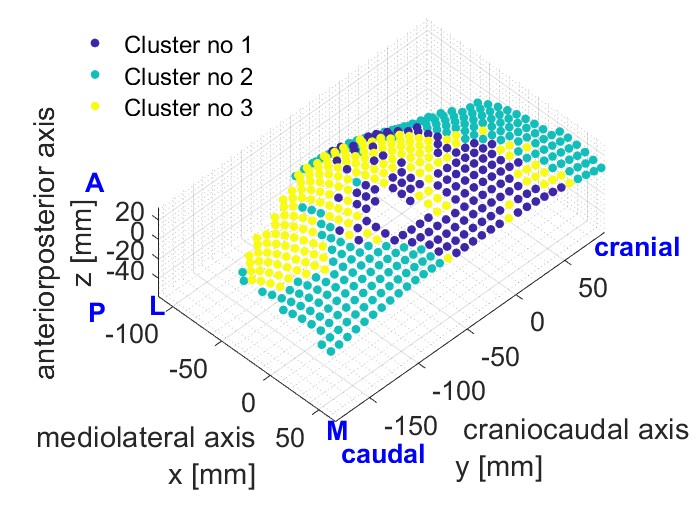}
    \caption{ {Clusters found by SOM marked on the abdominal wall surface}}
    \label{boxplot_max}
\end{subfigure}
\begin{subfigure}[t]{0.45\textwidth}  
\setcounter{subfigure}{5}
    \includegraphics[width=\textwidth]{ 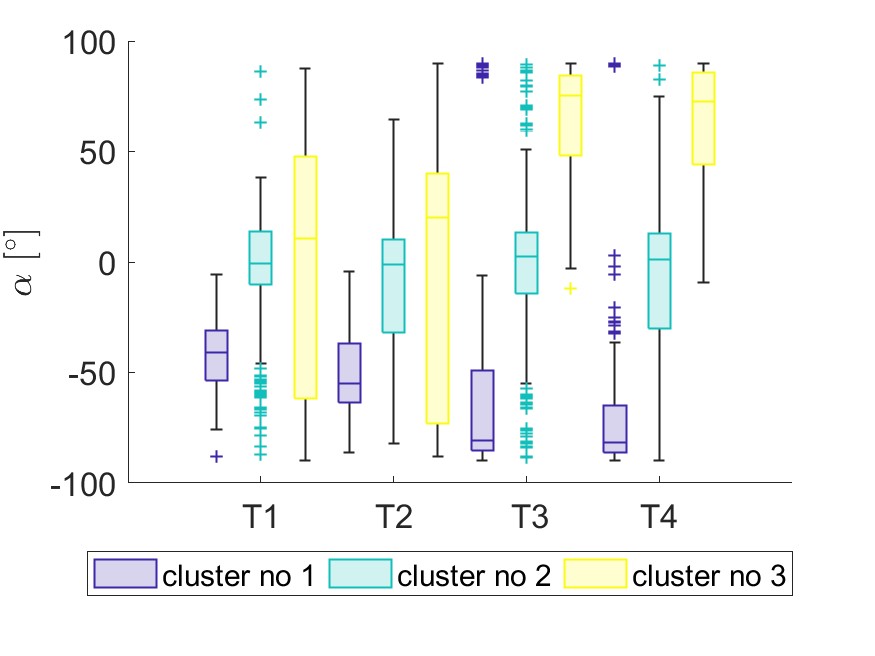}
    \caption{Boxplot of the principal direction $\alpha$ for each cluster  }
    \end{subfigure}
  
    \caption{Cluster results obtained by SOM in case of subject D12 (male, 56 years old, BMI 25.8 kg/m$^2$, intra-abdominal pressure 10 cmH$_2$O)} \label{fig_som_result_dic12}
\end{figure}

The component planes are compiled to create the U-matrix for the given subject (e.g. Figure \ref{fig_som_result_dic1}). Thus, we can explore the range of data for the respective variable and the influence it has on the clusters in the final U-matrix map. The U-matrix is built in a toroidal shape, whose 2D representation is presented on a rectangular map. Here the clusters are visible as darker areas separated by light borders. The results of multidimensional SOM analyses are presented in Figures \ref{fig_som_result_dic1} - \ref{fig_som_result_dic12}. These show the U-matrices generated by the SOM for each subject (a), the clusters identified by SOM (b), the same clusters mapped on the surface of the abdominal wall (c)  boxplots presenting the range of principal strains and their directions from the grid points of each cluster in T1--T4 (d--f). Half of the abdominal wall is shown in (c), which corresponds to the region of interest analysed by DIC. Blue symbols  {M, L,} P and A  denote the sides of the 3D view:  {medial, lateral}, posterior and anterior, respectively.
The clusters mapped on the abdominal wall geometry can be interpreted as zones of similar mechanical behaviour under pressure as indicated by the SOM, due to the three input vector quantities: maximum and minimum principal Lagrangian strains and the principal direction. In the case of subjects D1 to D8, SOM suggested two clusters, whereas in subjects D9--D12, three clusters were found.

The values of principal strains and directions in SOM specified clusters are shown in the form of boxplots. In each boxplot, the central line is the median, the edges of the box are the 25th and 75th percentiles (and the distance between them is interquartile range), and the whiskers extend to the furthest data points with a maximum length of 1.5 times the interquartile range, any points beyond them are outliers marked by a ‘+’.  {Additionally, the numerical values of the median are presented in Tables \ref{tab_app_d1}--\ref{tab_app_d12} in  \ref{appendix_stat}.}

\subsection{Assessment of the quality of the results obtained from SOM analysis}

\begin{figure}[ht!]\centering
    \includegraphics[width=\textwidth]{ 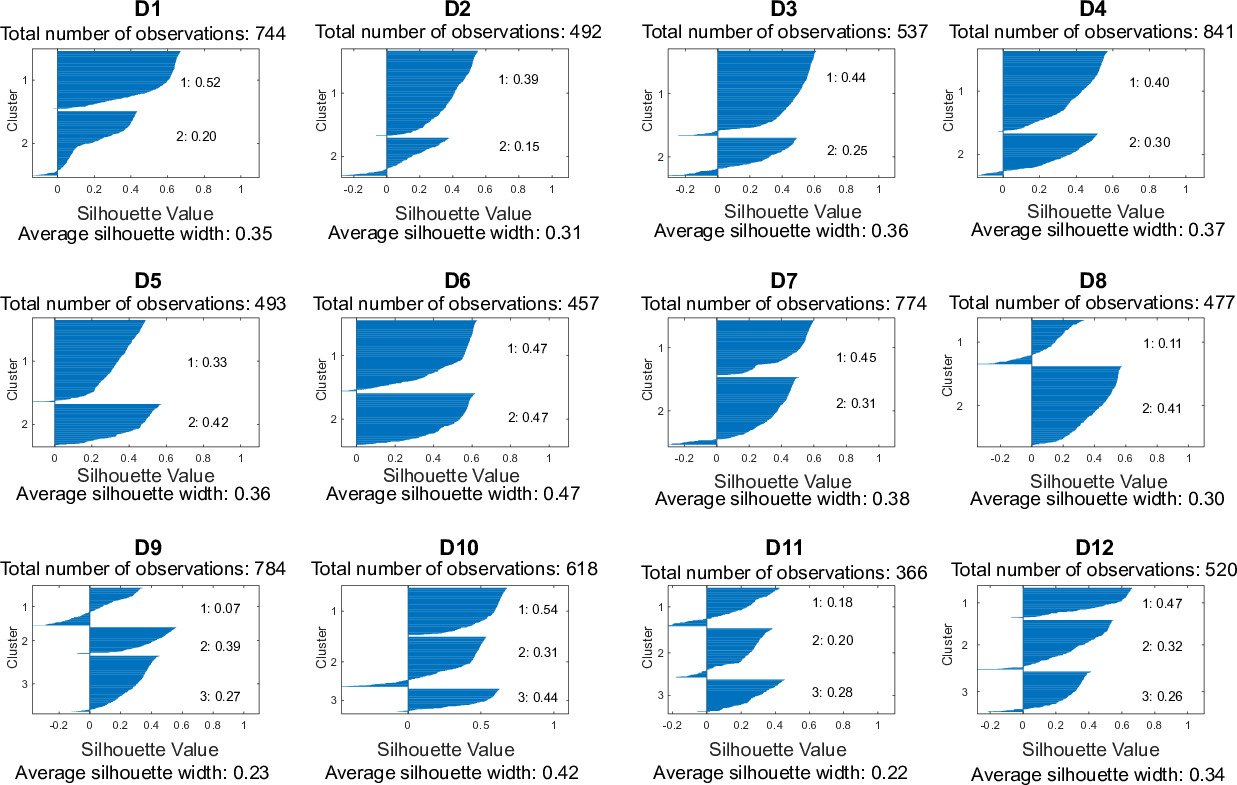}
    \caption{ {Silhouette plots of clusters obtained for patients datasets D1 to D12: total number of observations (on top of each plot); averaged silhouette width for objects in the cluster (next to the silhouette widths of objects classified to the cluster); averaged silhouette width for the entire dataset (below the plots)}}
    \label{DIC1-DIC12_silhouettes}
\end{figure}

 {Topographic error of our results range from 0.08 to 0.12 and quantization error from 0.18 to 0.22.}
 {Calculated silhouette plots for each object classified into respective clusters are shown in Figure \ref{DIC1-DIC12_silhouettes}. Although the average silhouette width for the entire dataset can be assessed as weak according to \cite{Kaufman1990} (perhaps due to data being collected from distant, side parts of the torso or zones close to the ribs), the average silhouette width for specific clusters shows an overall better classification. Namely, cluster no. 1 had the best silhouette for patients D1, D3, D4, D6, D7, D10 and D12. The best quality of cluster no 2 was achieved for subjects D5, D6 and D9. Overall, for the entire data set, the best average silhouette width was obtained for patients D6 with a score of 0.47 and D10 with a score of 0.42. In the future, different methods on cluster evaluation identified by SOM may be considered. Even if this overall result for the dataset is not ideal looking at the silhouette widths analysis, the U-matrices of SOM display better structure of clusters.}


\section{ {Discussion}}

\subsection{ {Clustering of the abdominal wall surface}}

 {Multi-layer anatomy of abdominal wall (Figure \ref{Abdominal wall}) suggests different mechanical behaviour of the lateral part of abdominal wall containing three muscles (external oblique, internal oblique and transversus abdominis) comparing to rectus abdominis muscle. The maps of parameters for isotropic non-linear material law obtained from inverse analysis of  \textit{in vivo} gathered data on rabbits by Simón-Allué et al. in \cite{simon2017towards} are coinciding with such anatomical division. In that study only the passive behaviour of the abdominal wall subjected to an inflation test was considered because anesthesia was used. However, Podwojewski et al. \cite{podwojewski2014mechanical} based on an \textit{ex vivo} experiment on humans,  concluded that the strain pattern on the inner surface of the abdominal wall reflects its anatomical structure to a much greater extent than the outer surface, where the strain field is more homogeneous.}

 {Alteration of the mechanical behaviour of the abdominal wall could also be expected along the longitudinal direction due to the following reasons: (1) rectus sheath has different mechanical properties and changes structure within arcuate line \cite{rath1997sheath}; (2) linea alba has different morphology and mechanical properties in the upper and lower part of the abdominal wall \cite{grabetael2005anisotropy}; (3) external oblique, internal oblique and transversus abdominis muscles thickness and fascicle orientation changes along the longitudinal direction of the abdominal wall \cite{urquhart2005regional}. It should be noted, that the layers of abdominal wall also differ in the fiber alignment. Rectus sheath \cite{whitehead2023characterisation} and linea alba, the components important in passive mechanical response of the abdominal wall, are stiffer in transverse direction, as claimed in \cite{astruc2018characterization}.}  {Taking into account the complexity of the abdominal wall, European Hernia Society classification divides hernia in terms of localisation into a medial (or midline) zone and a lateral zone, where both of the zones are also divided to subareas along longitudinal direction (medial part to 5 subzones, lateral to 3), \cite{muysoms2009classification}.}

%
As can be seen in the results, the clusters mapped on the abdominal wall geometry often indicate zones aligned to mediolateral axis of the human body (Figure \ref{fig_som_result_dic10}c). In some cases when there are more of them, clusters found in zones close to the rib and hip bones are skewed, as can be seen in the aforementioned figures (abdominal wall top and bottom).
In the case of Subjects D1 and D6 the network identified  zones separated by the navel line (one cluster above and one below the navel). Subjects D2, D3, D5, D7, D8, D9, D10, D11 and D12 have a cluster around the navel while the area around it is indicated as different. Subjects D4 and D7 show different results regarding to the cluster shape. In these cases, the  {clusters around the midline separated by the navel are distinguished}. So again the navel line separates zones detected as different. 

 {An important issue is also the active behaviour of the abdominal wall \cite{karami20233d}, that influences its mechanical response \cite{pavan2019effects}.}  {Our datasets taken to the SOM analysis reflect the mechanical response of breathing subjects and include both, exhalation and inhalation, phases. Therefore, the outcome may be different than those based only on passive behaviour of the abdominal wall. The clusters obtained by SOM differ between subjects and are not clearly corresponding to the anatomical regions. Nevertheless, in some subjects (D4, D5, D6, D8, D10--D12) the clusters mainly divided the abdominal wall in transverse direction, which may confirm distinct properties along the longitudinal direction, or be a results of boundary conditions, which are difficult to determine precisely. In case of other subjects (eg., D3 and D12), medial part is also distinguished by SOM from the lateral one.} 

 {When looking at all the subjects, one could expect a similar number and locations of clusters however, as often in biomechanics, the variability between subjects was observed. This may result from heterogeneity of the subjects group including female and male in various age and with different muscularity and body mass-indices (BMI), which all may cause differences in the outcomes. Additionally, different types of breathing amongst subjects may influence the results. } 

\subsection{ {Differences between clusters}}

As mentioned before, SOM analyses all input data simultaneously, which raises questions as to the ranges of individual variables in the identified clusters, and as to whether only one variable is sufficient to assess the evolution of deformations in the abdominal wall under pressure. In most subjects, the median of principal strains differ between clusters. Also the ranges of principal strains and their direction values, represented by the boxplots, vary between clusters. This means that one may find a range of dominating values in each cluster which could be used to fit the surgical mesh to a specific area. 
 The separation is not the same in all studied subjects. In some cases, the principal strain values vary more between clusters, while in other cases, their directions vary, but the principal strain values are relatively close. There are also cases (D4, D5 and D6) where all three analysed quantities differ significantly, especially under dialysis fluid pressure, unlike in most other cases, where the direction of maximum principal strains predominates. In view of this, it would be simplistic to specify one variable as the strongest factor in clustering and drawing conclusions on the basis of a single variable would be risky.
 Moreover, it may be observed that the key factors are different in different time steps. In some subjects, e.g. D2, D4, D5, D7 and D8, the maximum principal strain varies significantly under fluid pressure, while before filling these differences are relatively small. In D12, the principal strain does not differ much between its three clusters. Here, the direction of strains dominates in T3 and T4, while the values of principal strains overlap. The overlaps are small in the case of minimum principal strains (clusters 1 and 3), in other words, the minimum principal strain values differ more between the clusters than in the cases of maximum principal strains. 
 In case of subjects D3, D6, D7 and D10, the ranges of principal directions differ very much between clusters, their data ranges between the 25th and 75th percentile (between top and bottom boxplot edges) are completely separate, especially under the dialysis fluid pressure. Similar situations apply to  principal strain values in the case of subjects D2 and D5. 
In summary, it may be said that four time steps and different variables provide a deeper insight into abdominal wall deformation and evolution under pressure than can be observed on the basis of a single mechanical quantity.  

In the case of subjects with three SOM clusters, it is much more difficult to infer about the dominating values, since their ranges in different clusters sometimes overlap. The overlaps are neither particularly diverse nor uniform in all phases (T1--T4). In the case of subject D9, it should be noted that in cluster 2, the principal strain direction in stage T1 differs more from the others. In turn, the principal strain values in cluster 3 when the abdominal cavity is filled, differ markedly from the others.

The median of the principal directions between the clusters of most subjects changes. An exception is subject D5, where the difference in principal directions between clusters only occurs under the dialysis fluid pressure (T3, T4). The opposite was observed for subjects D1, D2 and D9, where the difference in the median decreases with loading. In most cases (D3, D4, D6, D7, D10 and D12) the principal directions differ significantly between clusters in most time steps. It should also be noted, that both dialysis fluid pressure and breathing affect the directions of principal strains. 

Dominating directions of principal strains in various zones of the abdominal wall should be considered when planning hernia repairs. This is because with such information, orthotropic or anisotropic implants may be positioned to function in a mechanically more compatible way with the living tissue. 
The ability to determine several abdominal wall clusters on the basis of principal strains and their directions may facilitate the use of an appropriate implant in the correct direction for the given hernia location.

\subsection{ {Comparison with contour maps  based on 1-variable-at-a-time approach}}

\begin{figure}[ht!]\centering
    \includegraphics[width=\textwidth]{ 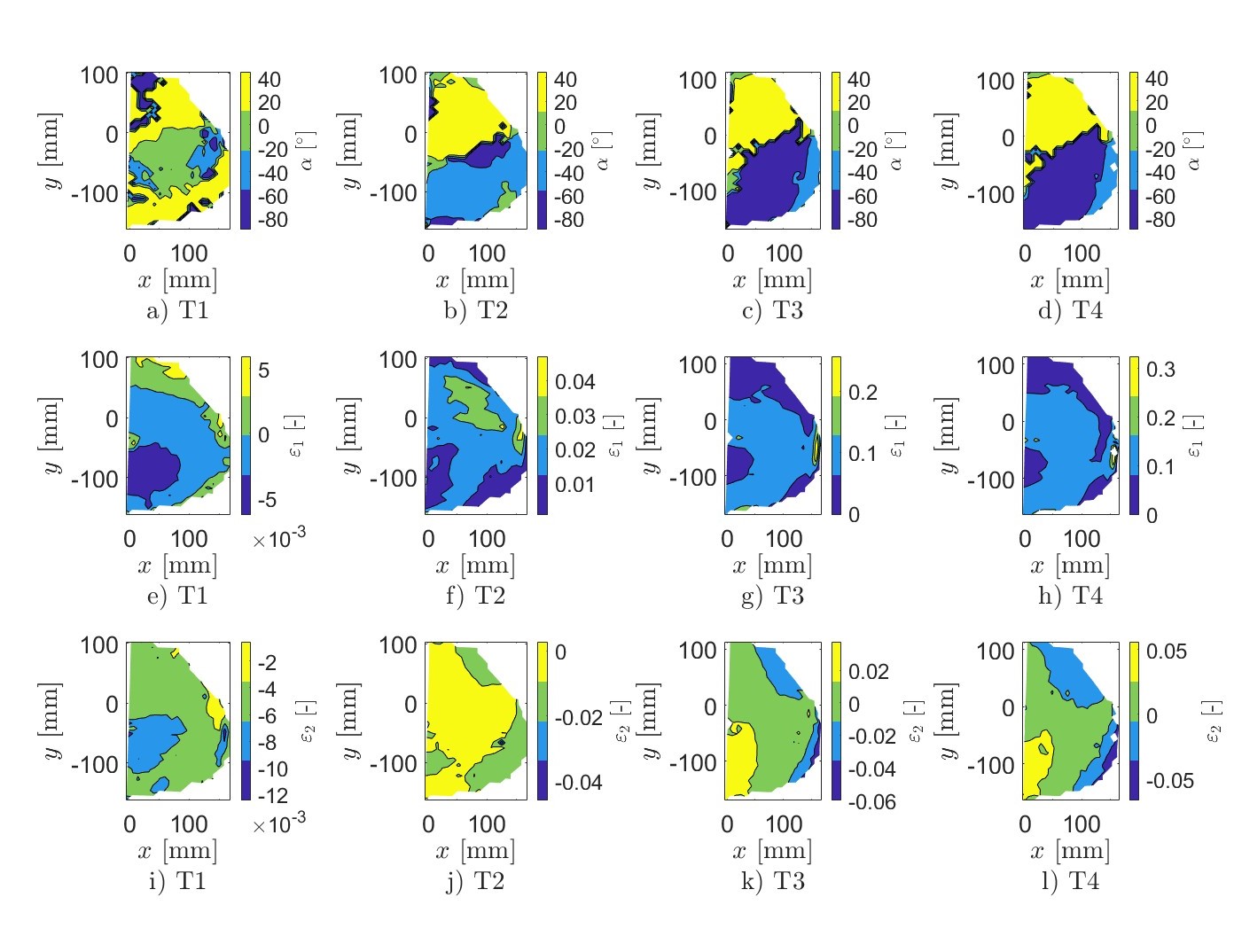}
    \caption{Contour maps with 3 levels of isolines of subject D6 {, where $x$, $y$ axes indicates coordinates on the abdominal wall and color scale indicates}: principal direction angle $\alpha$  {[$^\circ$]} a)--d), principal strains $\varepsilon_1$  {[-]} e)--h) and $\varepsilon_2$   {[-]} (i--l);  {---figure adapted from \cite{szepietowska2023fullfield}}}
    \label{figcontourD6}
\end{figure}

The results obtained in SOM analyses may be compared with the contour maps based on experimental data to see how they relate to the distributions of single variables (isolines). 
 The contour maps of the principal Lagrangian strain values $\varepsilon_1$, $\varepsilon_2$ and their directions angle  $\alpha$ observed on the abdominal walls of subject D6 in the four time steps (T1--T4) is shown in Figure \ref{figcontourD6}. A detailed analysis of the contours for all 12 subjects is presented in \cite{szepietowska2023fullfield}. The contour maps contain isolines on the x-y plane of half of the abdominal wall surface, showing the range of values observed in different areas of the wall.

As discussed in \cite{szepietowska2023fullfield}, the shapes of the areas separated by the isolines are different for each subject and among the analysed variables even with regard to one subject. Thus, the use of SOM, which allows for the simultaneous inclusion of many variables, may be beneficial in this application.
In the case of subject D6 (Figure \ref{figcontourD6}), it may be said that although all the data used in SOM were normalised from 0 to 1, there is some similarity between the contour maps and the SOM clusters. This may indicate the higher influence of a certain variable on clustering.  In this case, the variable is the distribution of the direction of principal strains ($\alpha$). The two clusters resemble two ($\alpha$) zones on the contour maps. However, other variables, most notably the distribution of the principal strain values $\varepsilon_1$ and $\varepsilon_2$, are sometimes different.   On the contour maps, they may take a semicircular shape, generally overlapping both clusters.

Contour maps for other subjects are presented in \cite{szepietowska2023fullfield}. A comparison of those contour maps and SOM clusters reveals that the most influential variable differs considerably between other subjects. In some cases, there are similarities between one or two variables in the contour maps and SOM clusters. For example, the variables $\varepsilon_1$ and $\alpha$ are similar in the case of one cluster for D1, as are variables $\varepsilon_1$ and $\varepsilon_2$ in one cluster for D8. There are also cases where two clusters look similar to the contour maps, as in D5 and D12 - two clusters for two principal strains and in the case of D7 with two clusters for the angle and maximum principal strain. In some cases, the similarities concern certain variables in only some of the time steps. Moreover, there are cases (D9 and D10) where the similarity concerns one variable in one cluster. The above observations concerning three variables $\alpha$, $\varepsilon_1$ and $\varepsilon_2$ in various time steps, suggest that certain areas of the abdomen may behave similarly under fluid pressure. 
Another observation is that SOM offers a fuller synthesis of the experimental data and shows a more complex correlation between the variables in different time steps than the contour maps.

%

\subsection{Limitations}
This study has some limitations that should be taken into account. The considered loading conditions are limited by the medical procedure. Hence, in order to obtain more general results, the analysis should be extended to more diverse physiological loading ranges and types. These could not be conducted due to safety concerns for the subjects (patients). Moreover,  {although surgical meshes used for hernia repair are attached to the inner layers of the abdominal wall, }the measurements were performed on the outer surface (skin) of the abdominal wall, which can be easily subjected to non-invasive optical measurements.
 { Podwojewski et al. in \cite{podwojewski2014mechanical} showed by \textit{in vitro} study that different pattern and values of strains can be observed on external and internal surface of the abdominal wall and contribution of every layer of the abdominal wall to the mechanical response of the whole structure was studied in \cite{tran2014contribution}}. Then, the mechanical behaviour of the whole complex structure could only be approximated.
Finally, more subjects should be examined, especially as in our case, female subjects were underrepresented in the tested group.

\section{Conclusions}

This study is based on full field measured data obtained from \textit{in vivo} tests on human subjects. It refers to the strain field of the human abdominal wall under changing pressure. 

An interesting phenomenon observed from the experiment is the changing  directions of principal strains on the surface of a pressurised abdominal wall. This probably means that different components of the abdominal wall become predominate during the passive and active work of abdominal muscles, depending on the intra-abdominal pressure level.
Such changes complicate analysis of different stages of simultaneous loading and necessitate the use of appropriate tools for the study of multidimensional data. For this purpose, in this study, we used Self-Organising Maps (SOM).  
SOM were applied to identify areas of human abdominal wall characterised by similar deformation state under pressure. The areas were represented by clusters of points on the abdominal wall. The resultant clustering showed that the deformation state varies depending on the given abdominal region and on the given subject.  {This variability indicates the need for a personalized approach to abdominal wall reconstruction procedures. However, to obtain more detailed indications, a larger test group is needed. }

The presented research may support the investigation of interaction between native tissue and prosthetic implants that would withstand physiological loading conditions, providing non-homogeneous and anisotropic mechanical properties similar to those of the human tissues, suitable for specific parts of the abdominal wall. This is particularly important because many surgical meshes, especially those characterised by increased stiffness compared to human tissues, cause postoperative pain and discomfort that have a strong impact on the patient's quality of life.

The use of Self-Organising Maps for the analysis of the experimental data may shed some light on the identification of mechanical properties of complex anisotropic and non-homogeneous materials by indicating areas of similar mechanical behaviour and thus, simplify the process.  {This can be achieved by identifying material parameters only for certain regions of the abdomen rather than for each datapoint. Then, the proposed methodology can be used as a first step in patient-specific abdominal wall characterization procedure.}

The study shows the strength of the proposed methodology using SOM to the analysis of deformation state of living human abdominal wall under pressure.
 { In this paper, we focused on patients with mostly healthy abdominal wall. The knowledge about mechanical behaviour of healthy abdominal wall is important as it can be a reference in finding the best treatment procedures. In the future, SOM analysis could also be applied to patients with hernia. Although the strain field depends not only on mechanical properties, but also on loading and boundary conditions, the knowledge about zones with a similar range of strains and the directions of principal strains  can facilitate the selection of an appropriate implant and its orientation in the abdominal wall. This will support  the development of implants tailored to a specific hernia location and even a specific person and would be a step forward to personalised medicine for the treatment of abdominal hernias}.

Moreover, the presented novel approach may be useful in analysis of full-field strain data from different deformation states in other applications. For example, the proposed SOM analysis of deformation fields, could also be applied to other soft tissue mechanics based on \textit{in vivo} experiments as well as other difficult to analyse anisotropic and nonlinear materials.

\section*{Acknowledgements}

We would like to thank Professor Monika Lichodziejewska-Niemierko and the staff of the Peritoneal Dialysis Unit Department of Nephrology Transplantology and Internal Medicine at the Medical University of Gdańsk and Fresenius Nephrocare  (dr Piotr Jagodzi\'nski, nurses Ms Gra\.zyna Szyszka and Ms Ewa Malek) for their help in obtaining the input data.

This work was supported by the National Science Centre (Poland) [grant No. UMO-2017/27/B/ST8/02518]. Calculations were carried out partially at the Academic Computer Centre in Gdansk.


\appendix

\section{ {Summary of cluster statistics}} \label{appendix_stat}

 {Median of strains $\varepsilon_1$, $\varepsilon_2$, $\varepsilon_{xx}$, $\varepsilon_{xx}$ and angle $\alpha$ in each cluster of each subject are presented in Tables \ref{tab_app_d1}--\ref{tab_app_d12}. }

\begin{table}[htbp]
  \centering
   \caption{ {Median values obtained for each cluster of subject D1}}
    \begin{tabular}{lrrrrrrrrrrr}
   \hline
     {} & \multicolumn{5}{c}{ { cluster no 1}} &  {} & \multicolumn{5}{c}{ { cluster no 2}} \\
\cmidrule{2-6}\cmidrule{8-12}          & \multicolumn{1}{c}{ {$\varepsilon_1$}} & \multicolumn{1}{c}{ {$\varepsilon_2$}} & \multicolumn{1}{c}{ {$\varepsilon_{xx}$}} & \multicolumn{1}{c}{ {$\varepsilon_{yy}$}} & \multicolumn{1}{c}{ {$\alpha$}} &  {} & \multicolumn{1}{c}{ {$\varepsilon_1$}} & \multicolumn{1}{c}{ {$\varepsilon_2$}} & \multicolumn{1}{c}{ {$\varepsilon_{xx}$}} & \multicolumn{1}{c}{ {$\varepsilon_{yy}$}} & \multicolumn{1}{c}{ {$\alpha$}}  \\
         & \multicolumn{1}{c}{ {[-]}} & \multicolumn{1}{c}{ {[-]}} & \multicolumn{1}{c}{ {[-]}} & \multicolumn{1}{c}{ {[-]}} & \multicolumn{1}{c}{ {[$^\circ$]}} &  {} & \multicolumn{1}{c}{ {[-]}} & \multicolumn{1}{c}{ {[-]}} & \multicolumn{1}{c}{ {[-]}} & \multicolumn{1}{c}{ {[-]}} & \multicolumn{1}{c}{ {[$^\circ$]}}  \\
\cmidrule{2-6}\cmidrule{8-12}
     {T1} &  {0.006} &  {-0.001} &  {-0.001} &  {0.006} &  {-80.2} &  {} &  {0.007} &  {0.000} &  {0.000} &  {0.007} &  {78.4} \\
     {T2} &  {0.011} &  {0.003} &  {0.004} &  {0.010} &  {-75.4} &  {} &  {0.019} &  {0.008} &  {0.008} &  {0.018} &  {71.1} \\
     {T3} &  {0.030} &  {0.004} &  {0.007} &  {0.026} &  {-72.9} &  {} &  {0.041} &  {0.012} &  {0.014} &  {0.039} &  {-78.5} \\
     {T4} &  {0.042} &  {0.004} &  {0.008} &  {0.039} &  {-77.3} &  {} &  {0.053} &  {0.014} &  {0.016} &  {0.050} &  {-75.0} \\
    \hline
    \end{tabular}%
  \label{tab_app_d1}%
\end{table}%

\begin{table}[htbp]
  \centering
   \caption{ {Median values obtained for each cluster of subject D2}}
    \begin{tabular}{lrrrrrrrrrrr}
   \hline
     {} & \multicolumn{5}{c}{ { cluster no 1}} &  {} & \multicolumn{5}{c}{ { cluster no 2}} \\
\cmidrule{2-6}\cmidrule{8-12}          & \multicolumn{1}{c}{ {$\varepsilon_1$}} & \multicolumn{1}{c}{ {$\varepsilon_2$}} & \multicolumn{1}{c}{ {$\varepsilon_{xx}$}} & \multicolumn{1}{c}{ {$\varepsilon_{yy}$}} & \multicolumn{1}{c}{ {$\alpha$}} &  {} & \multicolumn{1}{c}{ {$\varepsilon_1$}} & \multicolumn{1}{c}{ {$\varepsilon_2$}} & \multicolumn{1}{c}{ {$\varepsilon_{xx}$}} & \multicolumn{1}{c}{ {$\varepsilon_{yy}$}} & \multicolumn{1}{c}{ {$\alpha$}}  \\
         & \multicolumn{1}{c}{ {[-]}} & \multicolumn{1}{c}{ {[-]}} & \multicolumn{1}{c}{ {[-]}} & \multicolumn{1}{c}{ {[-]}} & \multicolumn{1}{c}{ {[$^\circ$]}} &  {} & \multicolumn{1}{c}{ {[-]}} & \multicolumn{1}{c}{ {[-]}} & \multicolumn{1}{c}{ {[-]}} & \multicolumn{1}{c}{ {[-]}} & \multicolumn{1}{c}{ {[$^\circ$]}}  \\
\cmidrule{2-6}\cmidrule{8-12}
     {T1} &  {0.014} &  {0.003} &  {0.013} &  {0.004} &  {-4.4} &  {} &  {0.007} &  {-0.001} &  {0.002} &  {0.004} &  {-48.9} \\
     {T2} &  {0.029} &  {0.009} &  {0.028} &  {0.012} &  {-12.8} &  {} &  {0.021} &  {0.007} &  {0.018} &  {0.009} &  {-23.0} \\
     {T3} &  {0.068} &  {0.025} &  {0.060} &  {0.034} &  {-22.4} &  {} &  {0.043} &  {0.008} &  {0.031} &  {0.023} &  {-40.6} \\
     {T4} &  {0.089} &  {0.034} &  {0.079} &  {0.046} &  {-26.1} &  {} &  {0.063} &  {0.016} &  {0.054} &  {0.029} &  {-27.8} \\
    \hline
    \end{tabular}%
  \label{tab_app_d2}%
\end{table}%

\begin{table}[htbp]
  \centering
   \caption{ {Median values obtained for each cluster of subject D3}}
    \begin{tabular}{lrrrrrrrrrrr}
   \hline
     {} & \multicolumn{5}{c}{ { cluster no 1}} &  {} & \multicolumn{5}{c}{ { cluster no 2}} \\
\cmidrule{2-6}\cmidrule{8-12}          & \multicolumn{1}{c}{ {$\varepsilon_1$}} & \multicolumn{1}{c}{ {$\varepsilon_2$}} & \multicolumn{1}{c}{ {$\varepsilon_{xx}$}} & \multicolumn{1}{c}{ {$\varepsilon_{yy}$}} & \multicolumn{1}{c}{ {$\alpha$}} &  {} & \multicolumn{1}{c}{ {$\varepsilon_1$}} & \multicolumn{1}{c}{ {$\varepsilon_2$}} & \multicolumn{1}{c}{ {$\varepsilon_{xx}$}} & \multicolumn{1}{c}{ {$\varepsilon_{yy}$}} & \multicolumn{1}{c}{ {$\alpha$}}  \\
         & \multicolumn{1}{c}{ {[-]}} & \multicolumn{1}{c}{ {[-]}} & \multicolumn{1}{c}{ {[-]}} & \multicolumn{1}{c}{ {[-]}} & \multicolumn{1}{c}{ {[$^\circ$]}} &  {} & \multicolumn{1}{c}{ {[-]}} & \multicolumn{1}{c}{ {[-]}} & \multicolumn{1}{c}{ {[-]}} & \multicolumn{1}{c}{ {[-]}} & \multicolumn{1}{c}{ {[$^\circ$]}}  \\
\cmidrule{2-6}\cmidrule{8-12}
     {T1} &  {0.018} &  {0.003} &  {0.006} &  {0.015} &  {-71.5} &  {} &  {0.010} &  {0.000} &  {0.003} &  {0.006} &  {21.0} \\
     {T2} &  {0.031} &  {0.005} &  {0.010} &  {0.027} &  {-71.9} &  {} &  {0.019} &  {0.001} &  {0.007} &  {0.011} &  {40.8} \\
     {T3} &  {0.074} &  {0.004} &  {0.013} &  {0.067} &  {-77.5} &  {} &  {0.040} &  {0.002} &  {0.014} &  {0.027} &  {27.5} \\
     {T4} &  {0.087} &  {0.002} &  {0.012} &  {0.080} &  {-77.3} &  {} &  {0.043} &  {0.000} &  {0.017} &  {0.031} &  {28.1} \\
    \hline
    \end{tabular}%
  \label{tab_app_d3}%
\end{table}%

\begin{table}[htbp]
  \centering
   \caption{ {Median values obtained for each cluster of subject D4}}
    \begin{tabular}{lrrrrrrrrrrr}
   \hline
     {} & \multicolumn{5}{c}{ { cluster no 1}} &  {} & \multicolumn{5}{c}{ { cluster no 2}} \\
\cmidrule{2-6}\cmidrule{8-12}          & \multicolumn{1}{c}{ {$\varepsilon_1$}} & \multicolumn{1}{c}{ {$\varepsilon_2$}} & \multicolumn{1}{c}{ {$\varepsilon_{xx}$}} & \multicolumn{1}{c}{ {$\varepsilon_{yy}$}} & \multicolumn{1}{c}{ {$\alpha$}} &  {} & \multicolumn{1}{c}{ {$\varepsilon_1$}} & \multicolumn{1}{c}{ {$\varepsilon_2$}} & \multicolumn{1}{c}{ {$\varepsilon_{xx}$}} & \multicolumn{1}{c}{ {$\varepsilon_{yy}$}} & \multicolumn{1}{c}{ {$\alpha$}}  \\
         & \multicolumn{1}{c}{ {[-]}} & \multicolumn{1}{c}{ {[-]}} & \multicolumn{1}{c}{ {[-]}} & \multicolumn{1}{c}{ {[-]}} & \multicolumn{1}{c}{ {[$^\circ$]}} &  {} & \multicolumn{1}{c}{ {[-]}} & \multicolumn{1}{c}{ {[-]}} & \multicolumn{1}{c}{ {[-]}} & \multicolumn{1}{c}{ {[-]}} & \multicolumn{1}{c}{ {[$^\circ$]}}  \\
\cmidrule{2-6}\cmidrule{8-12}
     {T1} &  {0.012} &  {0.006} &  {0.011} &  {0.007} &  {16.9} &  {} &  {0.008} &  {0.001} &  {0.003} &  {0.006} &  {-70.9} \\
     {T2} &  {0.036} &  {0.018} &  {0.031} &  {0.020} &  {11.9} &  {} &  {0.026} &  {0.003} &  {0.019} &  {0.010} &  {-36.8} \\
     {T3} &  {0.054} &  {0.029} &  {0.046} &  {0.036} &  {30.3} &  {} &  {0.029} &  {0.005} &  {0.020} &  {0.016} &  {-55.9} \\
     {T4} &  {0.069} &  {0.040} &  {0.053} &  {0.052} &  {39.3} &  {} &  {0.038} &  {0.007} &  {0.027} &  {0.019} &  {-48.8} \\
    \hline
    \end{tabular}%
  \label{tab_app_d4}%
\end{table}%

\begin{table}[htbp]
  \centering
   \caption{ {Median values obtained for each cluster of subject D5}}
    \begin{tabular}{lrrrrrrrrrrr}
   \hline
     {} & \multicolumn{5}{c}{ { cluster no 1}} &  {} & \multicolumn{5}{c}{ { cluster no 2}} \\
\cmidrule{2-6}\cmidrule{8-12}          & \multicolumn{1}{c}{ {$\varepsilon_1$}} & \multicolumn{1}{c}{ {$\varepsilon_2$}} & \multicolumn{1}{c}{ {$\varepsilon_{xx}$}} & \multicolumn{1}{c}{ {$\varepsilon_{yy}$}} & \multicolumn{1}{c}{ {$\alpha$}} &  {} & \multicolumn{1}{c}{ {$\varepsilon_1$}} & \multicolumn{1}{c}{ {$\varepsilon_2$}} & \multicolumn{1}{c}{ {$\varepsilon_{xx}$}} & \multicolumn{1}{c}{ {$\varepsilon_{yy}$}} & \multicolumn{1}{c}{ {$\alpha$}}  \\
         & \multicolumn{1}{c}{ {[-]}} & \multicolumn{1}{c}{ {[-]}} & \multicolumn{1}{c}{ {[-]}} & \multicolumn{1}{c}{ {[-]}} & \multicolumn{1}{c}{ {[$^\circ$]}} &  {} & \multicolumn{1}{c}{ {[-]}} & \multicolumn{1}{c}{ {[-]}} & \multicolumn{1}{c}{ {[-]}} & \multicolumn{1}{c}{ {[-]}} & \multicolumn{1}{c}{ {[$^\circ$]}}  \\
\cmidrule{2-6}\cmidrule{8-12}
     {T1} &  {0.004} &  {-0.006} &  {0.000} &  {-0.004} &  {-22.3} &  {} &  {0.004} &  {-0.031} &  {-0.003} &  {-0.026} &  {-23.1} \\
     {T2} &  {0.007} &  {-0.005} &  {0.004} &  {-0.003} &  {-20.2} &  {} &  {0.010} &  {-0.014} &  {0.007} &  {-0.011} &  {-11.1} \\
     {T3} &  {0.023} &  {-0.004} &  {0.018} &  {0.000} &  {-18.2} &  {} &  {0.112} &  {0.010} &  {0.017} &  {0.099} &  {74.2} \\
     {T4} &  {0.029} &  {-0.004} &  {0.024} &  {0.003} &  {-19.0} &  {} &  {0.126} &  {0.016} &  {0.025} &  {0.111} &  {73.5} \\
    \hline
    \end{tabular}%
  \label{tab_app_d5}%
\end{table}%

\begin{table}[htbp]
  \centering
   \caption{ {Median values obtained for each cluster of subject D6}}
    \begin{tabular}{lrrrrrrrrrrr}
   \hline
     {} & \multicolumn{5}{c}{ { cluster no 1}} &  {} & \multicolumn{5}{c}{ { cluster no 2}} \\
\cmidrule{2-6}\cmidrule{8-12}          & \multicolumn{1}{c}{ {$\varepsilon_1$}} & \multicolumn{1}{c}{ {$\varepsilon_2$}} & \multicolumn{1}{c}{ {$\varepsilon_{xx}$}} & \multicolumn{1}{c}{ {$\varepsilon_{yy}$}} & \multicolumn{1}{c}{ {$\alpha$}} &  {} & \multicolumn{1}{c}{ {$\varepsilon_1$}} & \multicolumn{1}{c}{ {$\varepsilon_2$}} & \multicolumn{1}{c}{ {$\varepsilon_{xx}$}} & \multicolumn{1}{c}{ {$\varepsilon_{yy}$}} & \multicolumn{1}{c}{ {$\alpha$}}  \\
         & \multicolumn{1}{c}{ {[-]}} & \multicolumn{1}{c}{ {[-]}} & \multicolumn{1}{c}{ {[-]}} & \multicolumn{1}{c}{ {[-]}} & \multicolumn{1}{c}{ {[$^\circ$]}} &  {} & \multicolumn{1}{c}{ {[-]}} & \multicolumn{1}{c}{ {[-]}} & \multicolumn{1}{c}{ {[-]}} & \multicolumn{1}{c}{ {[-]}} & \multicolumn{1}{c}{ {[$^\circ$]}}  \\
\cmidrule{2-6}\cmidrule{8-12}
     {T1} &  {-0.001} &  {-0.004} &  {-0.003} &  {-0.003} &  {33.7} &  {} &  {0.001} &  {-0.002} &  {-0.002} &  {0.001} &  {65.5} \\
     {T2} &  {0.018} &  {0.004} &  {0.014} &  {0.007} &  {-22.2} &  {} &  {0.026} &  {0.008} &  {0.012} &  {0.023} &  {61.8} \\
     {T3} &  {0.108} &  {0.026} &  {0.047} &  {0.087} &  {-64.0} &  {} &  {0.077} &  {0.013} &  {0.018} &  {0.073} &  {78.6} \\
     {T4} &  {0.133} &  {0.032} &  {0.060} &  {0.110} &  {-63.2} &  {} &  {0.106} &  {0.016} &  {0.022} &  {0.099} &  {78.7} \\
    \hline
    \end{tabular}%
  \label{tab_app_d6}%
\end{table}%

\begin{table}[htbp]
  \centering
   \caption{ {Median values obtained for each cluster of subject D7}}
    \begin{tabular}{lrrrrrrrrrrr}
   \hline
     {} & \multicolumn{5}{c}{ { cluster no 1}} &  {} & \multicolumn{5}{c}{ { cluster no 2}} \\
\cmidrule{2-6}\cmidrule{8-12}          & \multicolumn{1}{c}{ {$\varepsilon_1$}} & \multicolumn{1}{c}{ {$\varepsilon_2$}} & \multicolumn{1}{c}{ {$\varepsilon_{xx}$}} & \multicolumn{1}{c}{ {$\varepsilon_{yy}$}} & \multicolumn{1}{c}{ {$\alpha$}} &  {} & \multicolumn{1}{c}{ {$\varepsilon_1$}} & \multicolumn{1}{c}{ {$\varepsilon_2$}} & \multicolumn{1}{c}{ {$\varepsilon_{xx}$}} & \multicolumn{1}{c}{ {$\varepsilon_{yy}$}} & \multicolumn{1}{c}{ {$\alpha$}}  \\
         & \multicolumn{1}{c}{ {[-]}} & \multicolumn{1}{c}{ {[-]}} & \multicolumn{1}{c}{ {[-]}} & \multicolumn{1}{c}{ {[-]}} & \multicolumn{1}{c}{ {[$^\circ$]}} &  {} & \multicolumn{1}{c}{ {[-]}} & \multicolumn{1}{c}{ {[-]}} & \multicolumn{1}{c}{ {[-]}} & \multicolumn{1}{c}{ {[-]}} & \multicolumn{1}{c}{ {[$^\circ$]}}  \\
\cmidrule{2-6}\cmidrule{8-12} 
     {T1} &  {0.003} &  {-0.009} &  {0.001} &  {-0.004} &  {-24.3} &  {} &  {0.006} &  {-0.007} &  {0.000} &  {-0.002} &  {-43.9} \\
     {T2} &  {0.048} &  {0.002} &  {0.005} &  {0.044} &  {79.4} &  {} &  {0.017} &  {-0.002} &  {0.008} &  {0.007} &  {-38.0} \\
     {T3} &  {0.041} &  {0.003} &  {0.006} &  {0.036} &  {80.2} &  {} &  {0.017} &  {-0.002} &  {0.008} &  {0.007} &  {-34.1} \\
     {T4} &  {0.095} &  {-0.002} &  {0.004} &  {0.085} &  {81.5} &  {} &  {0.040} &  {-0.002} &  {0.012} &  {0.019} &  {-30.6} \\
    \hline
    \end{tabular}%
  \label{tab_app_d7}%
\end{table}%

\begin{table}[htbp]
  \centering
   \caption{ {Median values obtained for each cluster of subject D8}}
    \begin{tabular}{lrrrrrrrrrrr}
   \hline
     {} & \multicolumn{5}{c}{ { cluster no 1}} &  {} & \multicolumn{5}{c}{ { cluster no 2}} \\
\cmidrule{2-6}\cmidrule{8-12}          & \multicolumn{1}{c}{ {$\varepsilon_1$}} & \multicolumn{1}{c}{ {$\varepsilon_2$}} & \multicolumn{1}{c}{ {$\varepsilon_{xx}$}} & \multicolumn{1}{c}{ {$\varepsilon_{yy}$}} & \multicolumn{1}{c}{ {$\alpha$}} &  {} & \multicolumn{1}{c}{ {$\varepsilon_1$}} & \multicolumn{1}{c}{ {$\varepsilon_2$}} & \multicolumn{1}{c}{ {$\varepsilon_{xx}$}} & \multicolumn{1}{c}{ {$\varepsilon_{yy}$}} & \multicolumn{1}{c}{ {$\alpha$}}  \\
         & \multicolumn{1}{c}{ {[-]}} & \multicolumn{1}{c}{ {[-]}} & \multicolumn{1}{c}{ {[-]}} & \multicolumn{1}{c}{ {[-]}} & \multicolumn{1}{c}{ {[$^\circ$]}} &  {} & \multicolumn{1}{c}{ {[-]}} & \multicolumn{1}{c}{ {[-]}} & \multicolumn{1}{c}{ {[-]}} & \multicolumn{1}{c}{ {[-]}} & \multicolumn{1}{c}{ {[$^\circ$]}}  \\
\cmidrule{2-6}\cmidrule{8-12} 
     {T1} &  {0.011} &  {-0.001} &  {0.000} &  {0.010} &  {-55.9} &  {} &  {0.018} &  {0.003} &  {0.004} &  {0.017} &  {-74.5} \\
     {T2} &  {0.029} &  {0.005} &  {0.015} &  {0.020} &  {-38.1} &  {} &  {0.040} &  {0.018} &  {0.022} &  {0.035} &  {-63.8} \\
     {T3} &  {0.050} &  {0.016} &  {0.031} &  {0.038} &  {-47.4} &  {} &  {0.074} &  {0.036} &  {0.043} &  {0.063} &  {-65.4} \\
     {T4} &  {0.061} &  {0.018} &  {0.035} &  {0.046} &  {-47.5} &  {} &  {0.093} &  {0.045} &  {0.053} &  {0.076} &  {-65.0} \\
    \hline
    \end{tabular}%
  \label{tab_app_d8}%
\end{table}%

\begin{table}[htbp]
  \centering
   \caption{ {Median values obtained for each cluster of subject D9}}
    \begin{tabular}{lrrrrrrrrrrr}
   \hline
     {} & \multicolumn{5}{c}{ { cluster no 1}} &  {} & \multicolumn{5}{c}{ { cluster no 2}} \\
\cmidrule{2-6}\cmidrule{8-12}          & \multicolumn{1}{c}{ {$\varepsilon_1$}} & \multicolumn{1}{c}{ {$\varepsilon_2$}} & \multicolumn{1}{c}{ {$\varepsilon_{xx}$}} & \multicolumn{1}{c}{ {$\varepsilon_{yy}$}} & \multicolumn{1}{c}{ {$\alpha$}} &  {} & \multicolumn{1}{c}{ {$\varepsilon_1$}} & \multicolumn{1}{c}{ {$\varepsilon_2$}} & \multicolumn{1}{c}{ {$\varepsilon_{xx}$}} & \multicolumn{1}{c}{ {$\varepsilon_{yy}$}} & \multicolumn{1}{c}{ {$\alpha$}}  \\
         & \multicolumn{1}{c}{ {[-]}} & \multicolumn{1}{c}{ {[-]}} & \multicolumn{1}{c}{ {[-]}} & \multicolumn{1}{c}{ {[-]}} & \multicolumn{1}{c}{ {[$^\circ$]}} &  {} & \multicolumn{1}{c}{ {[-]}} & \multicolumn{1}{c}{ {[-]}} & \multicolumn{1}{c}{ {[-]}} & \multicolumn{1}{c}{ {[-]}} & \multicolumn{1}{c}{ {[$^\circ$]}}  \\
\cmidrule{2-6}\cmidrule{8-12} 
     {T1} &  {0.003} &  {-0.002} &  {-0.001} &  {0.002} &  {-77.3} &  {} &  {0.004} &  {-0.002} &  {-0.001} &  {0.003} &  {83.4} \\
     {T2} &  {0.007} &  {0.000} &  {0.005} &  {0.003} &  {-27.5} &  {} &  {0.006} &  {0.002} &  {0.004} &  {0.003} &  {-7.0} \\
     {T3} &  {0.023} &  {0.003} &  {0.020} &  {0.007} &  {3.6} &  {} &  {0.023} &  {0.003} &  {0.021} &  {0.005} &  {6.5} \\
     {T4} &  {0.030} &  {0.007} &  {0.025} &  {0.010} &  {-1.7} &  {} &  {0.029} &  {0.005} &  {0.026} &  {0.007} &  {1.3} \\
\hline
& \multicolumn{5}{c}{ { cluster no 3}} &  {} &       &       &       &       &  \\
\cmidrule{2-6}          & \multicolumn{1}{c}{ {$\varepsilon_1$}} & \multicolumn{1}{c}{ {$\varepsilon_2$}} & \multicolumn{1}{c}{ {$\varepsilon_{xx}$}} & \multicolumn{1}{c}{ {$\varepsilon_{yy}$}} & \multicolumn{1}{c}{ {$\alpha$}}  &  {} &       &       &       &       &  \\
  & \multicolumn{1}{c}{ {[-]}} & \multicolumn{1}{c}{ {[-]}} & \multicolumn{1}{c}{ {[-]}} & \multicolumn{1}{c}{ {[-]}} & \multicolumn{1}{c}{ {[$^\circ$]}}  &  {} &       &       &       &       &  \\
  \cmidrule{2-6}    {T1} &  {0.002} &  {-0.004} &  {-0.002} &  {0.001} &  {-62.4} &       &       &       &       &       &  \\
     {T2} &  {0.011} &  {0.005} &  {0.009} &  {0.008} &  {-28.3} &       &       &       &       &       &  \\
     {T3} &  {0.035} &  {0.015} &  {0.034} &  {0.017} &  {5.2} &       &       &       &       &       &  \\
     {T4} &  {0.046} &  {0.024} &  {0.044} &  {0.026} &  {5.0} &       &       &       &       &       &  \\
\cmidrule{1-6}    \end{tabular}%
  \label{tab_app_d9}%
\end{table}%

\begin{table}[htbp]
  \centering
  \caption{ {Median values obtained for each cluster of subject D10}}
    \begin{tabular}{lrrrrrrrrrrr}
   \hline
     {} & \multicolumn{5}{c}{ { cluster no 1}} &  {} & \multicolumn{5}{c}{ { cluster no 2}} \\
\cmidrule{2-6}\cmidrule{8-12}          & \multicolumn{1}{c}{ {$\varepsilon_1$}} & \multicolumn{1}{c}{ {$\varepsilon_2$}} & \multicolumn{1}{c}{ {$\varepsilon_{xx}$}} & \multicolumn{1}{c}{ {$\varepsilon_{yy}$}} & \multicolumn{1}{c}{ {$\alpha$}} &  {} & \multicolumn{1}{c}{ {$\varepsilon_1$}} & \multicolumn{1}{c}{ {$\varepsilon_2$}} & \multicolumn{1}{c}{ {$\varepsilon_{xx}$}} & \multicolumn{1}{c}{ {$\varepsilon_{yy}$}} & \multicolumn{1}{c}{ {$\alpha$}}  \\
         & \multicolumn{1}{c}{ {[-]}} & \multicolumn{1}{c}{ {[-]}} & \multicolumn{1}{c}{ {[-]}} & \multicolumn{1}{c}{ {[-]}} & \multicolumn{1}{c}{ {[$^\circ$]}} &  {} & \multicolumn{1}{c}{ {[-]}} & \multicolumn{1}{c}{ {[-]}} & \multicolumn{1}{c}{ {[-]}} & \multicolumn{1}{c}{ {[-]}} & \multicolumn{1}{c}{ {[$^\circ$]}}  \\
\cmidrule{2-6}\cmidrule{8-12}
     {T1} &  {0.011} &  {-0.008} &  {0.010} &  {-0.007} &  {0.5} &  {} &  {0.024} &  {0.008} &  {0.014} &  {0.018} &  {-39.5} \\
     {T2} &  {0.019} &  {-0.012} &  {0.017} &  {-0.011} &  {-0.3} &  {} &  {0.040} &  {0.011} &  {0.021} &  {0.031} &  {-49.9} \\
     {T3} &  {0.044} &  {-0.031} &  {0.042} &  {-0.030} &  {1.5} &  {} &  {0.138} &  {0.032} &  {0.049} &  {0.127} &  {-77.6} \\
     {T4} &  {0.056} &  {-0.028} &  {0.048} &  {-0.023} &  {-1.0} &  {} &  {0.159} &  {0.032} &  {0.053} &  {0.143} &  {-78.3} \\
\hline
& \multicolumn{5}{c}{ { cluster no 3}} &  {} &       &       &       &       &  \\
\cmidrule{2-6}          & \multicolumn{1}{c}{ {$\varepsilon_1$}} & \multicolumn{1}{c}{ {$\varepsilon_2$}} & \multicolumn{1}{c}{ {$\varepsilon_{xx}$}} & \multicolumn{1}{c}{ {$\varepsilon_{yy}$}} & \multicolumn{1}{c}{ {$\alpha$}}  &  {} &       &       &       &       &  \\
  & \multicolumn{1}{c}{ {[-]}} & \multicolumn{1}{c}{ {[-]}} & \multicolumn{1}{c}{ {[-]}} & \multicolumn{1}{c}{ {[-]}} & \multicolumn{1}{c}{ {[$^\circ$]}}  &  {} &       &       &       &       &  \\
  \cmidrule{2-6}
\cmidrule{2-6}     {T1} &  {0.018} &  {0.004} &  {0.008} &  {0.014} &  {58.9} &       &       &       &       &       &  \\
     {T2} &  {0.039} &  {0.009} &  {0.014} &  {0.032} &  {66.3} &       &       &       &       &       &  \\
     {T3} &  {0.097} &  {0.016} &  {0.024} &  {0.093} &  {77.8} &       &       &       &       &       &  \\
     {T4} &  {0.132} &  {0.023} &  {0.030} &  {0.119} &  {78.3} &       &       &       &       &       &  \\
\cmidrule{1-6}    \end{tabular}%
  \label{tab_app_d10}%
\end{table}%

\begin{table}[htbp]
  \centering
  \caption{ {Median values obtained for each cluster of subject D11}}
    \begin{tabular}{lrrrrrrrrrrr}
   \hline
     {} & \multicolumn{5}{c}{ { cluster no 1}} &  {} & \multicolumn{5}{c}{ { cluster no 2}} \\
\cmidrule{2-6}\cmidrule{8-12}          & \multicolumn{1}{c}{ {$\varepsilon_1$}} & \multicolumn{1}{c}{ {$\varepsilon_2$}} & \multicolumn{1}{c}{ {$\varepsilon_{xx}$}} & \multicolumn{1}{c}{ {$\varepsilon_{yy}$}} & \multicolumn{1}{c}{ {$\alpha$}} &  {} & \multicolumn{1}{c}{ {$\varepsilon_1$}} & \multicolumn{1}{c}{ {$\varepsilon_2$}} & \multicolumn{1}{c}{ {$\varepsilon_{xx}$}} & \multicolumn{1}{c}{ {$\varepsilon_{yy}$}} & \multicolumn{1}{c}{ {$\alpha$}}  \\
         & \multicolumn{1}{c}{ {[-]}} & \multicolumn{1}{c}{ {[-]}} & \multicolumn{1}{c}{ {[-]}} & \multicolumn{1}{c}{ {[-]}} & \multicolumn{1}{c}{ {[$^\circ$]}} &  {} & \multicolumn{1}{c}{ {[-]}} & \multicolumn{1}{c}{ {[-]}} & \multicolumn{1}{c}{ {[-]}} & \multicolumn{1}{c}{ {[-]}} & \multicolumn{1}{c}{ {[$^\circ$]}}  \\
\cmidrule{2-6}\cmidrule{8-12} 
     {T1} &  {0.003} &  {-0.002} &  {0.002} &  {-0.001} &  {-22.4} &  {} &  {0.014} &  {0.005} &  {0.007} &  {0.010} &  {26.4} \\
     {T2} &  {0.012} &  {0.004} &  {0.006} &  {0.011} &  {64.2} &  {} &  {0.022} &  {0.010} &  {0.015} &  {0.016} &  {-36.3} \\
     {T3} &  {0.015} &  {0.010} &  {0.012} &  {0.013} &  {38.6} &  {} &  {0.031} &  {0.013} &  {0.024} &  {0.017} &  {-20.8} \\
     {T4} &  {0.029} &  {0.014} &  {0.015} &  {0.027} &  {68.7} &  {} &  {0.040} &  {0.016} &  {0.031} &  {0.026} &  {-35.4} \\
\hline
& \multicolumn{5}{c}{ { cluster no 3}} &  {} &       &       &       &       &  \\
\cmidrule{2-6}          & \multicolumn{1}{c}{ {$\varepsilon_1$}} & \multicolumn{1}{c}{ {$\varepsilon_2$}} & \multicolumn{1}{c}{ {$\varepsilon_{xx}$}} & \multicolumn{1}{c}{ {$\varepsilon_{yy}$}} & \multicolumn{1}{c}{ {$\alpha$}}  &  {} &       &       &       &       &  \\
  & \multicolumn{1}{c}{ {[-]}} & \multicolumn{1}{c}{ {[-]}} & \multicolumn{1}{c}{ {[-]}} & \multicolumn{1}{c}{ {[-]}} & \multicolumn{1}{c}{ {[$^\circ$]}}  &  {} &       &       &       &       &  \\
  \cmidrule{2-6}
     {T1} &  {0.006} &  {0.000} &  {0.005} &  {0.001} &  {23.1} &  {} &       &       &       &       &  \\
     {T2} &  {0.023} &  {0.013} &  {0.015} &  {0.021} &  {69.5} &  {} &       &       &       &       &  \\
     {T3} &  {0.031} &  {0.021} &  {0.025} &  {0.026} &  {44.3} &  {} &       &       &       &       &  \\
     {T4} &  {0.048} &  {0.030} &  {0.033} &  {0.046} &  {66.6} &  {} &       &       &       &       &  \\
\cmidrule{1-6}
      \end{tabular}%
  \label{tab_app_d11}%
\end{table}%

\begin{table}[htbp]
  \centering
  \caption{ {Median values obtained for each cluster of subject D12}}
    \begin{tabular}{lrrrrrrrrrrr}
   \hline
     {} & \multicolumn{5}{c}{ { cluster no 1}} &  {} & \multicolumn{5}{c}{ { cluster no 2}} \\
\cmidrule{2-6}\cmidrule{8-12}          & \multicolumn{1}{c}{ {$\varepsilon_1$}} & \multicolumn{1}{c}{ {$\varepsilon_2$}} & \multicolumn{1}{c}{ {$\varepsilon_{xx}$}} & \multicolumn{1}{c}{ {$\varepsilon_{yy}$}} & \multicolumn{1}{c}{ {$\alpha$}} &  {} & \multicolumn{1}{c}{ {$\varepsilon_1$}} & \multicolumn{1}{c}{ {$\varepsilon_2$}} & \multicolumn{1}{c}{ {$\varepsilon_{xx}$}} & \multicolumn{1}{c}{ {$\varepsilon_{yy}$}} & \multicolumn{1}{c}{ {$\alpha$}}  \\
         & \multicolumn{1}{c}{ {[-]}} & \multicolumn{1}{c}{ {[-]}} & \multicolumn{1}{c}{ {[-]}} & \multicolumn{1}{c}{ {[-]}} & \multicolumn{1}{c}{ {[$^\circ$]}} &  {} & \multicolumn{1}{c}{ {[-]}} & \multicolumn{1}{c}{ {[-]}} & \multicolumn{1}{c}{ {[-]}} & \multicolumn{1}{c}{ {[-]}} & \multicolumn{1}{c}{ {[$^\circ$]}}  \\
\cmidrule{2-6}\cmidrule{8-12} 
     {T1} &  {0.017} &  {0.007} &  {0.012} &  {0.012} &  {-40.9} &  {} &  {0.005} &  {-0.003} &  {0.005} &  {-0.002} &  {-0.7} \\
     {T2} &  {0.054} &  {0.028} &  {0.039} &  {0.042} &  {-54.9} &  {} &  {0.023} &  {-0.002} &  {0.018} &  {0.001} &  {-1.1} \\
     {T3} &  {0.088} &  {0.052} &  {0.055} &  {0.086} &  {-80.6} &  {} &  {0.038} &  {0.011} &  {0.029} &  {0.016} &  {2.6} \\
     {T4} &  {0.120} &  {0.068} &  {0.071} &  {0.118} &  {-81.8} &  {} &  {0.055} &  {0.016} &  {0.041} &  {0.023} &  {1.2} \\

\hline
& \multicolumn{5}{c}{ { cluster no 3}} &  {} &       &       &       &       &  \\
\cmidrule{2-6}          & \multicolumn{1}{c}{ {$\varepsilon_1$}} & \multicolumn{1}{c}{ {$\varepsilon_2$}} & \multicolumn{1}{c}{ {$\varepsilon_{xx}$}} & \multicolumn{1}{c}{ {$\varepsilon_{yy}$}} & \multicolumn{1}{c}{ {$\alpha$}}  &  {} &       &       &       &       &  \\
  & \multicolumn{1}{c}{ {[-]}} & \multicolumn{1}{c}{ {[-]}} & \multicolumn{1}{c}{ {[-]}} & \multicolumn{1}{c}{ {[-]}} & \multicolumn{1}{c}{ {[$^\circ$]}}  &  {} &       &       &       &       &  \\
  \cmidrule{2-6}
     {T1} &  {0.017} &  {0.006} &  {0.009} &  {0.013} &  {10.8} &       &       &       &       &       &  \\
     {T2} &  {0.047} &  {0.022} &  {0.031} &  {0.033} &  {20.0} &       &       &       &       &       &  \\
     {T3} &  {0.085} &  {0.040} &  {0.049} &  {0.069} &  {75.2} &       &       &       &       &       &  \\
     {T4} &  {0.106} &  {0.050} &  {0.064} &  {0.086} &  {72.5} &       &       &       &       &       &  \\
      \cmidrule{1-6}
    \end{tabular}%
  \label{tab_app_d12}%
\end{table}%

\bibliography{mybibfile}

\end{document}